\PassOptionsToPackage{pdfpagelabels=false}{hyperref}
\documentclass[fleqn,usenatbib]{mnras}

\usepackage{newtxtext,newtxmath}
\usepackage{rotating}

\usepackage[T1]{fontenc}
\usepackage{ae,aecompl}

\usepackage{graphicx}	% Including figure files
\usepackage{amsmath}	% Advanced maths commands
\usepackage{siunitx}
\usepackage{booktabs}
\usepackage{multirow}
\usepackage{color, colortbl}
\usepackage[ruled,vlined]{algorithm2e}
\usepackage{lineno}
\let\vec\mathbf

\newcommand{\maglim}{\textsc{MagLim}\xspace}
\newcommand{\Balrog}{\textsc{Balrog}\xspace}
\newcommand{\deltaz}{\Delta z}

\newcommand{\mcal}{\textsc{metacalibration}}

\newcommand{\sqdeg}{{\rm deg}^2}

\newcommand{\vecp}{\ensuremath{\mathbf{p}}}

\newcommand{\vecs}{\ensuremath{\mathbf{s}}}
\newcommand{\sys}{${\rm Sys}(\textbf{s})$}

\newcommand{\wur}{\bar {w}_{\rm{ur}}}
\newcommand{\nz}{\textit{n(z)}\xspace}

\usepackage{xcolor}
\definecolor{Gray}{gray}{0.9}
\definecolor{myorange}{RGB}{255, 104, 51}

\usepackage{ulem}

\usepackage{xspace} %This allows to add extra space when needed in new commands

\newcommand{\redmagic}{RedMaGiC\ }

\title[DES Y3 Redshift Calibration]{Dark Energy Survey Year 3 Results: Redshift Calibration of the \maglim Lens Sample from the combination of SOMPZ and clustering and its impact on Cosmology}

\author[Giannini et al.]{
\parbox{\textwidth}{
\Large G.~Giannini$^{1,2,3}$\thanks{E-mail: giulia.giannini15@gmail.com}, 
A.~Alarcon$^{4}$,
M.~Gatti$^{5}$,
A.~Porredon$^{6,7}$,
M.~Crocce$^{8,9}$,
G.~M.~Bernstein$^{5}$,
R.~Cawthon$^{10}$,
C.~S{\'a}nchez$^{3}$,
C.~Doux$^{5}$,
J.~Elvin-Poole$^{6,7}$
% ...
\begin{center} (DES Collaboration) \end{center}
}
}

\date{Accepted XXX. Received YYY; in original form ZZZ}

\pubyear{2022}

\begin{document}
\label{firstpage}
\pagerange{\pageref{firstpage}--\pageref{lastpage}}

\maketitle

% Abstract of the paper (max 250 words)
\begin{abstract}
We present an alternative calibration of the \maglim lens sample 
redshift distributions from the Dark Energy Survey (DES) first three years of data (Y3). The new calibration is based on a combination of a Self-Organising Maps based scheme and clustering redshifts to estimate redshift distributions and inherent uncertainties, which is expected to be more accurate than the original DES Y3 redshift calibration of the lens sample. We describe in detail the methodology, we validate it on simulations and discuss the main effects dominating our error budget. The new calibration is in fair agreement with the fiducial DES Y3 \nz calibration, with only mild differences ($<3\sigma$) in the means and widths of the distributions. We study the impact of this new calibration on cosmological constraints, analysing DES Y3 galaxy clustering and galaxy-galaxy lensing measurements, assuming a $\Lambda$CDM cosmology. We obtain $\Omega_{\rm m} =   0.30\pm 0.04$, $\sigma_8  = 0.81\pm 0.07 $ and $S_8 =  0.81\pm 0.04$, which implies a $\sim 0.4\sigma$ shift in the $\Omega_{\rm}-S_8$  plane compared to the fiducial DES Y3 results, highlighting the importance of the redshift calibration of the lens sample in multi-probe cosmological analyses.
\end{abstract}


%%%%%%%%%%%%%%%%%%%%%%%%%%%%%%%%%%%%%%%%%%%%%%%%%%

%%%%%%%%%%%%%%%%% BODY OF PAPER %%%%%%%%%%%%%%%%%%

%_____________________INTRODUCTION_________________________
\section{Introduction}
\label{sec:intro}

The Dark Energy Survey (DES, \citealt{Flaugher2015}) is currently the largest photometric galaxy survey to date, spanning 5000 $\sqdeg$ of the southern hemisphere and having detected hundreds of millions of galaxies. Together with other ongoing and future galaxy surveys (e.g., Kilo-Degree Survey KIDS, \citealt{Kuijken2015}; Hyper Suprime-Cam HSC, \citealt{Aihara2018}; Vera Rubin Observatory Legacy Survey of Space and Time (LSST), \citealt{Abell2009}; Euclid, \citealt{Laureijs2011}), DES can achieve competitive constraints on cosmological parameters by studying both the spatial distribution of the detected galaxies and by measuring the tiny distortions in their observed shapes due to gravitational lensing effects induced by the large scale structure of the Universe. For instance, the analysis of the first three years (Y3) of DES data \citep{y3-3x2ptkp} placed tight constraints on cosmological parameters combining three different measurements of the two-point (3x2pt) correlation functions that involved galaxy positions and measured galaxy shapes. These measurements are namely:
\begin{enumerate}
\item Cosmic shear, i.e. the 2-point correlation function of galaxy shapes; the DES Y3 measurements \citep*{y3-cosmicshear1,y3-cosmicshear2} involve the angular correlation of $10^8$ galaxy shapes from the weak lensing sample \citep*{y3-shapecatalog}, divided into four tomographic bins. We refer to this as the ``source'' sample.
\item Galaxy clustering: the 2-point correlation function of the positions of bright galaxies (which we refer to as the ``lens'' sample) \citep{y3-galaxyclustering};
\item Galaxy-galaxy lensing: the cross-correlation function of galaxy shapes and the position of the galaxies of the lens sample \citep{y3-gglensing}.
\end{enumerate}

The modelling of each of these correlation functions requires knowledge of the redshift distributions (from hereafter \nz) of the two samples (lens and source galaxies), which have to be estimated with great accuracy in order to avoid biased cosmological results \citep{Huterer2006,Cunha2012,benjamin2013,huterer2013,dessv-photoz,Hildebrandt2017,desy1-photoz,Joudaki2019,Hildebrandt2020,tessore}. The optimal solution would be to avail ourselves of spectroscopic observations, providing an accurate redshift measurement of each targeted galaxy. Unfortunately, it is not feasible to obtain said spectra other than for a small fraction of the science sample, due to the required time and cost of the observing campaign. Cosmological surveys like DES therefore have to use for their redshift estimation measurements only a few, noisy, broad-band fluxes, requiring inventive methods to create robust and unbiased redshift calibration pipelines.

For the DES Y3 3x2pt analysis, two different lens samples were used. The first sample is defined by selecting luminous red galaxies through the \redmagic algorithm \citep{Rozo2016}, which retains galaxies with high quality photometric redshift, by fitting each galaxy to a red-sequence template. The galaxies passing the \redmagic selection have, however, a low number density, and the final sample comprises roughly 3,000,000 galaxies. The second sample slightly compromises on the redshift accuracy to the benefit of a larger number density. The \maglim\ sample \citep{y3-2x2maglimforecast} is a magnitude-limited sample with a number density more than 3 times greater than RedMaGiC, comprising roughly 10,000,000 galaxies. In the fiducial DES 3x2pt \citep{y3-3x2ptkp} and 2x2pt analyses \citep{y3-2x2ptaltlensresults} that rely on the \maglim\ sample, the redshift distributions of the sample have been characterised using the machine learning photometric redshift code Directional Neighbourhood Fitting (DNF, \citealt{DNF2016}). In its current implementation, the DNF code provides per-galaxy redshift estimates using nearest neighbour techniques. The redshift distributions were then further calibrated using clustering redshift (hereafter WZ), which relies on cross-correlation measurements with spectroscopic samples \citep{y3-lenswz}. This calibration step also placed uncertainties on the redshift distribution estimates, which were modelled by ``shifting'' and ``stretching'' the redshift distributions. 

This work presents an additional and more sophisticated calibration of the redshift distributions of the lens sample, and studies the impact of these new redshift distribution estimates on the cosmological constraints using DES Y3 galaxy clustering and galaxy-galaxy lensing measurements (2x2pt). In particular, we adopt an approach similar to the one adopted to characterise the redshift distributions of the DES Y3 weak lensing (WL) sample, presented in \cite*{y3-sompz,y3-sourcewz}. This methodology also combines photometric and clustering constraints to produce redshift estimates, and it is more powerful than the fiducial redshift calibration adopted for the lenses for a number of reasons. The photometric information is used to produce redshift estimates using a self-organizing-map-based scheme (hereafter SOMPZ), which allows a meticulous control over all the (known) potential sources of uncertainties affecting the estimates. The SOMPZ method works by leveraging the DES deep fields, which have deeper observations with additional photometric bands and overlap with many-band redshift surveys available. It is possible to reproduce realistic selection functions in the deep fields from the injection of galaxies into actual DES images using the sophisticated image simulation tool \Balrog\ \citep{y3-balrog}.
The SOMPZ method provides an ensemble of redshift \nz\ for a given galaxy sample, which captures the uncertainties in the redshift distributions at all orders (i.e., not only in the mean or width of the distributions). The clustering constraints are then incorporated through a rigorous joint likelihood framework where the clustering data is forward modelled as a function of the input \nz, and the specific WZ systematics are marginalized over. This scheme allows to draw \nz\ samples conditioned on both clustering and photometric measurements, improving the \nz\ estimates by correctly taking into account the significance of the information provided by each source of information. This combined approach has proven to be more robust than SOMPZ or WZ applied individually \citep*{y3-sourcewz}, as the combination exploits the complementarity of both methods and reduces the overall \nz\ uncertainty.

The paper is organised as follows. In section \ref{sec:data} we introduce all the samples used in this work, both on data and simulations. Simulated samples are used to validate the methodology. Section \ref{sec:methodology} summarises the SOMPZ+WZ methodology adopted in this paper, also outlining the differences with the ``standard'' SOMPZ+WZ methodology used to model the DES Y3 source redshift distributions \citep{y3-sompz,y3-sourcewz}. Section \ref{sec:uncertainty} is devoted to the characterisation of the method's uncertainties. Section \ref{sec:results} presents the redshift distributions \maglim\ sample produced using the techniques described in this work. Section \ref{cosmo_results} describes the impact of this new redshift calibration on cosmological parameters estimation and compares it to the ``fiducial'' constraints obtained using the DNF+WZ redshift calibration \citep{y3-2x2ptaltlensresults}. In Appendix \ref{app:buzz_maglim} we provide details on the construction of the \maglim sample in simulations. Appendix \ref{sec:resultssims} complements the paper with a validation of the methodology in simulations. In Appendix \ref{app:params} are listed the values of parameters and the prior functions used in the cosmological inference; Appendix \ref{sampling} discusses the impact of different redshift uncertainties marginalisation techniques on the cosmological parameters estimation.

\section{Data}
\label{sec:data}

We describe in this section the data and simulated products used in this work. The samples used in this work are the following:

\begin{itemize}
    \item the DES \maglim sample, used as lenses in the DES cosmological analysis. Characterising its redshift distribution is the main goal of this work;
    \item the DES deep field samples, which are observed in small fields by DES with deeper observations than wide field ones and where information from additional photometric bands are available. Deep fields are a key element of the SOMPZ methodology;
    \item the DES \Balrog\ sample; this sample consists of software-injected deep field galaxies into DES wide field images and is a key element of the SOMPZ methodology;
    \item the ``redshift'' samples, which are a collection of either spectroscopic or multi-band photometric samples collected by other surveys in the DES deep field region. The redshift samples are a key element in the SOMPZ methodology;
    \item BOSS/eBOSS spectroscopic galaxy catalogues; these are galaxies with spectroscopic redshift overlapping with the DES wide field footprint used for the WZ measurement;
    \item the DES WL sample, used as sources in the DES cosmological analysis; we use the WL sample here when presenting the impact of \maglim SOMPZ redshift distributions on the cosmological analysis results.
\end{itemize}

All of these samples in data have also been reproduced in simulation for testing purposes. 

\subsection{DES Year 3 Data}  
DES \citep{Flaugher2015} is a five broadband ($grizY$) photometric survey that mapped roughly $5000 \,\sqdeg$ of the southern sky, using a 570 megapixel camera \citep[DECam;][]{Flaugher2015} mounted on the 4 meter Blanco telescope at the Cerro Tololo Inter-American Observatory (CTIO) in Chile. In this work we use data from the first three years (out of six) of observations (Y3), which were taken from August 2013 to February 2016. The DES Data Management (DESDM) team was in charge of processing the raw images \citep{Sevilla2011,Morganson2018,DES_DR1}; full details are provided in \cite{y3-gold} and \cite*{y3-shapecatalog}. The main catalog upon all the DES samples are built is the DES gold catalog, obtained using observations in the \textit{griz} bands. Objects belonging to the gold catalog have passed a number of selections aimed at removing objects in problematic regions of the sky or anomalous detections (e.g., objects with pixels affected by saturation or truncation issues). The gold catalog consists of 388 millions objects \citep{y3-gold}. Each object comes with morphological and photometric measurements based on two different pipelines, the Multi-Object Fitting pipeline (MOF) and the Single-Object Fitting pipeline (SOF). The former performs a simultaneous multi-object, multi-epoch, multi-band fit to estimate morphology and photometric information; the latter does not perform the multi-object fit when it comes to crowded objects. The DES Y3 SOF implementation is faster and less prone to fit failures compared to the MOF pipeline, and it does not suffer from any significant loss in terms of accuracy \citep{y3-gold}.

\subsection{\maglim sample} 
The main galaxy sample considered in this work is the \maglim sample. The \maglim sample is a subset of the DES gold catalog and consists of bright galaxies selected with an \textit{ad-hoc} selection that optimises the number density and the redshift accuracy of the sample \citep{y3-2x2maglimforecast}. The \maglim sample spans the full DES Y3 wide field footprint, for a total of $\sim4143 $ deg$^2$. SOF magnitudes in the $riz$ bands\footnote{We exclude the $g$-band as its photometry is known to be affected by PSF estimation issues \citep{y3-piff}.} are used for the selection and photometry. The selection is meant to be linear in redshift and magnitude, and reads
\begin{equation}\label{eq:maglim_sel}
    i < 4 * z_{\rm mean} + 18 \\
    i > 17.5,
\end{equation}
where $m_i$ the i-band SOF magnitude and $z_{\rm mean}$ is a per-object redshift estimate from the photo-$z$ code DNF \citep{DNF2016}; see also next subsection). The sample is then further limited to the redshift range $0.2 < z_{\rm mean} < 1.05 $. This leads to a sample that ranges from $18.8< i_{\rm mag} <22.2$ %\giulia{add plot with the magnitude distribution of the \maglim sample, and maybe a comparison in the same plot with magnitude distribution of the WL sample.} 
The \maglim sample is divided into 6 tomographic bins using DNF $z_{\rm mean}$ and considering the following bin edges: $[0.2, 0.4, 0.55, 0.7, 0.85, 0.95, 1.05]$, with a total of a 10,716,506 galaxies, distributed across bins as summarised in Table \ref{tab:maglim_ndens}. The \maglim sample is used as lens sample in the galaxy-galaxy lensing and galaxy clustering measurements of the DES Y3 2x2 cosmological analysis \citep{y3-2x2ptaltlensresults}.

\begin{table}
    \centering
    \begin{tabular}{c c c c c}
        \hline
        \hline
        Bin & z range & N galaxies & n density & ${\rm C}_{\rm flux}$\\
        \hline
        1 & [0.20, 0.40] \, & \, 2 236 473 \, & \, 0.150 \, & \, 0.43 \\
        2 & [0.40, 0.55] \, & \, 1 599 500 \, & \, 0.107 \, & \, 0.30 \\
3 & [0.55, 0.70] \, & \, 1 627 413 \, & \, 0.109 \, & \, 1.75 \\
4 & [0.70, 0.85] \, & \, 2 175 184 \, & \, 0.146 \, & \, 1.94 \\
5 & [0.85, 0.95] \, & \, 1 583 686 \, & \, 0.106 \, & \, 1.56 \\
6 & [0.95, 1.05] \, & \, 1 494 250 \, & \, 0.100 \, & \, 2.96 \\

\hline
    \end{tabular}
    \caption{
    Summary of the \maglim sample. We have outlined for each tomographic bin the redshift range (selected using DNF $z_{mean}$), the number of galaxies, the number density, and the magnification coefficient as measured in \citealt{y3-2x2ptmagnification}
    }
    \label{tab:maglim_ndens}
\end{table}

%The limiting magnitudes are from DES DR1 paper.
\subsubsection{DNF}\label{sec:DNF}
The photo-$z$ code DNF (Directional Neighborhood Fitting) is used to define the \maglim selection and to define the \maglim tomographic bins. The DNF algorithm computes a point estimate $z_{\rm mean}$ of redshift of the galaxies by performing a fit to a hyper-plane in color and magnitude space using up to 80 nearest neighbors taken from a reference sample made of spectroscopic galaxies with secure redshift information. For this purpose, a large number of spectroscopic catalogs collected by \cite{Gschwend2018} has been used, including spectra from SDSS DR4 \citep{Abolfathi2017}, OzDES \citep{Lidman2020},  VIPERS \citep{Garilli2014}, and from the PAU spectro-photometric catalog \citep{Eriksen2019}. The total number of spectra used for training is $\sim10^5$. DNF also provides a redshift estimate $z_{\rm DNF}$ drawn from the redshift PDF for each individual galaxy, although only the quantity $z_{\rm mean}$ (used for the selection and for the binning) is of interest in this work.

\subsection{Deep Fields sample}

The Deep fields catalog is a key element of the SOMPZ methodology. We provide here a few key details, but we refer the reader to \cite{y3-deepfields} for extensive details and the characterisation of the sample.

This work uses four different deep fields, i.e., E2, X3, C3 and COSMOS (COS) covering 3.32, 3.29, 1.94, and 1.38 square degrees, respectively. Each deep field has undergone a scrupulous masking procedure aimed at removing artefacts (e.g., cosmic rays, meteors, saturated pixels, etc.). Considering the final unmasked area overlapping with the UltraVISTA and VIDEO near-infrared (NIR) surveys \citep{McCracken2012,Jarvis2013}, which is needed to provide photometric information in additional bands, we are left with 5.2 square degrees of area for a total of 267,229 galaxies with measured $u$, $g$, $r$, $i$, $z$, $J$, $H$, $K_s$ photometry with limiting magnitudes 24.64, 25.57, 25.28, 24.66, 24.06, 24.02, 23.69, and 23.58. Note that deep field galaxies have deeper photometry and photometry available in more bands compared to the wide field galaxies; this is key for a good performance of the SOMPZ method as it reduces the color-redshift degeneracy. 

\subsection{\Balrog\ sample}
The \Balrog\ sample is another key element of the SOMPZ methodology. It is used to relate galaxies with given deep photometry to observed galaxies with wide field photometry, which are noisier. To this aim we rely on \Balrog\ \citep{Suchyta2016}, a software which injects ``fake'' galaxies into real images. For this analysis, \Balrog\ was used to inject deep field galaxies into the broader wide field footprint \citep{y3-balrog}. After injecting galaxies into images, the output \Balrog\ images are passed into the DES Y3 photometric pipeline and injected galaxies are detected equivalently to real galaxies, yielding multiple realisations of each injected galaxy. The \Balrog\ sample spans $\sim$20\% of the DES Y3 footprint. We further select injected galaxies using the \maglim selection. We then construct a matched catalog matching \Balrog\ injected wide field \maglim galaxies with their deep field counterparts, for a total of 351,165 galaxies with both deep and wide photometric information. The resulting catalog is called the \Balrog\ sample.

\begin{figure}
    \centering
    \includegraphics[width=\linewidth]{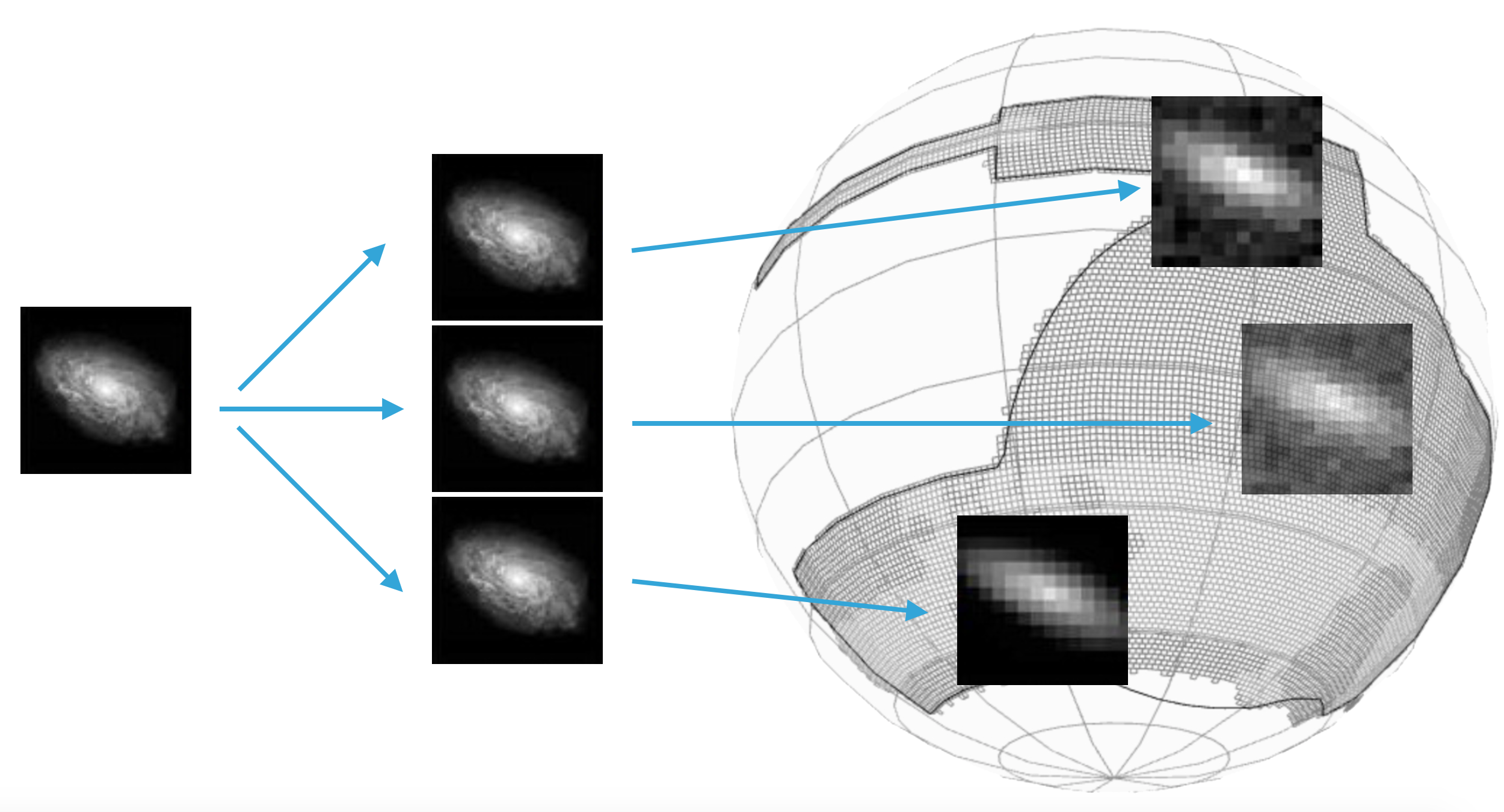}
    \caption{Scheme illustrating the operation of \Balrog: the practically noiseless deep fields galaxies are injected many times in DES real wide field images; those dichotomous images are then processed through the fiducial DES detection pipeline, to construct a sample containing several noisy representations of the same deep galaxies.}
    \label{fig:balrog}
\end{figure}

\subsection{Redshift Samples} \label{sec:redshift_samples}

The redshift samples used for the SOMPZ section of the analysis consist of galaxies with secure redshift information (either spectroscopic or high quality multi-band photometric) observed in the deep fields. These samples are key to characterise the redshifts of the deep field sample and, in turn, to transfer the redshift information to the wide field \maglim sample. 

We consider three separate redshift selections, similarly to what has been used in source sample redshift characterisation \citep*{y3-sompz}:
\begin{itemize}
    \item a collection of spectra from a number of different public and private spectroscopic samples, from the spectroscopic compilation by \cite{Gschwend2018}. We have not restricted ourselves to a few, selected surveys as in the case of the DES Y3 weak lensing sample  \citep*{y3-sompz}, where only zCOSMOS \citep{Lilly09zcosmos}, C3R2 \citep{Masters2017,C3R2_DR2}, VVDS \citep{vvds}, and VIPERS \citep{scodeggio2018} were considered, because due the bright nature of the \maglim sample we would mostly select high signal-to-noise galaxies. Furthermore, using more spectra from different surveys allow us to simultaneously reduce the shot noise and improve the completeness of the sample, while minimising the impact of possible outliers;
    \item multi-band photo-z galaxies from the COSMOS field; in particular, we used the \texttt{COSMOS2015} 30-band photometric redshift catalog \citep{Laigle2016}, which is equipped with narrow, intermediate and broad bands covering the IR, optical, and UV regions of the electromagnetic spectrum;
    \item redshifts from the \texttt{PAUS+COSMOS} 66-band photometric redshift catalog \citep{Alarcon2020}, which adds 40 narrow band filters from the PAU Survey.
\end{itemize}

We match these redshift catalogs to our deep field galaxies and keep only those that are selected at least once into our \maglim\ selection according to their \Balrog\ injections. 
Due to the bright nature of the \maglim sample, the number of galaxies in our final redshift samples is greatly reduced: for the SPC sample, for example, the unique total number of galaxies passes from 118745 to 17718, a reduction of around $85\%$.

In some cases, the same galaxy might have redshift information from multiple surveys. Following \cite*{y3-sompz}, we created three slightly different redshift samples, where in case of multiple information from different surveys we use as fiducial the redshift from a specific survey. The samples are:

\begin{itemize}
    \item 1) \textit{SPC}, where in case of multiple information available we first use the spectroscopic catalog (S), then \texttt{PAUS+COSMOS} (P), and finally \texttt{COSMOS2015} (C); 
    \item 2) \textit{PC}, where we rank first the \texttt{PAUS+COSMOS} catalog  before \texttt{COSMOS2015}, and we do not include spectroscopic redshifts; 
    \item 3) \textit{SC}: where we first use the spectroscopic catalog before \texttt{COSMOS2015}, but we do not include the \texttt{PAUS+COSMOS} catalog.
\end{itemize}

Table \ref{tab:specz} summarises the number of unique galaxies appearing in each of the three redshift samples, before and after performing the \maglim\ sample selection. The fiducial ensemble of redshift distributions is generated by marginalizing over all three of these redshift samples (SPC, PC, SC) with equal prior, which in practice is achieved by simply merging the $n(z)$ samples produced from the three redshift samples, creating a three times larger pool of \nz. In such a way we marginalise over potential uncertainties and biases in the different redshift catalogs (S, P and C).

\begin{table}%[Redshift Samples]
    \centering
    \begin{tabular}{l c c c c c c c}
        \hline 
        \hline
          & \multicolumn{3}{c}{Raw }  & \multicolumn{3}{c}{After MagLim selection } \\
        \hline  
           & SPC     & SC      & PC & SPC     & SC      & PC \\
        \hline
         Spec-z  & 35826  &   35826 &      -     & 10429 & 10429 & - \\
         PAU     & 18780  &    -     &    28780  & 3950 & - & 7015 \\
         COSMOS  & 64139  &  82856  &    69686  & 3299 & 7231 & 3721 \\
         \textbf{Total}   & 118745 &  118682 &    98466  & 17678 & 17660 & 10736 \\

         \hline
    \end{tabular}
    \caption{Number of unique galaxies belonging to each of the three redshift catalogs (spectroscopic collection, COSMOS, and PAU) for each of the samples SPC (composed by galaxies from spectra, PAU, COSMOS in this order), SC (spectra, COSMOS), PC (PAU, COSMOS). The sample selection for the \maglim\ sample applied to the corresponding \Balrog\ injections reduces greatly the size of all samples. For more information, see Section \ref{sec:redshift_samples}.}
    \label{tab:specz}
\end{table}

\subsection{BOSS/eBOSS Galaxy catalogs}

The BOSS/eBOSS galaxy catalog is our reference sample for the WZ measurement. It consists of a number of spectroscopic samples from the Sloan Digital Sky Survey (SDSS, \citealt{Gunn2006,Eisenstein2011,Blanton2017}), and combines SDSS galaxies from BOSS (Baryonic Oscillation Spectroscopic Survey, \citealt{Smee2013,boss}) and from eBOSS (extended-Baryon Oscillation Spectroscopic Survey \citealt{eboss2016,dr16,Alam2020}). In particular, the BOSS sample includes the LOWZ and CMASS catalogs from the SDSS DR 12 \citep{reid16}, while we included the large-scale structure catalogs from emission line galaxies (ELGs \citealt{raichoor17}), luminous red galaxies (LRGs, \citealt{prakash16}) and quasi stellar objects (QSOs) (eBOSS in prep.) from eBOSS. { Following \cite{y3-sourcewz,y3-lenswz}, we stack together the different samples and use them as a single reference sample. We also create a single random catalog by stacking all the random catalogs of each individual samples.} {The BOSS/eBOSS sample is divided into 50 bins spanning the $0.1<z<1.1$ range of the catalog (width $\Delta z = 0.02$). } The number of galaxies for each sample are listed in Table \ref{table:eboss}, with the final sample consisting of 241,987 objects and covering an area ranging from $14$ to $17\%$ of the total DES footprint.

\begin{table}
\begin{center}
    \begin{tabular}{c c c c}
      \hline
      \hline
      \multicolumn{4}{c}{Spectroscopic Samples} \\
      \hline
      Name & Redshifts & $N_{\rm gal }$ & Area \\
      \hline
      LOWZ (BOSS) & $z \in [0.0,0.5]$ & $45671$ & $\sim860 \  \text{deg}^2$   \\   
      CMASS (BOSS) & $z \in [0.35,0.8]$ & $74186$ & $\sim860 \  \text{deg}^2$  \\ 
      LRG (eBOSS) & $z \in [0.6,1.0]$ & $24404$ & $\sim700 \  \text{deg}^2$  \\   
      ELG (eBOSS) & $z \in [0.6,1.1]$ & $89967$ & $\sim620 \  \text{deg}^2$  \\   
      QSO (eBOSS) & $z \in [0.8,1.1]$ & $7759$ & $\sim700 \  \text{deg}^2$  \\   
      \hline
    \end{tabular}
  \caption{List of the spectroscopic samples from BOSS/eBOSS overlapping with the DES Y3 footprint used as reference galaxies for clustering redshifts in this work.}
  \label{table:eboss}
  \end{center}
\end{table}

We note that estimates of the magnification coefficients are not available for BOSS/eBOSS galaxies. For our fiducial analysis we assumed magnification values for the BOSS/eBOSS sample to be set to zero. We are confident about this choice for the narrow shape of the \maglim tomographic bins, since the magnification is usually significant in the tails of the distribution when the clustering kernel due to selection effects is larger. We nonetheless verify in this work that our analysis is not very sensitive to the particular choice of the values of the magnification parameters (see Section \ref{sec:magnif}). 
\subsection{Weak Lensing catalog}

The DES Y3 WL sample is used in this work as source in the galaxy-galaxy lensing measurement with the \maglim sample. The WL sample is created using the \mcal\ pipeline (described and tested in \citealt{HuffMcal2017} and \citealt{SheldonMcal2017} and applied to the Y3 data in \citealt*{y3-shapecatalog}) and it is a subset of the gold catalog. The \mcal\ pipeline provides a per-galaxy self-calibrated shape measurement, which is free from shear and selection biases. An additional, small calibration based on image simulations \citep{y3-imagesims} accounts for blending and detection biases. The final catalog consists of $\sim$ 100 million galaxies, spanning the full DES Y3 wide field footprint and with an effective number density of $n_{\rm eff} = 5.59$ gal/arcmin$^{-2}$. The WL sample is divided into four tomographic bin using the SOMPZ method \citep*{y3-sompz}.

\subsection{Simulated Galaxy catalogs}\label{sec:buzz_}

Our methodology is thoroughly validated using simulated catalogs. In particular, we use one realisation of the sets of the Buzzard N-body simulations \citep{y3-simvalidation}. All the catalogs we used in data have their simulated counterparts, although we adopted some reasonable simplifications, when needed. We give here a brief summary of the Buzzard simulation and the simulated catalog we had to create for this project, i.e., the simulated \maglim sample. The simulated BOSS/eBOSS catalog description is provided in \cite*{y3-sourcewz}, whereas the simulated WL sample is described in \cite{y3-simvalidation}.

Buzzard is a synthetic galaxy catalog built starting from N-body lightcones produced by L-GADGET2 \citep{springel2005}. Galaxies are incorporated in the dark matter lightcones using the ADDGALS algorithm \citep{DeRose2018}. Buzzard spans 10313 square degrees. The cosmological parameters chosen are $\Omega_{\rm m} = 0.286$, $\sigma_8 = 0.82$, $\Omega_b = 0.047$, $n_s = 0.96 $, $h = 0.7$. The simulations are created starting from three lightcones with different resolutions and size (1050$^3$, 2600$^3$ and 4000$^3$ Mpc$^3 h^{-3}$ boxes and 1400$^3$, 2048$^3$ and 2048$^3$ particles), to accommodate the need of a larger box at high redshift. Halos are identified using the public code ROCKSTAR \citep{Behroozi2013} and they are populated with galaxies using ADDGALS \citep{DeRose2018}, which provides positions, velocities, magnitudes, SEDs and ellipticities. Galaxies are assigned their properties based on the relation between redshift, $r$-band absolute magnitude, and large-scale density from a subhalo abundance matching model \citep{Conroy2006,Lehmann2017} in higher resolution N-body simulations. SEDs are assigned to galaxies by imposing the matching with the SED-luminosity-density relationship measured in the SDSS data. SEDs are $K$-corrected and integrated over the DES filter bands to generate DES $grizY$ magnitudes. Ray-tracing is performed through the CALCLENS algorithm \citep{Becker2013}, to introduce lensing effects, in order to provide weak-lensing shear, magnification and lensed galaxy positions for the lightcone outputs. CALCLENS is run onto the sphere, masked with the DES Y3 footprint, using the HEALPix algorithm \citep{healpix} and is accurate to $\sim 6.4$ arcseconds.

\subsubsection{Simulated \maglim sample}

In order to define a simulated \maglim sample, the photo-$z$ code DNF has been run on a subset of the Buzzard simulations, restricted to \textit{i}-band magnitudes $i < 23$, so as to reduce the running time without affecting the final result (note that the \maglim selection presents a cut at $i < 22.2$). The goal is to attain similar number density and color distributions as in data. We provide more detailed information on the adaptation to the sample selection for Buzzard in Appendix \ref{app:buzz_maglim}.

%$fi = (yi+1-yi) /(xi+1-xi) *(cat['DNF_ZMEAN']-xi)+yi$. \mg{is this equation explained? you could explain it}
%\mg{ potresi pure fare vedere la comparison con la distribuzione nei dati}

\subsubsection{Simulated Deep catalog}
The simulated true fluxes from Buzzard are used as the deep measurements, but we further assign a realistic error by using the limiting flux for each mock deep band. We use the same uncertainties as in data, but as the Buzzard simulation has a different zero point, those values have to be converted in magnitude using zero point of 30, and then is converted to a flux uncertainty for a zero point of 22.5, which is the zero point of the Buzzard fluxes.
We do not differentiate between fields, as it has been proven in \cite*{y3-sompz} that this had no impact on the simulated redshift distribution.
The size of the sample is 968759 galaxies. We use the true redshift for the redshift sample and to compare our inferred redshift distributions to the true ones.

\subsubsection{Simulated \Balrog\ catalog}

We mimic the \Balrog\ algorithm by randomly selecting positions over the full Y3 footprint and run the corresponding error model on the galaxies of the simulated deep catalog to obtain noisy versions, according to the exposure times of each location.  The deep galaxies can be injected an arbitrary number of times and we set this at 10. Only the wide counterparts of the deep galaxies that respect the \maglim selection defined in the Buzzard simulation are then included in the sample, yielding the final number of 250193 galaxies.

%\subsubsection{Simulated BOSS/eBOSS Galaxy catalogs}%\mg{you can mention this is the same simulated sample used in the WZ paper.}%We are using the same simulated sample used in the validation of the clustering redshifts paper for the sources \cite{gattigiannini}.%To replicate the spectroscopic BOSS/eBOSS sample in simulations, we selected bright galaxies with similar sky coverage and redshift distribution as the ones in data.  We did not try to further match other properties of the sample, {e.g. the galaxy-matter bias likely differs from that of the real data.  We note that the WZ methodology corrects for the reference bias, so at no point in the analysis of the real data are we assuming that the simulations have the same bias.} \gbso{We note that even if the galaxy-matter bias of the BOSS/eBOSS sample selected in simulation might differ from the one measured in data, all the methods implemented in this paper will correct for this.} 
\section{Redshift Inference Methodology}
\label{sec:methodology}

We describe in this section the methodology adopted in this work to infer the redshift distributions of the lens sample. The methodology is similar to the one adopted for the weak lensing sample \citep*{y3-sompz} and relies on two key techniques:

\begin{itemize}
\item photometric classification with Self-Organising Maps (SOM), known as the SOMPZ method \citep{y3-sompzbuzzard,y3-sompz}. The SOMPZ method takes advantage of the deeper photometry of 8 bands (ugrizJHKs) available in the DES deep fields, where galaxies with high-quality redshifts can be accurately classified in the deep color space, to ensure small selection biases, and well characterised redshift estimates and uncertainties of DES wide field galaxies;
\item clustering-based or clustering redshift techniques (WZ), more established in cosmology (\citealt{Newman2008,Menard2013}). The redshift distributions calibration is based on angular correlation with a reference sample with high-quality redshift estimates. This method is affected by systematic biases different than photometric methods, which makes this combination interesting and improves the robustness of our redshift estimates. For example, it does not require the spectroscopic sample used for calibration to be representative of the target sample. On the other hand, the galaxy bias evolution of the galaxy samples is involved, and magnification effects have to be taken into account.
\end{itemize}
These two techniques are combined together to provide an estimate of the redshift distributions of the lens sample. Such a combination is powerful because it exploits the complementarity of the two methods, which are affected by two very different sets of biases and uncertainties. We provide the key ingredients of these two techniques in the following sections, followed by a description of how the two are combined together.

We note that this method is an alternate method compared to the one presented in \cite{y3-2x2ptaltlensresults,y3-lenswz}, which provides redshift estimates combining photometric estimates from the photo-$z$ code DNF \citep{DNF2016} and clustering constraints from \cite{y3-lenswz}. We delay the comparison between the two methods to section \ref{comparisondnfsompz}.

%The combination of these two is performed with a Hamiltonian Monte Carlo, that computes the likelihood of obtaining the clustering redshift data from a $n(z)$ distribution retrieved with SOMPZ. The finite product of this combination is an ensemble of redshift distributions whose variance captures the uncertainties each method carries, but also the differences in constraining power.
%The cosmological likelihood is estimated by sampling from this ensemble. We compared two approaches for the sampling: (i) we use the algorithm called \hyperrank \cite{hyperrank} that categorises each component of the ensemble, correctly marginalizing over the differences in the distributions for the cosmological likelihood; (ii) in a simpler approach, we select as fiducial shape the average of the ensemble distributions and consequently marginalise only over the mean. 

\begin{figure*}
\centering
\includegraphics[width=15cm]{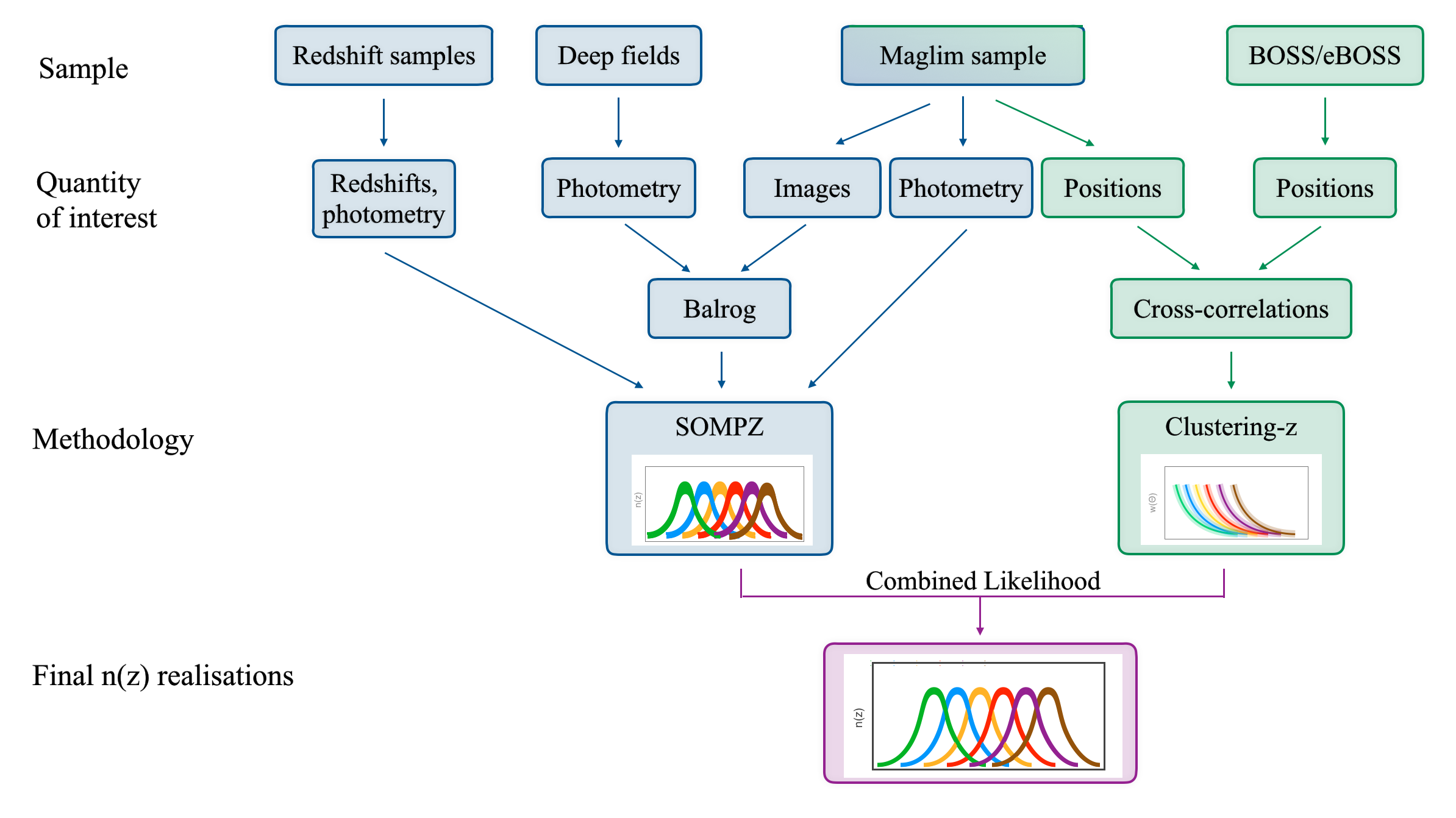}
\caption{
Flowchart illustrating the \maglim redshift distributions calibration scheme. The two methodologies included in the analysis are SOMPZ and clustering redshifts. Inspired by the flowchart in \citealt*{y3-sompz}.
}
\label{fig:flowchart}
\end{figure*}

\subsection{SOMPZ Methodology}

The SOMPZ methodology estimates wide field redshift distributions by exploiting a mapping between wide field galaxies and deep field galaxies with deeper and more precise photometry. Extracting the redshift information from deep, several band photometry in order to estimate the redshift of an observed wide field galaxy amounts to marginalizing over deep photometric information \citep{y3-sompzbuzzard}. Let us consider the probability distribution function for the redshift of a galaxy $p(z)$; let us assume such a probability to be conditioned on observed wide field color-magnitude $\mathbf{\hat{x}}$ and covariance matrix $\hat{\Sigma}$. %Let us further assume that such a probability is also conditioned on passing a selection function $\hat{s}$ which includes the choice of tomographic bin. 
The probability can be written by marginalizing over deep photometric color $\mathbf{x}$ as follows:
\begin{equation}
\label{eqn:marginal_pz}
    p(z|\mathbf{\hat{x}} , \mathbf{\hat{\Sigma}}) = \int d \mathbf{x} \ p(z| \mathbf{x}, \mathbf{\hat{x}}, \mathbf{\hat{\Sigma}}) p(\mathbf{x}|\mathbf{\hat{x}}, \mathbf{\hat{\Sigma}}).
\end{equation}
%
%
% \begin{equation}
% \label{eqn:marginal_pz}
%     p(z|\mathbf{\hat{x}} , \mathbf{\hat{\Sigma}}, \hat{s}) = \int d \mathbf{x} \ p(z| \mathbf{x}, \mathbf{\hat{x}}, \mathbf{\hat{\Sigma}}, \hat{s}) p(\mathbf{x}|\mathbf{\hat{x}}, \mathbf{\hat{\Sigma}} ,\hat{s})
% \end{equation}
%
The large dimensionality of this form prevents us from applying it to real situations. This problem can be circumvented by discretising the color space $\mathbf{x}$ and ($\mathbf{\hat{x}}$, $\mathbf{\hat{\Sigma}}$) in \textit{cells} $c$ and $\hat{c}$, each spanning a portion of the whole and representing a specific galaxy phenotype, respectively of the deep and wide field. The galaxy samples are arranged in cells/phenotypes using \textit{Self-Organizing Maps} (SOM) \citep{kohonensom}, which is an unsupervised machine learning technique used to produce a lower-dimensional representation of a complex data set, while preserving its core properties. The choice of the topology of the cells follows \cite{y3-sompzbuzzard}, where a two-dimensional representation of the color space was chosen as it ensures an immediate visualisation of the data not possible otherwise. Once we compressed our data in a more manageable set of information, we can write the $p(z)$ for the group of galaxies living in a particular wide cell $\hat{c}$. Since the \maglim\ tomographic bins $\hat{b}$ are already defined, we are going to construct one set of SOMs (one deep and one wide) for each bin. Assigning all galaxies belonging to a tomographic bin to a wide SOM is straight forward. In order to construct the deep SOM we have to use our \Balrog\ sample, consisting of all detected and selected \Balrog\ realisations of the galaxies in the wide field, each associated to its own ``noiseless'' replica in the deep sample. We therefore can assign to the deep SOM associated to a tomographic bin, galaxies whose \Balrog\ wide replica is selected in that specific wide bin. Therefore we can marginalize over deep field phenotypes $c$ as:
\begin{equation}
p(z|\hat{c}, \hat{b}) = \sum_{c} p(z|c, \hat{c}, \hat{b}) p(c|\hat{c}, \hat{b}).
\end{equation}
%
% \begin{equation}
% p(z|\hat{c}_{\hat{\rm b}}) = \sum_{c_{\hat{b}}} p(z|c_{\hat{b}}, \hat{c}_{\hat{b}}) p(c_{\hat{b}}|\hat{c}_{\hat{b}})
% \end{equation}
% %
At this point we want to marginalise over all wide cells $\hat{c}$ belonging to each tomographic bin. Again, we are computing $p(z|\hat{b})$ for each bin separately from different sets of SOMs:
%
% Note that $p(\hat{c}|\hat{s}) = p(\hat{c}|\hat{s},\hat{b})$ because all galaxies in a given cell $\hat{c}$ are assigned to exactly one bin $\hat{b}$. 
% \footnote{\giulia{Alex, I am not sure this is valid here as each wide som is a tomographic bin. -> } \aedit{Maybe write $p(\hat{c}|\hat{s},\hat{b})  \propto p(\hat{c}|\hat{s}), \forall \hat{c}\in\hat{b}$. Note that both $\sum_{\hat{c}}p(\hat{c}|\hat{s})=1$ and $\sum_{\hat{c}}p(\hat{c}|\hat{s},\hat{b})=1$, so they are proportional, not equal!}}
%
\begin{equation} 
    p(z|\hat{b}) \approx \sum_{\hat{c}} \sum_{c} p(z|c,\hat{c},\hat{b}) p(c|\hat{c},\hat{b}) p(\hat{c},\hat{b}).
\end{equation}
% \begin{equation} 
%     p(z|\hat{b}) \approx \sum_{\hat{c}_{\hat{b}}} \sum_{c_{\hat{b}}} p(z|c_{\hat{b}},\hat{c}_{\hat{b}})    p(c_{\hat{b}}|\hat{c}_{\hat{b}}) p(\hat{c}_{\hat{b}}).
% \end{equation}
% \begin{equation} 
%     p_{\hat{b}}(z) \approx \sum_{\hat{c}} \sum_{c} p(z|c,\hat{c}) p(c|\hat{c}) p(\hat{c})  \hfill    \forall \hat{b}
% \end{equation}
%
% \begin{equation} \label{eqn:bincond}
%     p(z|\hat{s}) \approx \sum_{\hat{c}} \sum_{c} p(z|c,\hat{c}, \hat{s}) p(c|\hat{c},\hat{s}) p(\hat{c}|\hat{s})
% \end{equation}
Unfortunately there are very few galaxies for each ($c$, $\hat{c}$) pair, and in many cases there are none. This makes the term $p(z|c, \hat{c})$ quite difficult to estimate. However, we can reasonably assume that the $p(z)$ for galaxies assigned to a given deep cell $c$ should not depend on the noisy wide photometry of that galaxy. Therefore we can relax the selection:
%
% \begin{equation} 
%     p(z|\hat{b}) \approx \sum_{\hat{c}_{\hat{b}}} \sum_{c_{\hat{b}}} p(z|c_{\hat{b}})    p(c_{\hat{b}}|\hat{c}_{\hat{b}}) p(\hat{c}_{\hat{b}}). \label{eqn:nobincond} 
% \end{equation} 
%
%
\begin{equation} 
    p(z|\hat{b}) \approx \sum_{\hat{c}} \sum_{c} p(z|c,\hat{b}) p(c|\hat{c},\hat{b}) p(\hat{c},\hat{b}).
\end{equation}\label{eqn:nobincond} 
% %
We use this approximation for our fiducial result. We obtain each of the terms appearing in Eq. \ref{eqn:nobincond} by placing galaxy samples to the SOM cells, as follows:

% In particular, we re-write \ref{eqn:nobincond} in such a way to highlight the relation of each term and the respective galaxy sample:

% %
% \begin{equation} \label{eqn:redshift_prob_samples}
%     p(z) \approx \sum_{\hat{c}} \sum_{c} \underbrace{p(z|c)}_\text{Redshift} \underbrace{p(c)}_\text{Deep} \underbrace{\frac{p(c,\hat{c})}{p(c)p(\hat{c})}}_\text{Balrog}  \underbrace{p(\hat{c})}_\text{Wide}, 
% \end{equation}
% %

\begin{itemize}
    \item $p(\hat{c})$ is computed collecting \textit{wide} field galaxies from the \maglim\ sample into a wide field SOM (one per tomographic bin);
    % \item $p(c)$ is obtained assigning the deep galaxy sample to a \textit{deep} SOM (again, one per tomographic bin);
    %\item $\frac{p(c,\hat{c})}{p(c)p(\hat{c})}$
    \item $p(c|\hat{c})$ is computed from the deep/\Balrog\ sample. It consists of all detected and selected \Balrog\ replicas of the deep galaxies injected in the wide field. We therefore can arrange the deep/\Balrog\ sample simultaneously into a wide and deep SOMs. We call this term the \textit{transfer function}. We weight the deep field galaxies according to their detection rate measured from \Balrog. An alternative to \Balrog\ would be using a sub-section of the wide field and deep fields overlap, giving us both deep and wide photometry for a limited number of galaxies. However, the area of overlap is small and the particular observing conditions found in this area will not be representative of the overall observing conditions found in the Y3 footprint as highlighted in \cite*{y3-sompz}.
    \item $p(z|c)$ is computed from the \textit{redshift} sample, which is a subset of the \textit{deep} sample, for which we have both credible redshifts, 8-band deep photometry, and thanks to \Balrog\ also wide-field realisations. 
\end{itemize}

% \begin{equation} \label{eqn:redshift_prob_samples}
%     p(z|\hat{b}, \hat{s}) \approx \sum_{\hat{c} \in \hat{b}} \sum_{c} \underbrace{p(z|c)}_\text{Redshift} \underbrace{p(c)}_\text{Deep} \underbrace{\frac{p(c,\hat{c})}{p(c)p(\hat{c})}}_\text{Balrog}  \underbrace{p(\hat{c})}_\text{Wide}, 
% \end{equation}
% %

% \begin{figure*}
% \centering
% \includegraphics[width=0.5\textwidth]{Figures/method.png}
% \caption{Visual representation of each term in the sompz inference methodology. (Top, Left): Wide SOM cells assigned to the second tomographic bin. (Middle, Left): Transfer Function for the selected wide SOM cell. (Bottom, Left): Three selected deep SOM cells with the highest contribution to the transfer function. (Right): the three inference terms of Equation \ref{eqn:nobincond}}. \label{fig:method}
% \end{figure*}

% %
% \begin{equation}
%     p(\hat{c}|\hat{s}) = \sum_{i \in \hat{c}} w_i R_i
% \end{equation}

% \begin{equation}
%     p(\hat{c}) \propto \sum_{i} \delta_{\hat{c},\hat{c}_i}/N_{i,inj}
% \end{equation}

% We again emphasise that the transfer function is computed from \Balrog\ realisations, not from the wide galaxy sample. In fact, only for the former both wide- and deep-field photometry are available.
% There is a sub-section of the deep galaxy sample which also has wide photometry, but these are very few galaxies and it would provide us with a noisy estimate. 

\subsubsection{SOM properties}

As in \cite{y3-sompzbuzzard} and \cite*{y3-sompz}, we use squared-shaped SOMs with $n$ cells for each side (for a total of $n \times n$ cells) and periodic boundaries, which makes the visualisation easier without compromising the efficiency. 
We parametrize the SOMs using luptitudes and lupticolors, following \cite{y3-sompzbuzzard}. Luptitudes are defined in \cite{lupton1999} as inverse hyperbolic sine transformation of fluxes:
\begin{equation}
    \mu = \mu_0 - a \sinh^{-1}{\frac{f}{2b}}\\   \mu_0 = m_0 - 2.5 \log b,      
\end{equation}
where \textit{m} are magnitudes, \textit{f} are fluxes, $a = 2.5 \log b$ and b is a softening parameter that defines at which scale luptitudes transition between logarithmic and linear behaviour.
For the deep SOM we compute 7 lupticolors with respect to the i-band
\begin{equation}
    \mu = (\mu_1-\mu_i, ..., \mu_7-\mu_i),
\end{equation}
where the index from 1-7 runs over the deep bands \textit{urgzJHK}. 
We avoid using the g-band for the wide field galaxies, as any observational systematics and chromatic effects are more evident in the g band. With only two lupticolors available in the wide SOM, we decided to add the i-band luptitude, as \cite{y3-sompzbuzzard} find empirically that addition of the luptitude improves the training performance:
\begin{equation}
    \mu = (\mu_i, \mu_r-\mu_i, \mu_z-\mu_i).
\end{equation}
The resolutions of the SOMs are 32x32 cells for the wide, and 12x12 cells for the deep. The reason behind the fewer cells in the deep SOM lies in the \maglim\ selection: the bright magnitude-redshift cuts must be applied also to the wide-component of the deep and redshift samples, and only the deep galaxies whose wide component is selected are included in the sample. This results in smaller deep and redshift samples covering a very small portion of the color space, compared to the weak lensing source sample \cite*{y3-sompz}. Also, reducing the number of cells means yielding more galaxies in each one. This is necessary in order to minimise the number of wide field galaxies assigned by the transfer function to a deep SOM cell with no redshift information. Reducing this number under $1\%$ is crucial to ensure that we get a correctly estimated redshift distribution for our sample. We note that shot noise caused by a small number of redshifts in a deep cell can play a significant role in biasing the estimate. We therefore performed a test to identify the optimal SOM size which would minimise these issues. We first computed several estimates in the Buzzard simulations using different resolutions for the deep SOM. We then evaluated which setting produced the smallest shift on the mean redshift with respect to the true value. 
As mentioned at the beginning of this section, SOMs require to be trained before being able to classify galaxies. After ensuring that the redshift samples and the \maglim\ sample span the same luptitude-lupticolor space (achieved using \Balrog\ to obtain the redshift samples wide photometry), we decided to use the redshift sample for the deep SOM training. We instead use the \maglim\ sample itself to train the wide SOM.

\subsection{WZ}

Clustering redshift is a widely used method (\citealt{Newman2008,Menard2013,Davis2017,Morrison2017,Scottez2018,Johnson2017}; \citealt*{Gatti2018}; \citealt{vandenBusch2020,Hildebrandt2020,y3-lenswz}; \citealt*{y3-sourcewz}) to infer or calibrate redshift distributions of galaxy samples. It relies on the assumption that the cross-correlation between two samples of objects is non-zero only in the case of overlap of the distribution of objects in physical space, due to their mutual gravitational influence.

Various implementations of the clustering redshift methodology differ in their details, but they all agree on one key aspect: the ``target'' sample (hereafter dubbed ``unknown'' sample), which has to be calibrated, has to be cross-correlated with a ``reference'' sample divided into thin redshift bins. The reference sample consists of  either high-quality photometric or spectroscopic redshift galaxies, and has to spatially overlap with the unknown sample.

Assuming linear galaxy-matter bias, we can express the clustering $w_{\rm ur}$ between the unknown sample and each of the reference sample thin bins as function of the separation angle $\theta$ between the unknown and reference sample:
%\mg{annoying, but when you write an equation, you need to use 'roman' indexes and suffixes {\rm}}
%
\begin{equation}
    w_{\rm ur}(\theta) = \int dz' n_{\rm r}(z') n_{\rm u}(z') b_{\rm r}(z') b_{\rm u}(z') w_{\rm DM}(\theta, z') + M(\theta),
\end{equation}
where $n_{\rm r}$ and $n_{\rm u}$ are the redshift distributions of the reference and unknown sample, $b_{\rm r}$ and $b_{\rm u}$ are the galaxy-matter biases of both samples, $w_{\rm DM}$ is the clustering of dark matter and $M(\theta)$ denotes contributions due to magnification. Note that we are assuming Limber approximation \citep{Limber}, but this has been shown to have no impact on the results \citep{McQuinn2013}. 

In our methodology, we use a single estimated value from the cross-correlation signal for each thin redshift bin. In practice, we do this by measuring the correlation function as a function of angular separation and then averaging it with a weight function to produce the single estimate:
\begin{equation}
\label{crosscor2}
\wur=\int_{\theta_{\min}}^{\theta_{\max}}d\theta\ W(\theta){w}_{\rm ur}(\theta) ,
\end{equation}
where $W(\theta) \propto \theta^{-1}$ is a weighting function \citep*{y3-sourcewz}. The integration limits in the integral in Eq.~\ref{crosscor2} are set to fixed physical scales (1.5 to 5 Mpc). 
%The advantage of this estimator rather than the \cite{} Landy & Szalay (1993) lies in the convenience of not having to create a random catalogue for the unknown sample, while producing negligible variations \cite{y3wz}. 

Since the $n_{\rm r}$ are binned in narrow bins we can approximate the number density of the sample of reference as a Dirac delta, and the revised expression becomes: 
\begin{equation}
    \label{eq:cl_easy}
    \wur \approx n_{\rm u} b_{\rm r} b_{\rm u} \bar{w}_{DM} + \bar{M}.
\end{equation}
The above equation relates the redshift distribution of the unknown sample to the measured clustering signal $\wur$. The galaxy-matter biases of the reference can be estimated from the autocorrelation of the reference sample. Usually the galaxy-matter bias of the unknown sample cannot be inferred and is treated as nuisance parameter. In this work, however, due to the relatively good redshift provided by DNF for the \maglim sample, we also use the autocorrelation of the latter as a prior for $b_{\rm u}$ (see section \ref{sec:wzunc}). The other terms in the above equation are the clustering of dark matter $\bar{w}_{\rm DM}$, which can be estimated from theory and it is not very sensitive to the cosmological parameters \citep*{y3-sourcewz}, and the magnification term, which is expected to have a little impact \citep*{y3-sourcewz} and can be estimated if magnification coefficients for the samples are provided.

%\subsubsection{Application to DES Year 3}

The angular scales considered have been chosen to span the physical interval between 1.5 and 5.0 Mpc. These bounds, applied to data as well as simulations, are selected so that the upper bound is below the range used for the galaxy clustering cosmological analyses, therefore granting the WZ likelihoods to be essentially independent of the assumed cosmology, and allowing us to produce n(z) samples in an MCMC chain that runs independently of the cosmological ones.
We perform the cross-correlations of \maglim\ with each of the 50 bins of width $\deltaz = 0.02$ of the BOSS/eBOSS catalog, which spans 0.1 < z < 1.1 as previously mentioned. We also weigh each galaxy of the \maglim\ sample by the clustering weights computed in \cite{y3-galaxyclustering}.

We use the \cite{DavisPeebles1983} estimator for the cross-correlation signal,
\begin{equation}
\label{DavisPeeblesestimator}
w_{\rm ur} (\theta) =\frac{N_{\rm Rr}}{N_{\rm Dr}}\frac{D_{\rm u}D_{\rm r}(\theta)}{D_{\rm u}R_{\rm r}(\theta)}-1,
\end{equation}
where $D_{\rm u}D_{\rm r}(\theta)$ and $D_{\rm u}R_{\rm r}(\theta)$ represent data--data and data--random pairs. The pairs are normalized through ${N_{\rm Dr}}$ and ${N_{\rm Rr}}$, which is the total number of galaxies in the reference sample and in the reference random catalog. The correlation estimates were computed using \texttt{treecorr}\footnote{https://github.com/rmjarvis/TreeCorr}.

\section{Characterization of Sources of Uncertainty}
\label{sec:uncertainty}

In this section, we present the characterisation of the systematic uncertainties of our methodology.
The dominant sources of uncertainties for the SOMPZ method are sample variance and shot noise. In the clustering redshift method, the main uncertainty is caused by the lack of prior knowledge on the redshift evolution of the galaxy-matter bias of the \maglim\ sample. This is modelled by a flexible systematic function, informed by a measurement of the \maglim\ auto-correlation function in data. Other, minor sources of uncertainties are related to magnification effects and the approximation of linear bias \citep*{y3-sourcewz}. We provide further details on each source of uncertainty in the following subsections. A full catalog-to-cosmology validation of the method (in simulations) is then presented in Appendix \ref{sec:resultssims}.

\subsection{SOMPZ uncertainties}\label{sompz_unc}
For the SOMPZ method we consider the following sources of uncertainty:

\begin{itemize}
    \item \textit{sample variance of the deep fields}: main uncertainty, caused by the limited area of the deep fields. We model the effect of  sample variance by means of the \textit{three step Dirichlet} (3sDir) analytical model described in \S \ref{sec:3sdir};%Accounted for by running the 3sDir analytical model described in \S \ref{sec:methodology};
    \item \textit{shot noise in the deep and redshift samples}: this is induced by the limited number of galaxies available in the deep and redshift samples. We model the effect of shot noise by means of the 3sDir analytical model described in \S \ref{sec:3sdir};
    \item \textit{SOMPZ method uncertainty}: this uncertainty stems from discretising the color space in the SOMPZ mapping. We do estimate its impact on the SOMPZ estimates by replicating the SOMPZ methods multiple times in simulations, and incorporate its effects by using Probability Integral Transforms (PITs) (\S~\ref{sec:sompzunc});
    \item \textit{photometric calibration}: related to uncertainties in the calibration of the deep fields zeropoint, it is accounted for in the SOMPZ estimates by means of PITs (\S~\ref{sec:zeropointunc}).
    \item \textit{redshift sample biases}: these biases stem from uncertainties and biases in the redshift estimates of the redshift samples. Their impact is accounted for in our methodology by marginalising over three different combinations of redshift samples (\S~\ref{sec:redshiftsampleunc});
    \item \textit{transfer function}: any bias induced by an erroneous estimation of the transfer function due to a size-limited \Balrog\ sample; we anticipate this to be negligible following the results from \cite*{y3-sompz} (\S~\ref{sec:balrogunc}).
\end{itemize}

In the following sections we will proceed to describe in detail how we account for each of the items listed above.

% \begin{figure}
%     \centering
%     \includegraphics[width=\linewidth]{Figures/unc_bar_mean_new.pdf}
%     \caption{Uncertainty in the mean redshift for the four identified sources of error, for the six tomographic bins, included in the final ensemble of n(z): zeropoint in red, the intrinsic uncertainty from SOMPZ in orange, sample variance in green, and the redshift sample uncertainty in blue.}
%     \label{fig:unc_bar}
% \end{figure}

\begin{table*}
    \centering
    \begin{tabular}{l c c c c c c}
        & & & Mean \\
        \hline
        \hline
        Uncertainty & Bin 1 & Bin 2 & Bin 3 & Bin 4 & Bin 5 & Bin 6\\
         & z $\in$ [0.2, 0.4] &  z $\in$ [0.4, 0.55] &  z $\in$ [0.55, 0.7] &  z $\in$ [0.7, 0.85] &  z $\in$ [0.85, 0.95] &  z $\in$ [0.95, 1.05] \\
      
        \hline
        Sample Variance \& shot noise & 0.015 & 0.010 & 0.010 & 0.008 & 0.009 & 0.009 \\
        SOMPZ method & 0.004 & 0.003& 0.005& 0.001 & 0.007 & 0.005\\
        Redshift samples & 0.009 & 0.001 & 0.006 & 0.003 & 0.004 & 0.007 \\
        Zeropoint & 0.008 & 0.007 & 0.004 & 0.005 & 0.005 & 0.005\\
        \hline
        SOMPZ & 0.315 $ \pm\ $ 0.015 & 0.445 $ \pm\ $ 0.010 & 0.630 $ \pm\ $ 0.010 & 0.776 $ \pm\ $ 0.008 & 0.895 $ \pm\ $ 0.009 & 0.983 $ \pm\ $  0.012 \\
    
        SOMPZ+WZ & 0.316 $ \pm\ $ 0.014 & 0.456 $ \pm\ $ 0.008 & 0.632 $ \pm\ $ 0.008 & 0.780 $ \pm\ $ 0.007 & 0.893 $ \pm\ $ 0.008 & 0.985 $ \pm\ $ 0.010\\
        \hline
        SOMPZ (with all unc) & 0.317 $ \pm\ $ 0.020 & 0.447 $ \pm\ $ 0.012 & 0.634 $ \pm\ $ 0.013 & 0.778 $ \pm\ $ 0.010 & 0.897 $ \pm\ $ 0.011 & 0.988 $ \pm\ $ 0.013\\
        
        SOMPZ+WZ (with all unc) & 0.315 $ \pm\ $0.016 & 0.463 $ \pm\ $ 0.010 & 0.633 $ \pm\ $ 0.009 & 0.781 $ \pm\ $ 0.008 & 0.893 $ \pm\ $ 0.009 & 0.990 $ \pm\ $ 0.012 \\
        \hline

        %0.0145 & 0.0100 & 0.0086 & 0.0084& 0.0093 & 0.0115\\
        \\
        & & & Width \\

        \hline
        \hline
        Sample variance \& shot noise & 0.007 & 0.005 & 0.003 & 0.003 & 0.004 & 0.009 \\
        SOMPZ method & 0.003 &  0.003 & 0.0007 & 0.0003 & 0.002 & 0.0001 \\
        Redshift samples & 0.001 & 0.005 & 0.0007 & 0.0006 & 0.0003 &  0.001 \\
        Zeropoint & 0.003 &  0.004 &  0.001 &  0.0004 & 0.001 & 0.001 \\
        \hline
        SOMPZ & 0.077 $ \pm\ $ 0.007 & 0.093 $ \pm\ $ 0.007 & 0.065 $ \pm\ $ 0.004 & 0.081 $ \pm\ $ 0.004 & 0.071 $ \pm\ $ 0.004 & 0.096 $ \pm\ $ 0.009 \\
        SOMPZ + WZ & 0.080 $ \pm\ $ 0.004 & 0.089 $ \pm\ $ 0.004 & 0.060 $ \pm\ $ 0.002 & 0.077 $ \pm\ $ 0.003 & 0.074 $ \pm\ $ 0.004 & 0.105 $ \pm\ $ 0.006\\
        \hline
        SOMPZ (with all unc) & 0.081 $ \pm\ $ 0.008 & 0.096 $ \pm\ $ 0.007 & 0.067 $ \pm\ $ 0.005 & 0.081 $ \pm\ $ 0.004 & 0.073 $ \pm\ $ 0.005 & 0.098 $ \pm\ $ 0.009\\
        SOMPZ + WZ (with all unc) & 0.080 $ \pm\ $ 0.005 & 0.081 $ \pm\ $ 0.005 & 0.060 $ \pm\ $ 0.002 & 0.073 $ \pm\ $ 0.003 & 0.074 $ \pm\ $ 0.004 & 0.101 $ \pm\ $ 0.007\\
        \hline
    \end{tabular}
    \caption{Summary of values for systematic uncertainties and center values for mean (top panel) and width (bottom panel) for the \nz\ distributions. The various components are computed as described in section \ref{sec:uncertainty} and as they are not completely independent it is not expected that they sum up to the total value. The values related to SOMPZ and SOMPZ+WZ refer to Figure \ref{fig:nz_hmc_dataa}, and include only the 3sDir uncertainty due to sample variance and shot noise (and the redshift samples uncertainty), because it was logistically not possible to add the SOMPZ method and the zeropoint sources of uncertainty before the combination with WZ. As a comparison, the ``SOMPZ (with all unc)'' includes all uncertainties. The final \nz\ which has been used in the cosmological analysis is the bottom line.  }
    \label{tab:unc}
\end{table*}

\subsubsection{Sample variance and shot noise (\textit{3sDir})}
\label{sec:3sdir}
Sample variance is the dominant uncertainty affecting our SOMPZ estimates, and stems from the limited size and area coverage of the redshift and deep samples, with respect to the whole wide field. The deep fields only cover $\sim 9 {\rm deg}^2$, which means we could be learning the color/redshift relation from a non-representative sample of the sky due to fluctuations in the matter density field; moreover, the finite size of the redshift sample can introduce shot noise effects, preventing a correct sampling of the quantities required for the redshift inference. %cannot sample well the true distribution for statistical reason

Generally the impact of sample variance can be evaluated estimating the redshift distributions in simulations multiple times using different line of sights for the deep fields (e.g. \citealt{Hildebrandt2017}, \citealt{Hildebrandt2020}; \citealt{desy1-photoz}; \citealt{y3-sompzbuzzard}; \citealt{wright2020}). Although we also performed a test where we evaluated the impact of sample variance using the Buzzard simulation, in our standard procedure we use the \textit{three step Dirichlet} (3sDir) approach \textit{3sDir} presented in \cite{Sanchez2020} and applied to the redshift calibration of the DES Year 3 source sample \citep*{y3-sompz}.

The 3sDir method consists of an analytical sample variance model predicting what the redshift-color distribution would be from the observed individual redshift and galaxy phenotypes (colors) of galaxies coming from smaller deep fields. Using this model we can build an ensemble of redshift distributions realisations whose fluctuations realistically represent the effect of sample variance. During the cosmological inference, by sampling over these realisations, one can effectively marginalise over the effect of sample variance. Here we provide a short description of the 3sDir method, but we direct the reader to \cite*{y3-sompz} and \cite{Sanchez2020} for more details. The 3sDir method assumes the probability $p(z,c)$ that galaxies belong to a redshift bin $z$ and color phenotype $c$ to be described by a probability histogram with coefficients $f_{zc}$ (with $\sum f_{zc} = 1$ and $ 0 \leq f_{zc} \leq 1$). Under this assumption, the expected number counts of galaxies in a deep SOM cell given the coefficients $f_{zc}$ are described by a multinomial distribution; if we assume a Dirichlet function for the prior on $f_{zc}$, the posterior of $f_{zc}$ given the observed number count will also be described by a Dirichlet function. Such a Dirichlet posterior can be used to draw samples and naturally accounts for the effect of shot noise in the data. The effect of sample variance can be introduced by tuning the width of the prior on $f_{zc}$, which does not change the expected value for $f_{zc}$ in the Dirichlet distribution, but does change its variance to simultaneously account for shot noise and sample variance.

If all the galaxies belonging to the redshift sample were independently drawn, then a Dirichlet distribution parametrized by the redshift sample counts in each couple of redshift bin $z$ and phenotype $c$, $N_{zc}$, would fully characterize $f_{zc}$. However, one subtlety is that sample variance correlates with redshifts; to increase the variance with the correct redshift dependence one can use the fact that two different phenotypes (deep SOM cells) overlapping in redshift are correlated due to the same underlying large-scale structure fluctuations. The 3sDir model assumes that phenotypes at the same redshift share the same sample variance, and therefore groups cells with similar redshifts in \textit{superphenotypes T}. One can then express the \textit{$f_{zc}$} as:
\begin{equation}
    f_{zc} = \sum{f_c^{zT}f_z^{T}f_T}.
\end{equation}
The 3sDir method consists of drawing values of these three sets of coefficients with three Dirichlet functions. In this way, it is possible to include a redshift-dependent variance while conserving the expected value of \textit{$f_{zc}$}.

% In DES Y3 the Redshift sample spans a smaller area than the whole deep field area, and the deep field carries important information regarding the distribution of colors, $p(c)$. We modify the 3sDir method to account for this, using the counts from the deep field to sample $\{f_T\}$ during step 1. Similarly, we update $\bar{\lambda}$ in step 1 to use a $\lambda_z$ that uses a sample variance theory prediction from the deep field area, $\Delta^{\D}_z$. This implementation is the most suitable for sampling the 3sDir likelihood jointly with the WZ likelihood in an Hamiltonian Monte-Carlo chain. 

The validation of the \textit{3sDir} method has been carried out in \cite*{y3-sompz}, applied to the weak lensing source sample. The only difference with this work stands in the fact we are performing the 3sDir estimation independently for each tomographic bin, due to their definition. 

As reported in Table \ref{tab:unc}, this uncertainty is dominant, both on the mean and width values of the \nz\ distributions, computed from the ensemble of realisations provided by the \textit{3sDir} method. 

\subsubsection{SOMPZ Method Uncertainty} \label{sec:sompzunc}

The SOMPZ method relies on the discretisation on the color space spanned by our deep field sample, and this is an approximation that can lead to small biases or additional uncertainties. In order to estimate these,  we compute our SOMPZ \nz\ a large number of times in the Buzzard simulations. In order to factor out sample variance, each time we randomly select patches of the Buzzard footprint to construct the mock deep fields. In this way, by averaging over all the final \nz\ realisations, we can produce an estimate of the \nz\ only minimally biased by sample variance, and test the agreement with the true \nz\ . Due to the computational cost of the SOMPZ pipeline, we decided to produce 300 \nz\ replicas. To perform this test, we assumed that the redshift sample would only be limited to one of our four fields, of the size of COSMOS. 

We computed the mean redshift offset of the ensemble with respect to the true value, for each tomographic bin. As reported in Table \ref{tab:unc}, these values are smaller than the effect of sample variance. These values are incorporated into our final \nz\ ensemble using the PIT method described in the following section, by additionally shifting each probability integral transform (used to correct for the zeropoint uncertainties) by a value drawn from a Gaussian centered at zero with standard deviation equal to the root-mean-square of the aforementioned mean offset values.  
\subsubsection{Deep Fields Photometric Calibration Uncertainty} \label{sec:zeropointunc}
% \begin{figure}
%     \centering
%     \includegraphics[width=\linewidth]{Figures/PIT_shifts.png}
%     \caption{Shifts on the average of the mean of the ensemble distributions before and after applying the PIT calibration show a negligible impact}
%     \label{fig:pit_shifts}
% \end{figure}
Although the uncertainty in the photometry of each individual galaxy is implicitly accounted for in the SOM training, the uncertainty on the photometric calibrations as a whole must be evaluated by testing how the measured $n(z)$ are affected by changes in the photometric zeropoint in each band. This is relevant for the deep fields, where the relatively precise fluxes are key to constraining reliable $p(z)$ in parts of parameter space that are not subject to selection biases. Ideally, this would be tested by rerunning the full analysis for an ensemble of perturbations of the photometric zeropoint according to the zeropoint uncertainty, but the computational requirements of the \Balrog\ injection procedure make this infeasible. Instead, we produce an analogous ensemble of realizations in simulations, where the \Balrog\ mock photometric survey is reduced to a computationally simpler procedure of adding Gaussian noise to true magnitudes. For each realization of this ensemble, we perturb all deep field magnitudes by a draw from a Gaussian whose width is determined by the photometric zeropoint uncertainty in the Y3 deep fields catalog in a specified band, as computed in \cite{y3-deepfields}. We then ``inject'' these perturbed deep field fluxes with a mock \Balrog\ procedure to generate wide field realizations of the galaxies and measure the corresponding $n(z)$. In this way we generate a full ensemble of $n(z)$ realisations reflecting the uncertainty in our redshift calibration due to the photometric calibration. We apply Probability Integral Transforms (PITs) as in \cite*{y3-sompz} to transfer the variation encoded in the ensemble from simulated $n(z)$ to our fiducial data result. Essentially, this process involves calculating the inverse cumulative distribution function (CDF)
% denoted as $F^{-1}_i$, 
for each simulated realization $n_i(z)$ in the ensemble. The PIT is then obtained by computing the difference between the CDF of each realization and the average CDF of the entire ensemble. To apply these transformations to the data, the PIT value is added to the inverse CDF of the fiducial data $n(z)$. The PIT resulting from a single draw of zero-point offsets is determined and collectively applied to all tomographic bins. More details on this new implementation of the PIT can be found in \cite{pitpz}.

% We begin by computing the inverse cumulative distribution function $F^{-1}$ for each of the ensemble.
% The Probability Integral Transform can be written:
% \textbf{finish this}
% For the zeropoint i computed the quadratic difference of the std of the means of the final pit ensemble and the hmc ensemble.

% Applied on the realisations after the combination with WZ.
% \begin{algorithm}
% \KwResult{An ensemble B of $n(z)$ about a fiducial $n_{\mathrm{data}}(z)$ with variation from an ensemble A of $n_{\mathrm{sims}}(z)$}
% Compute $F^{-1}_{\langle \cdot \rangle} \gets \langle F^{-1}_{i \in A} \rangle$\;
% Compute $F^{-1}_{\mathrm{data}}$\;
% \For{$i\in A$}{
%         Compute $F^{-1}_{i, \mathrm{sims}}$\;
%         Compute $\Delta_i \gets F^{-1}_i - F^{-1}_{\langle \cdot \rangle}$\;
%         Set $F^{-1}_{i, \mathrm{data}} \gets F^{-1}_{i, \mathrm{sims}} + \Delta_i$\;
%         Differentiate: $ n_{i, \mathrm{data}}(z)\gets \frac{d}{dz} F^{-1}_{i, \mathrm{data}}$\;}
% \end{algorithm}

\subsubsection{Redshift Sample uncertainty} \label{sec:redshiftsampleunc}

As mentioned in Section \ref{sec:redshift_samples}, we decided to choose three different catalogs to infer our redshift distributions from: a collection of spectroscopic surveys galaxies \citep{Gschwend2018}, PAU+COSMOS redshift as in \cite{Alarcon2020}, and COSMOS30 photometric redshifts \citep{Laigle2016}. The reason for availing ourselves of more than one catalog lies in the fact neither of these are exempt from systematic uncertainties: each survey uses different photometry, different model assumptions, and can be affected systematically by selection effects, incorrect templates, photometric outliers, etc.
Since there is a considerable overlap in the number of galaxies belonging to more than one of the redshift catalogs selected for this work, to account for the intrinsic biases we decided to build three samples which are combinations of the aforementioned catalogs. We ranked the redshift catalogs differently for each sample: if a galaxy has information from multiple origins, we assign the redshift from the highest ranked catalog. The three redshift samples \textit{SPC, PC, SC}, are described in Section \ref{sec:redshift_samples}.

%
%\begin{itemize}
%    \item \textit{SPC/OptPrime}: This sample ranks first the spectroscopic %catalog, then \texttt{PAUS+COSMOS}, and finally \texttt{COSMOS2015}. This %sample is designed to inform cosmological results that are mostly derived %from spectroscopic redshifts. 
%    \item \textit{PC/PhOpt}: This sample ranks first the \texttt{PAUS+COSMOS} %catalogue before \texttt{COSMOS2015}, and does not include spectroscopic %redshifts. In this way the cosmology will be maximally reliant on many-band %photometric redshifts.
%    \item \textit{SC/Opt}: This sample ranks first the spectroscopic catalogue  %before \texttt{COSMOS2015}, and does not include the \texttt{PAUS+COSMOS} %catalog. This sample does not exploits on the relatively new %\texttt{PAUS+COSMOS} data.
%\end{itemize}
% 

For each of these, we will perform the complete pipeline, and the final set of realisation will be constructed by an equal fraction $p(R)=1/3$ from each survey. By placing equal prior probability to each sample, this is equivalent as saying that we do not believe any of the samples is more likely to be correct. But note that for galaxies from which we have information from only one catalog, we are assuming that information to be true, and this is a caveat of this approach. 

\begin{figure}
    \centering
    \includegraphics[width=1.\linewidth]{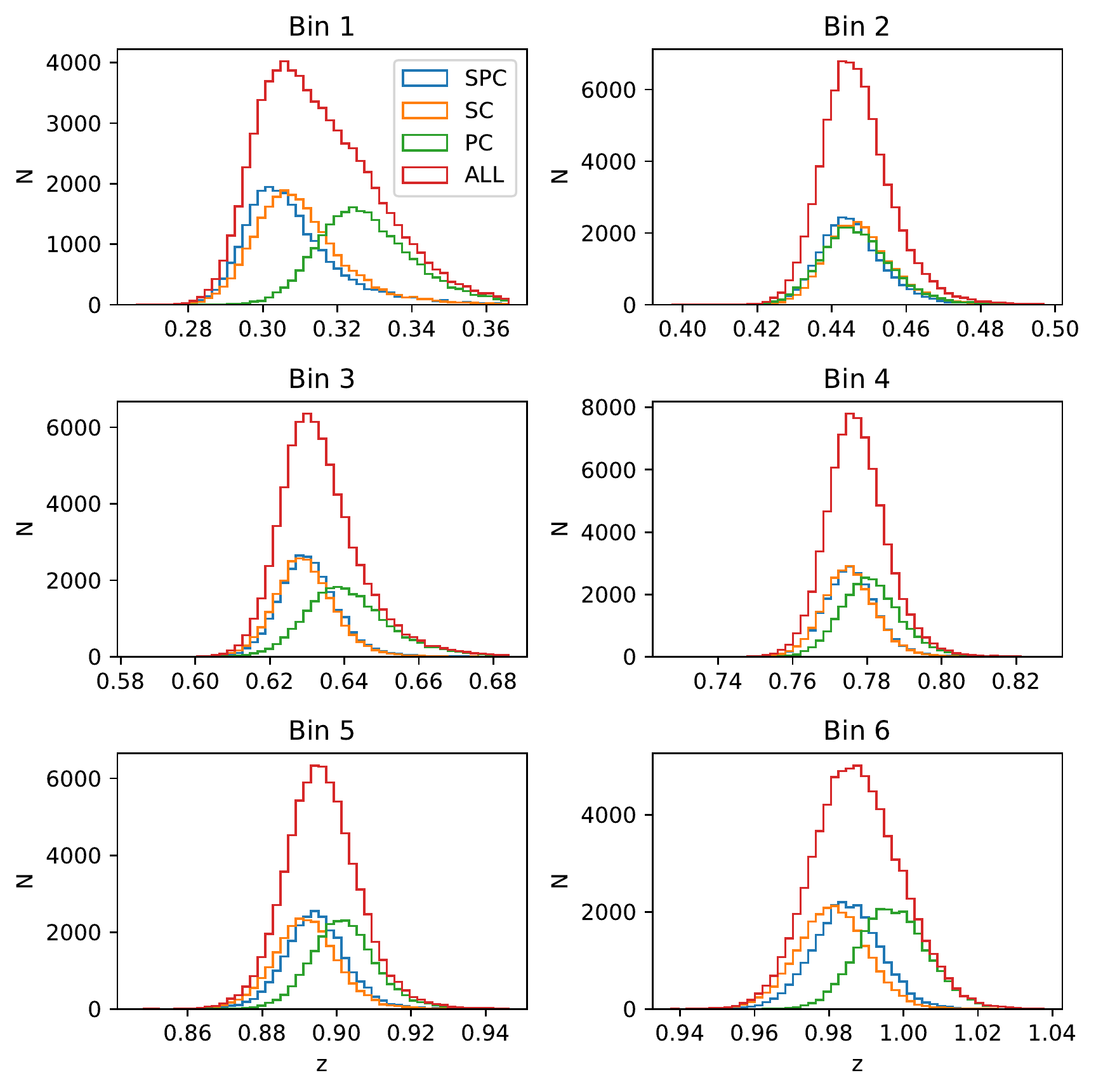}
    \caption{Uncertainty on the mean redshift represented by the number counts of the three redshift samples: SPC (prioritizes spectra, than PAU photo-z, then COSMOS30), PC (prioritizes  PAU photo-z, then COSMOS30) and SC (prioritizes spectra, then COSMOS30). In red the total uncertainty given by their combination.}
    \label{fig:redshift_means}
\end{figure}

\subsubsection{Transfer function uncertainty} \label{sec:balrogunc}

% \mg{remind the reader what's the transfer function, what's the potential impact of having it wrong, and how you'd test it}
One of the key points in this redshift calibration is the transfer function $p(c|\hat{c})$, the intermediate step necessary to assign redshifts from deep field galaxies to the whole wide field. If the transfer function is inaccurate, regardless of how a precise the color/redshift characterisation is in the deep SOM, it can bias the final \nz\ distributions. 
$p(c|\hat{c})$ depends on the observation conditions in that location, determining if the galaxy is detected or not. Observing conditions vary across the wide field, but for our analysis we are interested in redshift distributions estimated across all the footprint. \Balrog\ injects the same deep galaxies in random wide tiles, and despite these covering only around $\sim20\%$ of the DES footprint, in \cite*{y3-sompz} was verified that \Balrog\ is adequately sampling the observing conditions in the wide field. They boostrapped the sample by the injected position and recomputed 1000 different transfer functions. They concluded that the dispersion in the final \nz\ mean redshift from repeating the analysis using each time a different transfer function was completely negligible. Here we repeated that test, since our deep field sample has less galaxies and might impact differently the transfer function.% We proceeded to estimate this by bootstrapping the transfer function $p(c|\hat{c})$, and computed the spread of the mean redshift value of the resulting ensemble of \nz\.
We found that this is also negligible for our case, with variations on the \nz\ mean $<10^{-3}$ in each tomographic bin, and therefore decided not to propagate this in the final \nz\ estimate.

\subsection{WZ Uncertainties}\label{sec:wzunc}

\begin{figure*}\label{fig:sub2}
\includegraphics[width=0.45\linewidth]{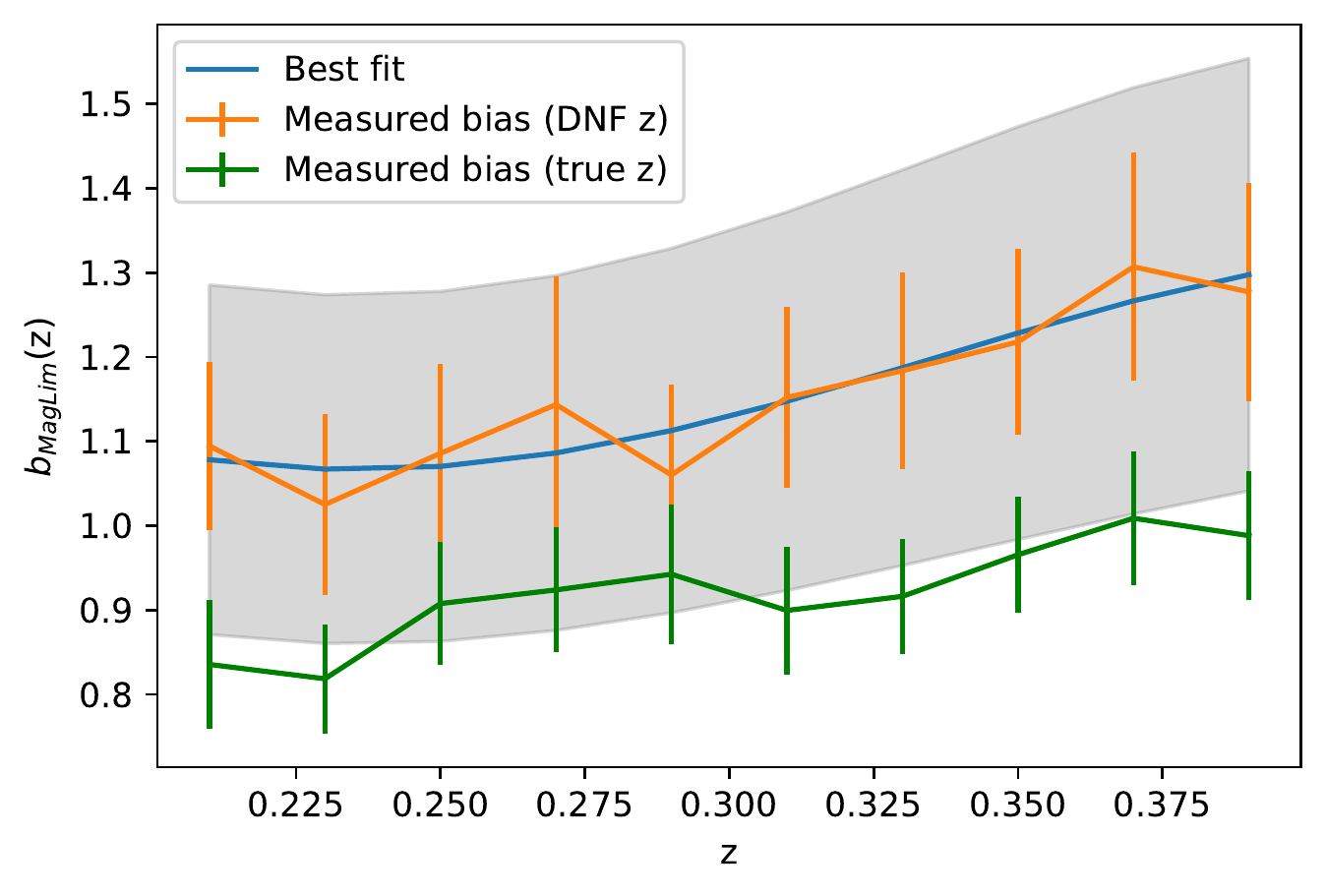}
\includegraphics[width=0.45\linewidth]{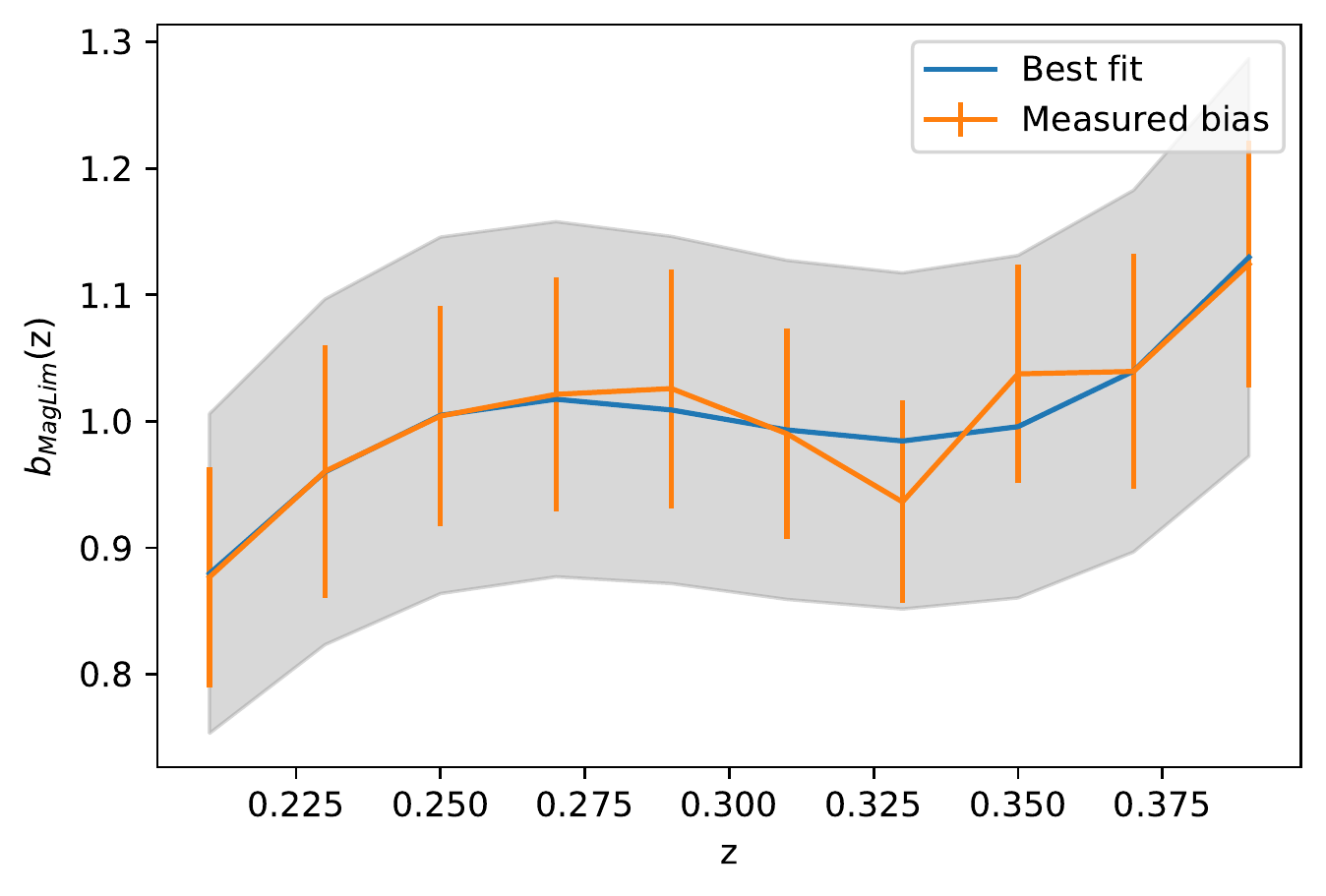}
\caption{\textit{Left panel}: galaxy-matter bias of Bin 1 or the \maglim\ sample (0.2 < z < 0.4) as estimated in simulation following the methodology outlined in Section \ref{sec:wzunc}. The green points are obtained by dividing the sample into thin bins using the true redshifts, while the orange ones are obtained by binning the sample using the DNF redshift estimates. The grey band encompasses the 68$\%$ confidence interval of the \sys\ function. \textit{Right panel}: galaxy-matter bias of Bin 1 of the \maglim\ sample (0.2 < z < 0.4) as measured from the data (orange points); the blue line shows the best-fitting \sys\ function, and the grey band encompasses its 68$\%$ confidence interval.}
\label{fig:autocorr_maglim}
\end{figure*}

The WZ systematic uncertainties have been identified and characterised in detail for the WL sample in \cite*{y3-sourcewz}. Namely, the systematic budget was found to be dominated by our lack of prior knowledge of the redshift evolution of the galaxy-matter bias of the unknown sample. This is also expected to be the case for the \maglim\ sample, although the amplitude of the effect might differ from the WL sample (ideally, since the \maglim\ redshift distributions are narrower, we might expect a smaller impact due to systematics slowly varying with redshift like the galaxy-matter bias of the unknown sample).

Similarly to \cite*{y3-sourcewz}, we model our systematics by means of a flexible function, \sys, which mostly captures the redshift evolution of the galaxy-matter of the unknown sample. The \sys\ function is parameterized by $\vecs=\{s_1,s_2,\ldots\}$ that we will marginalize over and is given by:
\begin{align}
\label{eq:model_syst_unc}
  \log[{\rm Sys} \left(z,\vecs\right)] & = \sum_{k=0}^M \frac{\sqrt{2k+1}}{0.85} s_k P_k(u), \\
  \label{eq:sys_rescale}
  u & \equiv 0.85 \frac{z-0.5(z_{\rm max}+z_{\rm min})}{(z_{\rm max}-z_{\rm min})/2},
\end{align}
with $P_k(z_i)$ being the $k$-th Legendre polynomial and $M=6$ is the maximum order. In this work, we set the prior $p(\vecs)$ to be a simple diagonal normal distribution, with the standard deviations $\{\sigma_{s0},\ldots,\sigma_{sM}\}$ and means informed by the measured auto-correlation of the \maglim\ sample.

In \cite*{y3-sourcewz}, such a systematic function was let to vary by the typical amplitude of the redshift evolution of the galaxy-matter bias of the WL sample we measured in simulations. In practice, this was achieved by imposing a Gaussian prior with zero mean $p(\vecs)$ on the coefficients $\vecs$ of the systematic function.  

In the case of the \maglim\ sample, we can use a more informative prior $p(\vecs)$ that uses the information we have from the data about the galaxy-matter bias evolution of the sample. In particular, we rely on the fact that the \maglim\ sample has good per-galaxy redshift estimates, which allows us to divide the sample in relatively small bins and measure the auto-correlation of such bins. This was not possible for WL sample, due to the poor per-galaxy redshift accuracy.

To this aim, we use DNF 1-point estimates $z_{\rm mean}$ to further divide the \maglim\ sample in bins of width of $\Delta_z = 0.02$, and we measure the auto-correlation of each bin. We note that the true width of each bin will be much larger than $\Delta_z = 0.02$, as the DNF photo-$z$ are uncertain. Under the approximation of negligible redshift evolution of the galaxy-matter bias of the \maglim\ sample over each thin bin, the measured autocorrelaton  can be related to the galaxy-matter bias by knowing how broad the true $n(z)$ distribution of each bin is \citep{Gatti2018,y3-lenswz}:
\begin{equation}
\label{eq:auto-correlation1}
    {w}_{\rm uu}(z_i) =  b_{\rm u}^2 (z_i) {w}_{\rm DM}(z_i) \int dz'  n_{\rm u,i}^2(z'), 
\end{equation}   
where $n_{\rm u,i}(z')$ is indeed the true distribution of the thin bin \maglim\ sample. Such a quantity is estimated using the PDF estimate from DNF $z_{\rm PDF}$.

From this measurement performed in data we can then retrieve the galaxy bias $b_{\rm u}(z)$ by inverting Eq. \ref{eq:auto-correlation1}. We fit the \sys\ function presented in Eq. \ref{eq:model_syst_unc} to the measured $b_{\rm u} (z)$ and obtain best-fit \textbf{s} values, which we show in Figure \ref{fig:autocorr_maglim}. These best-fit coefficients are then used as the mean value of the Gaussian prior $p(\textbf{s})$.
% From this measurement performed in data we can then retrieve the galaxy bias $b_u(z)$. We do extrapolate the prior mean by fitting the bias function with the same Sys function presented in Eq. \ref{eq:model_syst_unc} and use the coefficients of this best fit as priors in the likelihood. 
The best fitting \sys\ function to the data is shown in the right panel of Fig. \ref{fig:autocorr_maglim}.

To estimate the width of the prior p(\textbf{s}) we took a different approach. First, we estimate the bias evolution in simulations by dividing galaxies into thin redshift bins using: (i) the true redshifts from the simulation; and (ii) the photo-z estimated from the DNF code. When dividing the galaxies with the photo-z from DNF, we 
further correct the measured auto-correlation using Equation \ref{eq:auto-correlation1}. These measurements are shown in the left panel of Figure \ref{fig:autocorr_maglim}. The discrepancy between the measured bias evolution from photo-z (equivalent to the application with real data) relative to the measured bias evolution with true redshifts (equivalent to the truth) is a systematic bias. We use the sum in quadrature of this difference with the statistical uncertainty of the bias measurement as the prior width of $s_0$. For the higher order parameters we estimate the standard deviation of the prior by summing in quadrature the ratio between the two biases and the statistical uncertainty from the bias measurement in data. This allows to best capture the RMS variations of the bias function itself. As can be seen in Figure \ref{fig:autocorr_maglim}, the 68$\%$ confidence interval spanned by the \sys\ function both brackets the ideal and real world measurements. The values for the mean and width of the prior are displayed in Table \ref{tab:coeff}. Both the width of the prior on the 0-th and higher order coefficients are much tighter than in \cite*{y3-sompz}, where $s_0 = 0.6$ and $s_{1..4} = 0.15$. As already explained, the difference lies in the initial accuracy of the photo-z estimates, that enables the measurement of the auto-correlation of the galaxy sample in thin redshift bins. For the weak lensing source sample such information was not available, and therefore a more conservative prior was deemed appropriate. In the \maglim\ sample case instead, the greater accuracy on its photo-z allows to extract more information from the auto-correlation.

\begin{table}
    \centering
    \begin{tabular}{c c c c c c c}

    &&& Mean\\
    \hline
    \hline
    
        & Bin 1 & Bin 2 & Bin 3 & Bin 4 & Bin 5 & Bin 6 \\
        \hline
        $<s_0>$  & -0.028 & -0.085& -0.319 & -2.630 & -0.119& -2.249\\
        $<s_1>$ & 0.186&  0.559&    0.007 &  1.161& -1.660 &   0.819\\
        $<s_2>$ &0.046&  0.139& -0.120&  0.202&  0.134&   0.033\\
        $<s_3>$ &0.035& 0.105& -0.130&  0.314&  0.293&  0.174\\
        $<s_4>$ & 0.037&  0.111& -0.112& -0.197&  0.211&   0.279 \\
        $<s_5>$ &-0.062 & -0.189 & -0.203& -0.210&  1.408&   0.569\\
        \hline
         &&& Width\\

    \hline
    \hline
        & Bin 1 & Bin 2 & Bin 3 & Bin 4 & Bin 5 & Bin 6 \\
        \hline
        $\sigma_{s_0}$  & 0.107 & 0.216 & 0.123 & 0.072 & 0.067 & 0.198\\
        $\sigma_{s_{1..5}}$ & 0.029 & 0.053 & 0.041 & 0.052 & 0.081 & 0.044\\
         \hline
    \end{tabular}
    \caption{Means and widths of the Gaussian prior function $p(\vec{s})$ appearing in Eq. \ref{eq:wzlike}. }
    \label{tab:coeff}
\end{table}

% For the standard deviation of the prior we took a different approach. We performed the same measurement in simulations, and compared it with an auto-correlation computed dividing the sample into thin bins using the true redshift point-estimate available in Buzzard. The two galaxy-matter biases are displayed in the left panel of Fig. \ref{fig:autocorr_maglim}. We treated the 0-th order and the higher orders of the Sys function parameters differently, since we are mostly interested in the amplitude of the bias function, which is accounted by the 0-th parameter in the Sys function. In particular, for the higher orders we estimate the standard deviation of the prior by summing in quadrature the ratio of the two biases and the statistical uncertainty from the bias measurement in data. This allows to best capture the RMS variations of the bias function itself. For the 0-th order instead we summed in quadrature the difference of the two biases with the the statistical uncertainty from the bias measurement in data, to  better account for the bias seen in simulations between the ``ideal'' result (when the sample is divided into thin bins using exact redshifts) and the ``real world'' result  (when the sample is divided into thin bins using photo-$z$ redshifts). 

Last, we mention that an additional source of uncertainties for the WZ measurement is related to the impact of magnification. We do model magnification effects, but the accuracy of that model is limited by our knowledge of the magnification coefficients for the two samples. In particular, we do not have any prior knowledge of such a coefficients for the BOSS/eBOSS sample. Those coefficients are set to 0 for our fiducial analysis (on the contrary, estimates for the magnification coefficient of the \maglim\ sample are available). We expect magnification to have a small impact, based on tests performed in \cite*{y3-sourcewz}, but we nonetheless test in the following section the impact of having a non null magnification coefficient for the BOSS/eBOSS sample.
%For the p parameters of the p(p) prior we assumed the fiducial magnification parameter $C_{FLUX} = \alpha *2$ estimated from injected simulations (Balrog) and for $b_u$ a Gaussian prior with $(\mu, \sigma) = (1., 1.5).$

\subsection{Combination of SOMPZ and WZ}\label{wzlikelihood}

In order to combine SOMPZ and WZ constraints, we follow \cite*{y3-sourcewz} and write the clustering likelihood by forward modelling the full clustering signal as a function of the SOMPZ redshift distributions estimates $n(z)_{\rm pz}$. Moreover, we include the systematic function \sys\ introduced in the previous section, which describes the  uncertainties on the WZ measurement, mostly driven by the lack of knowledge of $b_{\rm u}$ and its redshift dependence:
%
%$\hat w_{\rm ur}[z_i; n_{\rm u}(z), b_{\rm r}(z), \alpha_{\rm r}(z), \vecs, \vecp]$.
%
\begin{multline}
\label{wursys}
{\hat w}_{\rm ur}(z_i) =  n(z)_{\rm pz}(z_i) b_{\rm r}(z_i)
 {w}_{\rm DM}(z_i) \times Sys(z_i,\vecs) + \\M(\alpha_{\rm u},\alpha_{\rm r},b_{\rm u},n(z)_{\rm pz}).
\end{multline}
In the above equation, the quantities $\alpha_{\rm u}(z_i)$ and $\alpha_{\rm r}(z_i)$ are the magnification coefficients for the unknown and reference samples. See \citealt*{y3-sourcewz} for full description of the magnification term $M$. The clustering of dark matter ${w}_{\rm DM}(z_i)$ is estimated from theory assuming fixed cosmology. We tested that varying cosmology has a negligible impact on our methodology. 

% To combine the SOMPZ and WZ constrains, we importance-sample $n(z)_{\rm pz}$ SOMPZ samples by assigning each a weight through the following likelihood:
The likelihood of the WZ data conditioned on the target \nz\ and all the systematic parameters reads as:
\begin{multline}%
\label{eq:wzlike}
  \mathcal{L}\left[{\rm WZ} | n_{\rm u}(z), b_{\rm r}(z), \alpha_{\rm r}(z), w_{\rm DM}(z)\right]
  \propto \\
  \int d\vecs\, d\vecp\, \exp\left[ -\frac{1}{2} (w_{\rm ur}-\hat w_{\rm ur})^T \Sigma_w^{-1} (w_{\rm ur}-\hat w_{\rm ur}) \right] p(\vecs) p(\vecp),
\end{multline}
were $\Sigma_w$ is the clustering covariance, estimated through jackknife, and $\vecp = {b_{\rm u},\alpha_{\rm u}}$. %Note that for efficiency reasons, instead of importance-sampling the SOMPZ realisations, 
We implemented a Hamiltonian Monte Carlo sampler (HMC) that simultaneously samples the SOMPZ and WZ likelihood. The HMC does directly take as input the SOMs output of the sample variance estimation (described in \ref{sec:3sdir}), and it perturbs selectively the number counts in the SOMs in such a way to produce realisations that are already more likely to match the clustering redshift data.

% In Appendix \ref{sec:autocorr} we show what is the impact on the redshift distributions of running the HMC with the more informative priors on Sys(s) we have from the auto-correlation.}

%nd instead of randomly selecting realisations to give as input to the joint likelihood, it perturbs selectively the number counts in the SOMs in such a way to produce realisations that are already more likely to match the clustering redshift data.

\section{Results in Data}\label{sec:results}

In this section, we present the final redshift distributions for the \maglim\ sample as obtained in data. We also compare the SOMPZ+WZ redshift distributions with the fiducial DNF+WZ estimates used for the same sample and adopted in the cosmological analysis presented in \cite{y3-2x2ptaltlensresults}. A complete validation of the method in simulations is presented in Appendix \ref{sec:resultssims}.

%mg{Describe the plot; quote how much the mean shifts by adopting such a bias model; quote ho much the uncertainty of the mean change by including the systematic functions; similarly for the width. Mention that for the WL sample the bias of the unknown had a $\sim$ 0.015 unc. on the mean; the fact you're using a tight prior informed by data for the \maglim\ sample should give you smaller systematic uncertainties than that. Mention that a similar approach has been adopted by \cite{y3-lenswz}, although he uses a different systematic function}.

%As already shown in Table \ref{table:unc}, adopting the bias model described in the previous sections at the combination level results in a very small shift of the mean redshift values for each bin, much smaller than the uncertainty associated with the mean itself; the exception is always Bin 2, although the shift still lies at the edge of the one-sigma range. This is due by the limited information WZ carries about the mean, given that the largest source of uncertainty is on the bias. For the width instead, the average value changes by an amount of the order of one sigma. The most interesting feature we can read from these results though, is the confirmation of the capacity of the WZ of reducing the uncertainties on the distributions. 

% \begin{figure}
%     \centering
%     \includegraphics[width=\linewidth]{Figures/widthmeanunc.png}
%     \caption{}
%     \label{fig:unc_bar2}
% \end{figure}

\begin{figure}
\centering
\includegraphics[width=\linewidth]{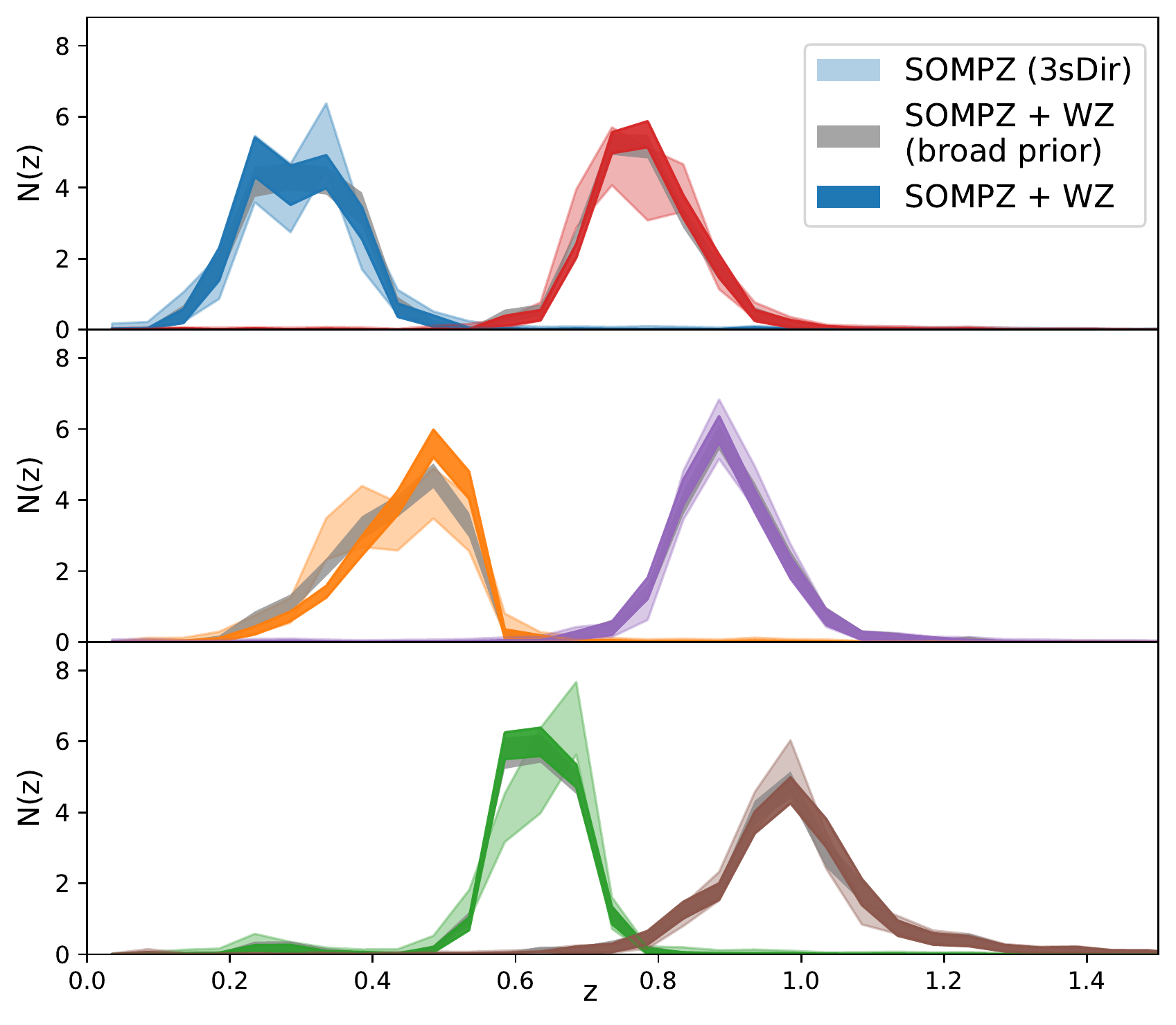}
\caption{3sDir distributions before (lighter shades) and after the combination with clustering-z (solid shades), and after the combination with clustering-z but using a broader prior on the parameters of the galaxy-matter bias function \sys (the same values of the width of the prior $p(\vec{s})$ that were used in \citealt{y3-sourcewz}). In the top row we have bins 1 and 4, in the middle row bins 2 and 5, and in the bottom rows bin 3 and 6. The bands represent the $1\sigma$ error from the central value. Note how the combination with WZ tightens the constraint on the shape of the \nz.}
\label{fig:nz_hmc_dataa}
\end{figure}

We first compare in Figure \ref{fig:nz_hmc_dataa} the redshift estimates obtained using the 3sDir method and the estimates obtained including the WZ information as described in section \ref{sec:uncertainty}. Due to logistics, the combination of the two methods was performed before incorporating the SOMPZ and zeropoint errors. As here we are just displaying the effect of the combination, we are showing only how the 3sDir uncertainty from sample variance and shot noise (from the three redshift samples) varies once we add the information from WZ. The combination of the two methods result in stronger constraints on the shape of the \nz , thanks to the complementarity in the information provided by each SOMPZ and WZ. Particularly, the WZ signal strongly correlates across adjacent bins, excluding large portions of possible \nz\ shapes allowed by the SOMPZ likelihood alone, which are affected by sample variance fluctuations from the small calibration fields, and resulting in a smoother distribution. The improvement on the uncertainty on the mean is more modest, but not null, as reported in Table \ref{tab:unc}. Usually, WZ data provides limited information on the mean redshift, especially compared to SOMPZ, as the systematic uncertainty on the galaxy bias evolution of the target sample is large and directly degenerate with the mean redshift, as is the case in \cite*{y3-sourcewz}. However, in this work we have included a tighter prior on the \sys\ function describing the galaxy bias evolution uncertainty by measuring it directly from the \maglim\ auto-correlation function. The addition of the WZ information has a modest impact on the values of the mean and width of the redshift distributions, at most at the 1$\sigma$ level (see Table \ref{tab:unc}); this is somewhat expected, as the WZ and SOMPZ information are independent, but consistent with each other.

\subsection{Comparison with DNF} \label{comparisondnfsompz}

We find it interesting to compare the final SOMPZ+WZ redshift distributions with the fiducial ones used for DES Y3, obtained using DNF photometric estimates and clustering constraints (hereafter DNF+WZ). Since the two sets of distributions have been obtained with two different methods, we also briefly discuss the major differences between the two pipelines. The DNF code presented in \ref{sec:DNF} produces per-galaxy redshift estimates; these are stacked to produce the redshift distributions for the lens samples. Then, following \cite{y3-lenswz}, a clustering redshift measurement is performed, using BOSS/eBOSS galaxies as reference sample, similarly to this work. %The WZ \nz are then fitted by a sum of Gaussian distributions. Finally, 
The DNF \nz are matched to the WZ-estimated \nz through a chi-square fitting; in particular, the DNF \nz are allowed to shift and stretch to improve the $\chi^2$. The maximum-a-posteriori values of the shift and stretch and related uncertainties obtained through this matching procedure are used as a prior for the DNF \nz shift and stretch used in the cosmological inference. 

Despite the DNF+WZ and SOMPZ+WZ methods using the same photometric and clustering measurements, the methodologies differ in a number of aspects:
%\begin{enumerate}[label=\arabic*),leftmargin=*,align=left]
% \begin{enumerate}[1)]
\begin{enumerate}
    \item \textbf{SOMPZ vs DNF uncertainties:} SOMPZ and DNF are both machine learning methods, but they are substantially different in spirit and implementation. DNF is a traditional supervised machine learning code where the likelihood (directional neighborhood) between wide field magnitudes/colors and redshift is learned from training with a subsample of galaxies with both reliable redshift information and measured wide field photometry. On the other hand, in SOMPZ machine learning is only used in an unsupervised fashion (without knowledge of redshift), to group self-similar parts of wide field magnitude/color space together. Then, these groups (wide SOM cells) are probabilistically related using Bayes theorem to the color-redshift relation measured empirically in the calibration deep fields, where much better information is available. The likelihood between each set of wide and deep field photometry is also measured empirically by injecting galaxies of the latter into images of the former. Furthermore, SOMPZ provides a comprehensive list of statistical as well as systematic uncertainties affecting the calibration samples which are rigorously propagated through the \nz. On the other hand, DNF only describes statistical uncertainties related to the residual differences to the closest training neighbors to the fitted hyperplane of the target galaxies.
    % Moreover, they use a slightly different training sample (e.g., DNF does not use multi-band photometric redshifts from COSMOS). DNF per-galaxy redshift estimates do not come with uncertainties, whereas the SOMPZ framework and the implementation adopted in this work accounts in detail for every (known) source of uncertainty that can impact the redshift distributions estimates.
    \item \textbf{Combination:} The clustering information is included and combined with the photometric estimates in a substantially different way. In this work, SOMPZ and WZ are combined by sampling from the joint posterior using the HMC method. No approximation is performed when combining the two likelihoods. On the other hand, matching DNF \nz to the WZ measurements it has been implicitly assumed that the DNF \nz estimates can only be biased at the level of their mean and width, and that inaccuracies in the higher order moments of the \nz can be neglected (or do not affect the matching procedure with the WZ measurements). However, if the DNF and WZ \nz estimates are substantially different beyond their first two moments, the matching might cause biases \citep{Gatti2018} also in the first and second moments. Furthermore, in the combination of the fiducial method, the DNF shape is only allowed to be modified by shifting and stretching it. Therefore the shift and stretch parameters are centered at the WZ values. This means that the photo-z priors for the cosmological inference only carry uncertainty from the WZ measurement, as this method does not propagate any systematic uncertainties related to uncertainty from the accuracy of DNF or the quality of its training sample photometry. In comparison, SOMPZ+WZ properly combines the statistical significance from SOMPZ and WZ yielding a final uncertainty that truly combines the information from each of them separately. Finally, the SOMPZ+WZ \nz samples also capture the uncertainties in the higher moments of the redshift distributions, whereas the DNF+WZ uncertainties are only relative to the mean and width.
    \item \textbf{WZ distribution tails:} The WZ measurements used to calibrate the DNF \nz have clipped tails, since the measurements were performed in a restricted redshift window to avoid biases related to un-modelled magnification effects in the tails of the redshift distribution. On the other hand, in this work, when combining the clustering information with SOMPZ estimates, we use the WZ measurements over all the redshift range, since we also marginalise over magnification effects.
    \item \textbf{WZ galaxy-matter bias:} The WZ measurements used in the DNF+WZ estimates are corrected for the redshift evolution of the galaxy-matter bias of the \maglim\ sample computed from auto-correlations measurements following Eq. \ref{eq:auto-correlation1} \citep{y3-lenswz}. As for this work we use the forward modelling approach described in Section \ref{sec:wzunc}, we instead do not correct directly for the bias, but from the \maglim\ auto-correlations we determine prior values of the parameters of our \sys, and then marginalise over possible bias functions in the sampling from the joint likelihood. We are therefore assuming an uncertainty on the galaxy-matter bias and validating the central value using SOMPZ data. 
\end{enumerate}

 \begin{table*}
     \centering
     \begin{tabular}{c c c c c c c c} 
          & & Bin 1 & Bin 2 & Bin 3 & Bin 4 & Bin 5 & Bin 6\\
         & & z $\in$ [0.2, 0.4] &  z $\in$ [0.4, 0.55] &  z $\in$ [0.55, 0.7] &  z $\in$ [0.7, 0.85] &  z $\in$ [0.85, 0.95] &  z $\in$ [0.95, 1.05] \\
          \hline
          \hline
          %\begin{turn}{90}MEAN \end{turn} 
          %<z> & SOMPZ & 0.309 & 0.451 & 0.617 & 0.762 & 0.883 & 0.967 \\
          <z> & SOMPZ+WZ & 0.315 $\pm$ 0.016 & 0.463 $\pm$ 0.010 & 0.633 $\pm$ 0.009 & 0.781 $\pm$ 0.008 & 0.893 $\pm$ 0.009 & 0.990 $\pm$ 0.012 \\
          & DNF+WZ & 0.292 $\pm$ 0.007 & 0.422 $\pm$ 0.011 & 0.616 $\pm$ 0.006 & 0.762 $\pm$ 0.006 & 0.887 $\pm$ 0.007 & 0.969 $\pm$ 0.008\\
          \hline 
          & $\Delta_{<z>}$ & 1.3 & 2.7 & 1.7 & 1.9 & 0.5 & 1.5 
          \vspace{0.2cm}\\
       
          \hline
          \hline
         $\sigma_z$ & SOMPZ+WZ & 0.080 $\pm$ 0.005 & 0.081 $\pm$ 0.005 & 0.060 $\pm$ 0.002 & 0.073 $\pm$ 0.003 & 0.074 $\pm$ 0.004 & 0.102 $\pm$ 0.007 \\
         & DNF+WZ & 0.078 $\pm$ 	0.005 & 0.094	$\pm$ 	0.007 & 0.055	$\pm$ 	0.003 & 0.062	$\pm$	0.003 & 0.075	$\pm$	0.004 & 0.080 $\pm$ 	0.007 \\
         \hline
         & $\Delta_{\sigma_z}$ & 0.2 & 1.6 & 1.3 & 2.2 & 0.3 & 2.3 \\

     \end{tabular}
     \caption{Values of mean and width of the SOMPZ+WZ final ensemble of distributions and the DNF estimate. The statistical difference $\Delta_{<z>}$ is computed by considering the uncertainties of both methods summed in quadrature, as in $\rm \Delta_{<z>} = (<z>_{SOMPZ} - <z>_{DNF}) / \sqrt{\sigma(<z>_{SOMPZ})^2 + \sigma(<z>_{DNF})^2}$. We refer to these as are lower limits. Because the WZ measurement is very similar in the two cases, and the uncertainties summed in quadrature are correlated and therefore we are likely underestimating $\Delta_{<z>}$.}
     \label{tab:dnf_values}
 \end{table*}

% \begin{figure*}
% \centering
% \includegraphics[width=0.5\linewidth]{Figures/nz_final_dnf.png}
% \caption{Final distributions including all sources of uncertainties, compared to the DNF zmc pdf distributions \label{fig:nz_final_dnf}}
% \end{figure*}

\begin{figure*}
\includegraphics[width=0.49\textwidth]{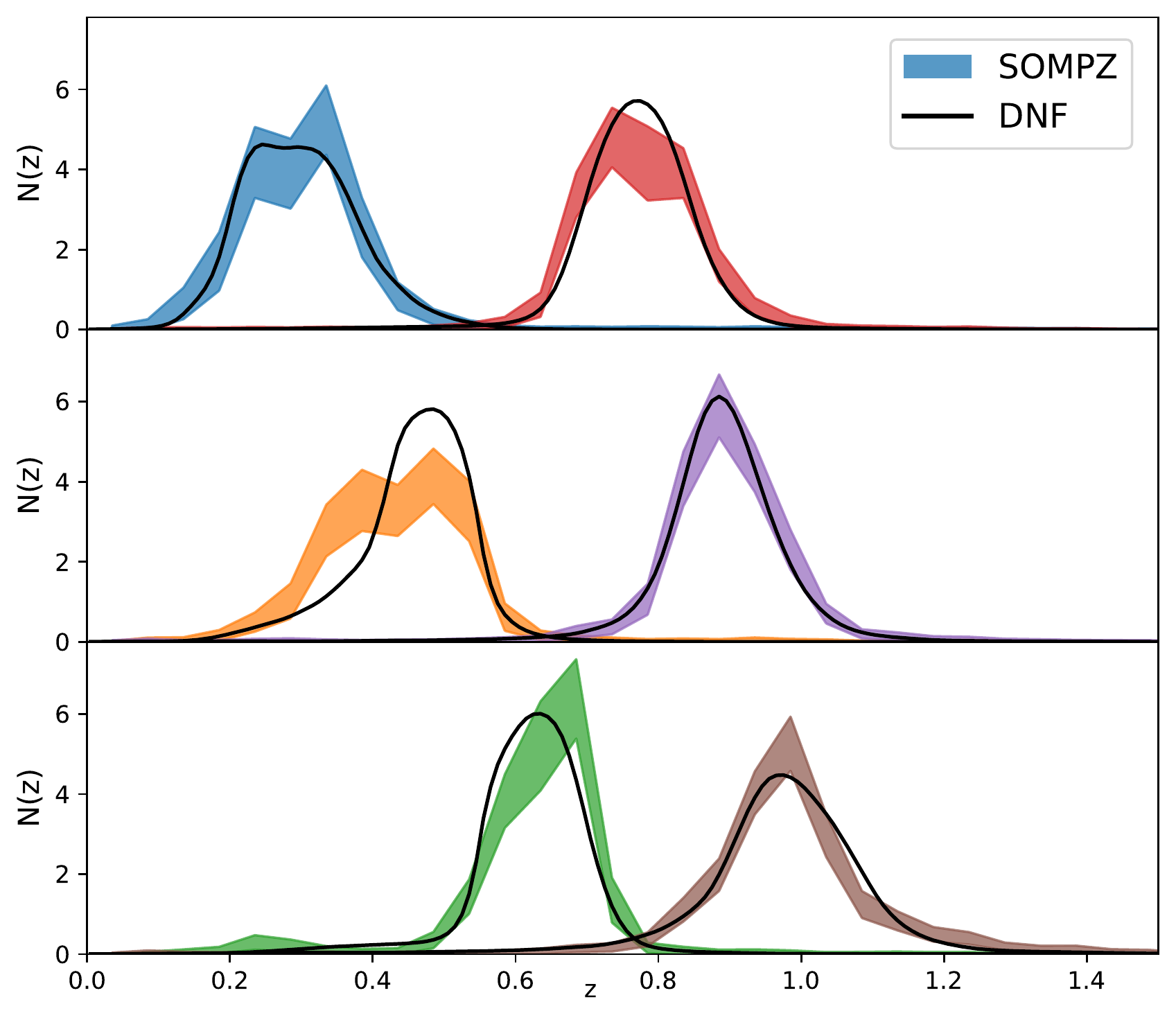}
\includegraphics[width=0.49\textwidth]{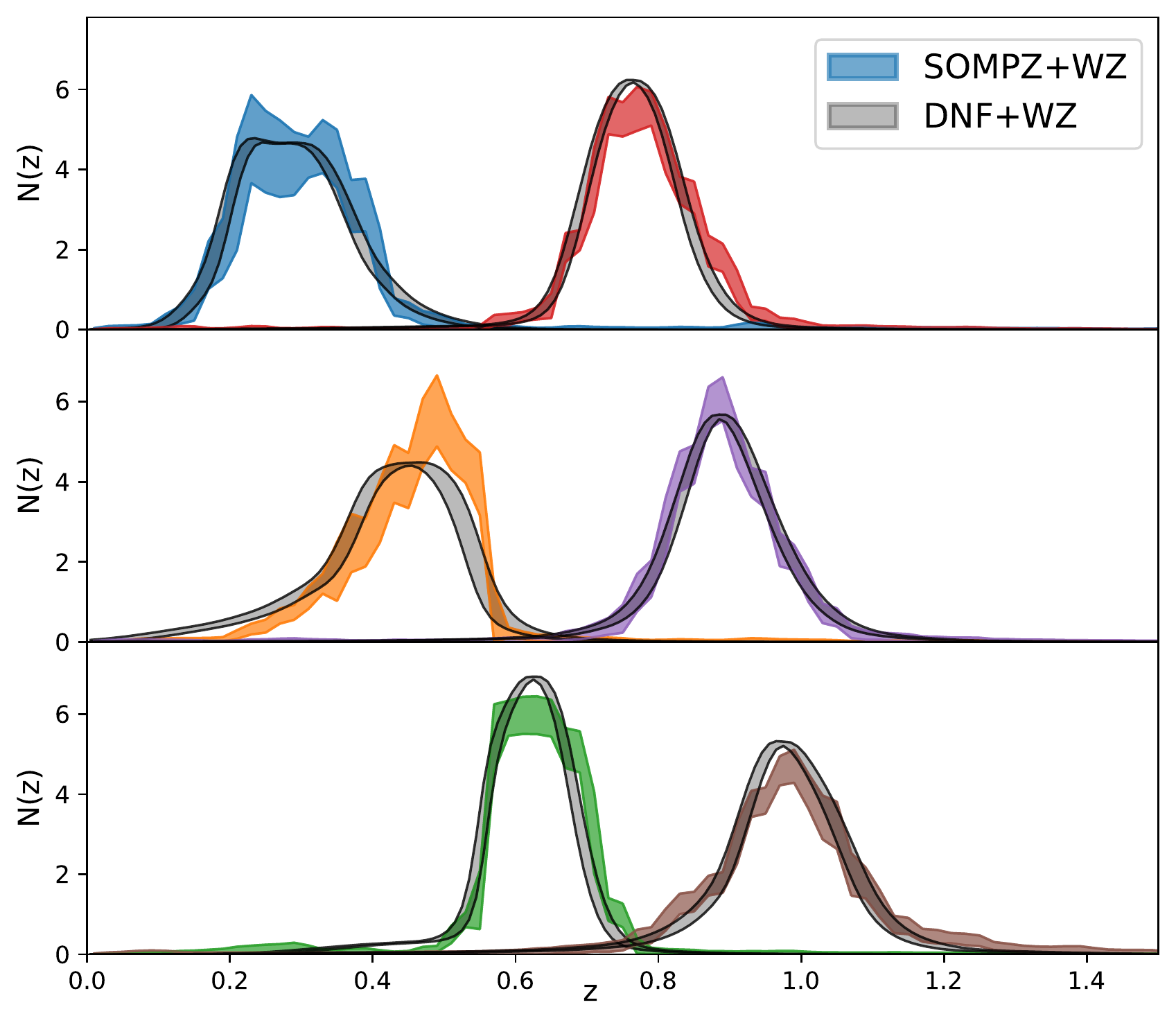}
     %\begin{subfigure}[b]{0.45\textwidth}
     %    \includegraphics[width=\textwidth]{Figures/pit_3sdir_dnf_final2.pdf}
     %    \label{fig:pit_3sdir_dnf}
     %\end{subfigure}

    % \begin{subfigure}[b]{0.45\textwidth}
     %    \includegraphics[width=\textwidth]{Figures/pit_dnf_final2.pdf}
    %     \label{fig:pit_dnf}
    % \end{subfigure}
     \caption{Left panel) Final \nz realisations obtained from the SOMPZ methodology alone compared to the fiducial DNF distribution for \maglim\ (in black). Right panel) Final \nz realisations obtained from both SOMPZ and WZ methodology compared to the fiducial DNF distribution for \maglim\ (grey bands) after shifting and stretching them to fit WZ measurement. Since in the inference the shift and stretch values are marginalised over, the uncertainties of the gray bands are obtained by sampling over the allowed ranges of shift and stretch defined by the prior, and applied respectively to the DNF estimate. Note that for a fairer comparison of the methods, the two remaining uncertainties were applied to the SOMPZ ensemble (zeropoint and SOMPZ intrinsic), to include all the SOMPZ-related uncertainties. For both plots, in the top row we have bins 1 and 4, in the middle row bins 2 and 5, and in the bottom row bins 3 and 6. .}
     \label{fig:dnf_vs_sompz_plot}
\end{figure*}

\begin{figure*}
\includegraphics[width=\textwidth]{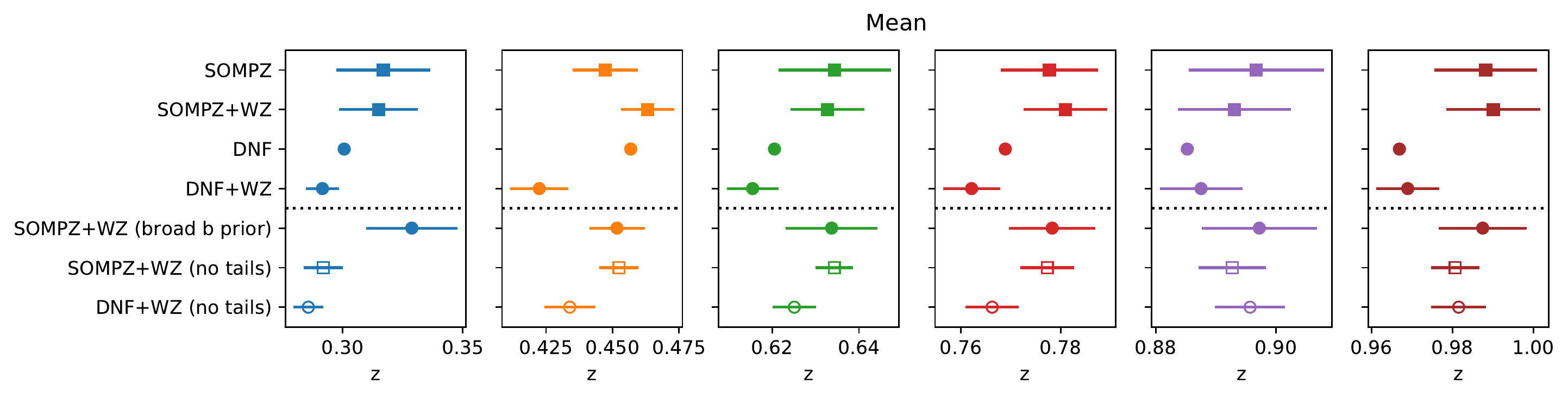}
\includegraphics[width=\textwidth]{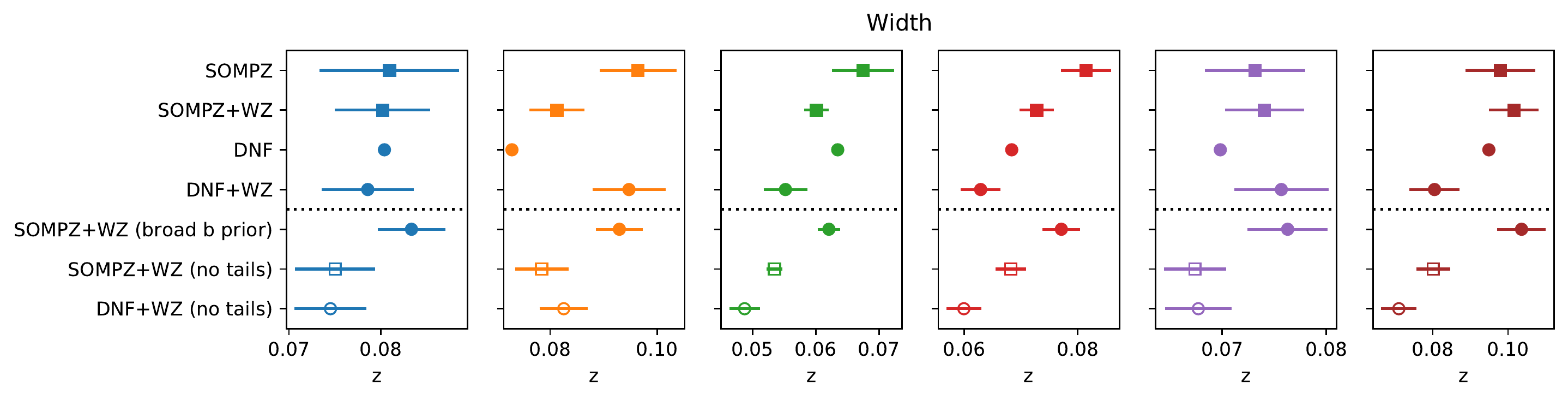}
     \caption{Visual representation of the uncertainties on mean (above) and width (below) of the redshift distributions estimated using the SOMPZ (square markers) and DNF (round markers) methods, before and after including the WZ information, for each tomographic bin. Below the dashed line is the comparison of the values computed in the redshift range used for the $\chi^2$ fit of the DNF estimate with the smoothed WZ \nz . 
     %For the methodology presented in this paper, the addition of WZ results in a shift on the mean much smaller than the standard deviation, while for the width in bin 2 and 3 the shift is around one sigma. For the fiducial method instead, the lack of uncertainty associated with the DNF estimate results in shift much larger, especially in bin 2. 
     }
     \label{fig:dnf_vs_sompz_unc}
\end{figure*}

We must highlight that in \cite{y3-lenswz,y3-2x2ptaltlensresults} several tests were performed to test the robustness of the DNF+WZ method. In particular, \cite{y3-lenswz} tested the performance of the clustering measurements in simulations, whereas \cite{y3-2x2ptaltlensresults} tested that matching DNF \nz to the WZ measurements was not introducing biases in the cosmological constraints, and that modelling only the uncertainties in the mean and width of the distributions was sufficient for the DES Y3 cosmological analysis. These tests should cover potential worries raised in points ii), iii) and iv) above for the DNF+WZ method. Having said this, any discrepancy between the SOMPZ+WZ \nz and the DNF+WZ \nz should boil down to the points listed above.

%and we they tested the methodology itself in the ideal scenario of having spectroscopic measurements for both the reference and unknown sample, only possible in simulations. Further tests involved different approaches for the galaxy-matter bias correction. The conclusion from these tests in simulations was that the DNF+WZ methodology had been deemed robust for the cosmological analysis. 

%Any discrepancy between the two results boils down to the points listed above.

In Figure \ref{fig:dnf_vs_sompz_plot},  the shapes and uncertainties of the two methodologies are compared, before and after the inclusion of WZ information, respectively in the left and right panel. Visually the DNF+WZ \nz look very similar to the SOMPZ+WZ ones, although some discrepancies can be noticed (e.g., in the second bin). We report in Table \ref{tab:dnf_values} the redshift means and widths of the two sets of distributions, and their agreement. The means and widths are also visually compared in Figure \ref{fig:dnf_vs_sompz_unc}. The agreement is computed assuming the uncertainties of the two methods to be uncorrelated, which is likely not true; therefore, the reported agreements are optimistic. Computing the level of correlation between the two redshift estimates is not trivial. The DNF+WZ estimates and uncertainties are driven only by the WZ measurements in the range where WZ measurements are available and magnification effects are negligible; the tails of the distribution, on the other hand, are described by the DNF estimates. The SOMPZ+WZ estimates receive contributions from both SOMPZ and WZ; if the SOMPZ method was to completely drive our estimates, then the SOMPZ+WZ and the DNF+WZ estimates could be assumed to be independent. This is likely the case for the mean redshift estimates, as we have seen that WZ is not particularly constraining on the mean redshift (see Figure \ref{fig:dnf_vs_sompz_unc}). The width estimates are inferred more by the WZ measurements, and this might indicate that our tensions are under estimated, because we know that the two calibration methods share part of the WZ information. With this in mind, large tensions between means/widths of the two methods might indicate that either that the DNF+WZ uncertainties are under estimated, or there are some real differences between the two methods (one or both are biased). The reported values in Table \ref{tab:dnf_values} does not point to dramatic differences between the two methods: the most extreme statistical distance is 2.7$\sigma$ between means of Bin 2, and 2.3$\sigma$ between widths of Bin 6. 
%never exceed 3$\sigma$, with some bins showing differences at the 2$\sigma$ level, which does not point to dramatic differences between the two methods. 1.6 and 1.3

From Table \ref{tab:dnf_values} we note that SOMPZ+WZ uncertainties on the mean are larger than the DNF+WZ ones, while uncertainties on the widths are comparable. This is due to the fact that the uncertainties in the mean redshifts for the SOMPZ estimates are very sensitive to contributions from outliers at high redshift. The DNF+WZ mean redshift estimates (and uncertainties), on the other hand, are driven by the match with the WZ measurements with clipped tails, i.e., they do not take into account uncertainties in the tails, and are therefore smaller. The fact that the modelling of the tails is different between the two methodologies is also responsible for the slightly higher mean redshifts of the SOMPZ+WZ estimates compared to the DNF+WZ estimates. If we restrict the comparison of the aforementioned quantities in redshift intervals that exclude the tails of the distributions, the match between SOMPZ+WZ and DNF+WZ improves (Figure \ref{fig:dnf_vs_sompz_unc}). We further investigate the importance of the tails on the cosmological constraints in Appendix \ref{sect:notails}, finding that, despite them being important, they do not drive the main difference between the SOMPZ+WZ and DNF+WZ constraints.

\subsubsection{Galaxy-matter bias prior from WZ auto-correlation}

We tested the impact on the $\Lambda$CDM cosmological parameters of using the same broad prior on the \sys\ function describing the galaxy-matter bias as was done for the WL sample \citep{y3-sourcewz}. In this work we used more informative values computed from the clustering auto-correlation of the \maglim\ sample, the application of which is explained in more detail in Section \ref{sec:wzunc}. 
It is particularly interesting to look at the shape of distributions, especially for bin 2. Figure \ref{fig:nz_hmc_dataa} shows in grey the 1-sigma bands for the case without using the auto-correlation, and leaving a much broader prior. While in most bins the difference is not appreciable, and the grey bands are very similar to the solid bands, in bin 2 there is an evident difference. It is therefore suggested that this implementation of the auto-correlation information used as priors in the SOMPZ+WZ combination is able to help us constraining the galaxy-matter bias value, in a way that otherwise would not have been possible with traditional methods. %in a redshift range that is historically complex like the $z = 0.6$. 
In figure \ref{fig:dnf_vs_sompz_unc} is shown the comparison over mean redshift and width of the distributions between SOMPZ+WZ with the more informative prior from the auto-correlation, against the broad prior (labelled as ``SOMPZ+WZ (broad prior)''). The means and widths are well compatible with the standard SOMPZ+WZ results, and for bins 2 and 3 they are slightly closer to the DNF+WZ results. Even in bin 2, where the shape of the \nz\ is substantially different, the values of mean and width do not differ greatly from the standard case, reinforcing the notion that mean and width alone are not sufficient to fully characterise redshift distributions of a lens sample.
% SHOW NZ WITH AND WITHOUT AUTOCORRELATION SOMETIME BEFORE THIS POINT

\section{Cosmological Results}\label{cosmo_results}

In this section, we show the constraints on cosmological and nuisance parameters obtained using the DES Y3 measurements for galaxy-galaxy lensing and galaxy clustering \citep{y3-gglensing,y3-galaxyclustering} (a.k.a. 2x2pt), and the \nz from this paper. As in \cite{y3-2x2ptaltlensresults}, we also include in our analysis an additional likelihood constructed with the Shear Ratio (SR) measurements \citep{y3-shearratio}. This exploits galaxy-galaxy lensing signal at small scales (< 6 Mpc/h) to provide further constraint to the redshift distributions and intrinsic alignment parameters. The ratio of a galaxy-galaxy lensing signal of each lens sample redshift bin computed with respect to two source sample bins results in a primarily geometric measurement, which has been proven a powerful method for constraining systematics and nuisance parameters. This adds independent information from SOMPZ and WZ to the source redshift calibration.
% provide additional constrain- ing power and validation by measuring the galaxy-galaxy lensing signal of a lens galaxy redshift bin at small scales. The ratio of this signal from two source bins reflects the ratio of mean lens- ing efficiencies of objects in those source bins with respect to the lens bin redshift. This, in turn, depends on the redshift distribution of the sources. Because this methodology utilizes lensing signals, it is virtually independent from SOMPZ and clustering redshifts. The methodology of this analysis is described fully in Sánchez, Prat et al. (2020a). Both the clustering and shear ratio redshift con- straints are derived from data on small angular scales, which allows the redshift constraints to remain largely statistically independent of cosmological constraints based on larger-scale signals.
%
\newline
The posterior distribution obtained follows the Bayes theorem:
\begin{equation}
    P(p|D,M) \propto \mathcal{L}(D,p,M)\Pi(p|M),
\end{equation}
where $\Pi(p|M)$ is the prior distribution for all the parameters of the model $M$.
For the cosmological inference we use the \texttt{CosmoSIS} pipeline \citep{cosmosis}, and we sample the parameter posteriors using the \texttt{PolyChord} sampler \citep{poly1,poly2}.

Our data vector $D = \{w(\theta), \gamma_t(\theta)\}$ is compared to theoretical predictions $T(p)= \{w(\theta,p), \gamma_t(\theta,p)\}$ in a Bayesian fashion, and the posterior of the parameters conditional on the data is evaluated
by assuming a Gaussian likelihood for the data:
\begin{equation}
    \log{\mathcal{L}} \propto -\frac{1}{2} (D-T(p))^T C^{-1} (D-T(p)),
\end{equation}
where $C$ is the measurement covariance. In our analysis, we vary 5 (or 6) cosmological parameters assuming a $\Lambda$CDM (or wCDM) cosmology: $\Omega_{\rm m}$, $\sigma_8$, $n_s$, $\Omega_{\rm b}$, $h_{100}$, and $w$ for the wCDM case. Moreover, we also marginalise over ``astrophysical'' nuisance parameters (describing intrinsic alignment effects and the galaxy-matter bias of the lens sample), and calibration parameters (redshift uncertainties, shear measurement uncertainties). In short, our setup (covariance, parameters varied, prior ranges, etc.) is the same as the one adopted in \cite{y3-2x2ptaltlensresults}, except for the redshift $n(z)$ and uncertainties priors of the lens sample, where the ones obtained in this work have been assumed, and other minor changes that we describe below. All modelling and analysis choices, together with the calculations of the theoretical two-point functions, are described in detail in \cite{krause}.

Our analyses were not "blinded", since this work occurred after the "unblinding" of the DES Y3 3x2pt results. We did not perform any cosmological analysis until the redshift distributions were frozen; no changes to the redshift distributions (and uncertainties prior) have been performed after looking at the cosmological constraints. To ensure the robustness of our final estimates, we adopted a $p$-value criteria on the best-fitting models to our data vector. Following  \cite{y3-2x2ptaltlensresults}, we required the goodness-of-fit $p-$value on unblinded data vectors was larger than 1 per cent. The goodness-of-fit has been computed using the Predictive Posterior Distribution (PPD, \citealt{Doux2021}) and adopted in the main DES Y3 3x2pt analysis. The PPD methodology derives a calibrated probability-to-exceed $p$; in the case of goodness-of-fit tests, this is achieved by drawing realisations of the data vector for parameters drawn from the posterior under study which are then compared to actual observations. The distance metric ($\chi^2$) is computed in data space, which is then used to compute the $p$-value. 

Concerning the redshift uncertainties, as it is the primary goal of this work, we proceeded using the fiducial DES Y3 methodology: we parametrize the redshift uncertainties with two parameters for each tomographic bin, that modify a fiducial \nz distribution with a shift on the mean and a stretch on the width. The fiducial \nz is estimated by averaging the SOMPZ+WZ $n(z)$ realisations. The Gaussian priors on the mean and stretch parameters are centered at the mean and width of the fiducial \nz, while the Gaussian priors width are measured from the variance in the mean and width of the \nz ensemble. This parametrization can be compared directly to the fiducial DES Y3 2x2pt analysis \citep{y3-2x2ptaltlensresults}. In Appendix \ref{sampling} we describe an alternative marginalisation of the redshift uncertainties, by  marginalising over the full sets of \nz realisations provided by the SOMPZ+WZ method. In principle, this latter method describes better the redshift uncertainties of our method. However, we find that the currently available techniques that marginalise over the full ensemble of realisations during cosmology inference are prohibitively computationally expensive. Therefore we defer its application to future work.

Besides the different \nz, we also ran a few analyses where we marginalised over magnification parameters of the lens samples over wide priors. This is different from \cite{y3-2x2ptaltlensresults}, where magnification parameters have been fixed. 

For the fiducial DES Y3 2x2pt analysis, the p-value from the data-model $\chi^2$ using all six bins of \maglim\ was not sufficient to pass the 1 per cent criteria. After a series of tests the consensus was that the two highest redshift tomographic bins were responsible for worsening the fit. Therefore the analysis in \cite{y3-2x2ptaltlensresults} included only the first 4 \maglim\ bins. Here, we perform the analyses using all the 6 bins of the \maglim\ sample, but also using only the first 4 bins, to verify if the same applies also to this work using different redshift distributions.

In particular, we consider the following scenarios:

\begin{itemize}
    \item $\Lambda$CDM ($w$CDM); 4 and 6 lens bins, fixed magnification. This is the fiducial analysis that mirrors the one presented in \cite{y3-2x2ptaltlensresults}. Five (six) cosmological parameters are varied, including $\Omega_{\rm m}$, $\sigma_8$, $n_{\rm s}$, $\Omega_{\rm b}$, $h_{\rm 100}$ (and $w$ for the $w$CDM case). Intrinsic alignment, shear measurement and redshift uncertainties parameters (of both lenses and sources) and galaxy-matter linear biases of the lenses also are marginalised over. The magnification coefficients of the lens sample, however, are fixed to the values estimated from \texttt{Balrog} \citep{y3-balrog}. Uncertainties in the redshift distributions of the lens sample are modelled as a shift and stretch in the distributions.
    \item $\Lambda$CDM  ($w$CDM); 4  and 6 lens bins, free magnification. Same parameters as the ones above, but magnification parameters are marginalised over using Gaussian priors. This is an additional setup considered only after analysing the results from the aforementioned fixed magnification setup. 
    % \item $\Lambda$CDM  ($w$CDM); 4  and 6 lens bins, free magnification. Same as the ones above, but magnification parameters are marginalised over using wide flat priors.
\end{itemize}

% We also consider a couple of minor extra variations to the analysis choices below, which will be introduced when needed in the fiducial 2x2pt analysis.

In what follows, we will also quote results in terms of the $S_8$ parameter, defined as $S_8 \equiv \sigma_8(\Omega_{\rm m}/0.3)^{0.5}$. In Table \ref{tab:table_results} we summarise the best fit values of $S_8$, $\Omega_m$, $\sigma_8$, $w$, and the computed PPD goodness-of-fit p-value for all the different analyses.

\subsection{\texorpdfstring{$\Lambda$}\ CDM results}

\begin{figure*}
\includegraphics[width=0.45\textwidth]{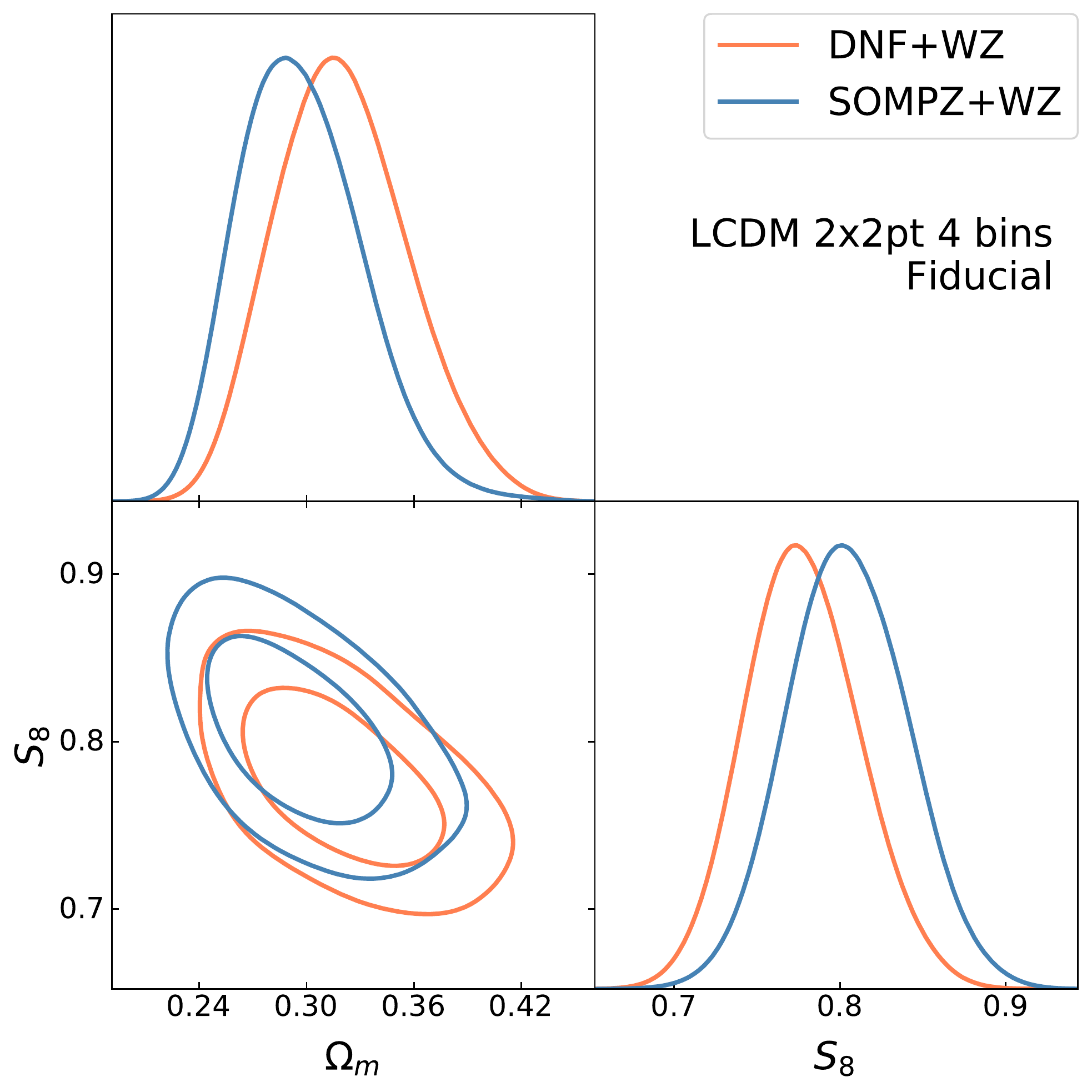}
\includegraphics[width=0.45\textwidth]{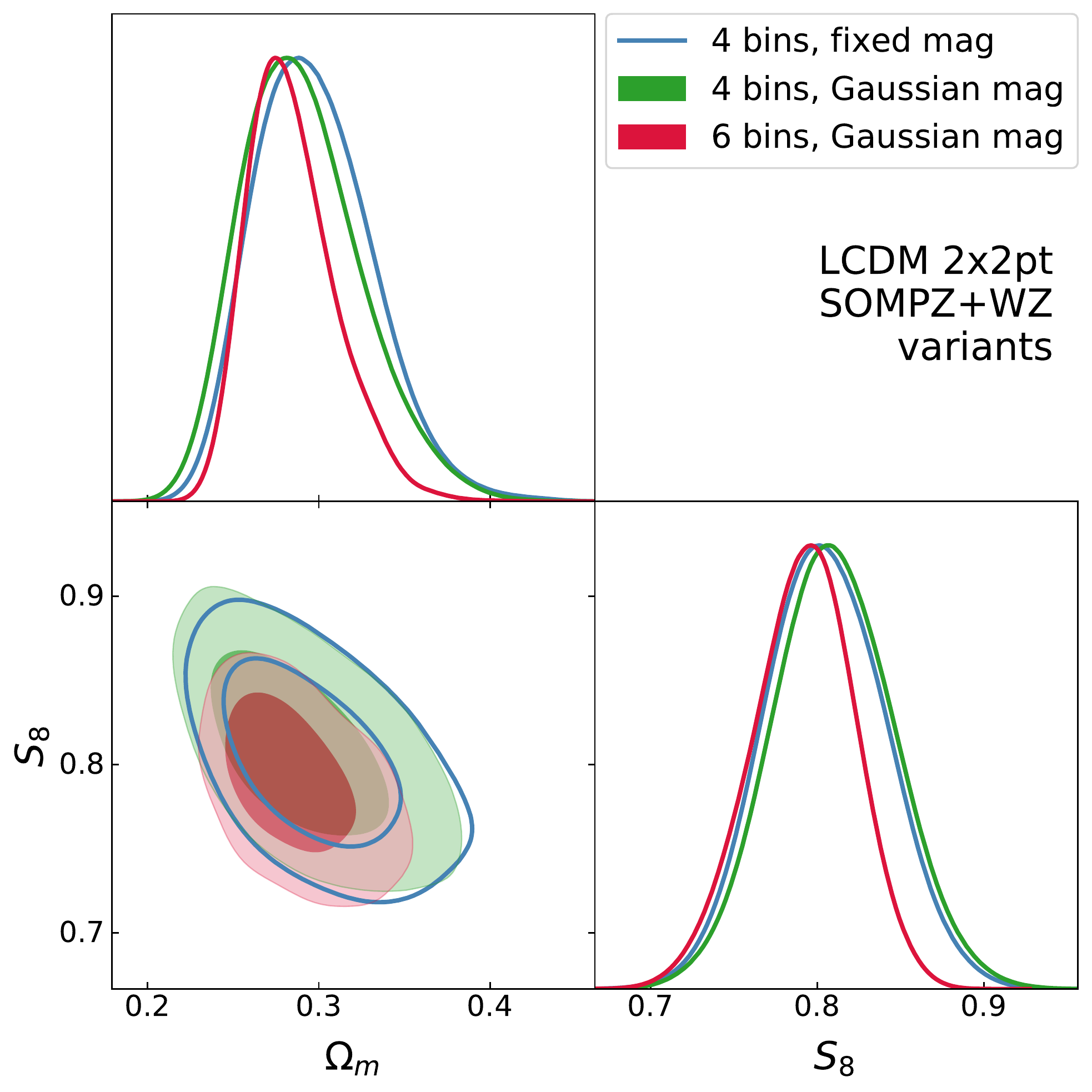}
     \caption{\textit{Left panel}: Posterior distributions of the cosmological parameters $\Omega_{\rm m}$, and $S_8$ for the $\Lambda$CDM analysis involving 4 bins and fixed magnification parameters. The ``fiducial'' posteriors have been obtained using the DNF+WZ redshift distributions, and they are compared to the ones obtained using the SOMPZ+WZ redshift distributions. \textit{Right panel}: Posterior distributions of the cosmological parameters $\Omega_{\rm m}$, and $S_8$ for the $\Lambda$CDM analysis for three different cases: 1) 4 bins and fixed magnification parameters (the blue contours in the two plots share the same analysis choices); 2) 4 bins and marginalised over magnification parameters (in solid green); 3) 6 bins and marginalising over magnification parameters (in solid red). The 2D marginalised contours in both of these figures show the 68 per cent and 95 per cent confidence levels.}
     \label{fig:contours}
\end{figure*}

\begin{table*}
%\tiny
\caption {Constraints on the cosmological parameters  $\Omega_{\rm m}$, $S_8$, and $\sigma_8$. For each parameter we report the mean of the posterior and the 68 per cent confidence interval. We also report the PPD goodness-of-fit $p$-value and the probability of the parameter difference (computed over the full parameter space) between the analyses considered in this work and \textit{Planck} TTTEEE0 lowl lowE \citep{aghanim2020planck}. The fiducial results from this work is reported in bold in the first row, while the official, fiducial results of DES Y3 are reported in bold in the second to last row.}
\centering
\begin{tabular}{cccccccccc}

\nz & Model & bins & Magnif. & $\Omega_{\mathrm{m}}$ & $S_8$  & $\sigma_8$ & $w$ & $p$-value & \textit{Planck}\\ %[0.1cm]

\hline
%\rule{{0pt}}{{4ex}} \multirow{4.5}{*}{\rotatebox{90}{\textbf{Fiducial}}}  & {2nd moments} & $ 0.799\pm0.015$  & $0.21\pm0.04$  & $0.98\pm0.10$ & 14.0/9.0  & 0.12   \\ 
%\rule{{0pt}}{{4ex}} & {3rd  moments} & $0.72\pm0.05$  & $0.33\pm0.16$  & $0.73\pm0.16$  & 12.7/10.3  & 0.26    \\ 
%\rule{{0pt}}{{4ex}}  & {2nd + 3rd moments} & $0.784\pm0.013$  & $0.27\pm0.03$  & $0.83\pm0.05$ & 10.9/9.1  & 0.29  \\ 

\textbf{SOMPZ+WZ} &  \textbf{$\Lambda$CDM} & \textbf{4} & \textbf{Fixed}                    & $\mathbf{0.30\pm0.04}$  & $\mathbf{0.81\pm0.04}$   & $\mathbf{0.81\pm0.07}$ &   $\mathbf{-}$ & $\mathbf{0.029}$  & $\mathbf{1.15\sigma}$ \\ 
SOMPZ & & & & & & \\
(broad prior)& $\Lambda$CDM & 4  & Fixed                    & $0.31\pm0.04$  & $0.76\pm0.06$   &  $0.76\pm0.09$ &   - & 0.037  & - \\

SOMPZ+WZ &  $\Lambda$CDM & 4  & Gauss.         & $0.29\pm0.04$   & $0.81\pm0.04$   & $0.83\pm0.08$   & - & 0.035    & 1.11$\sigma$ \\ 

SOMPZ+WZ &  $\Lambda$CDM & 6  & Fixed                    & -   & -  & - &   - & 0.008 & - \\

% SOMPZ &  $\Lambda$CDM & 4 bins & Gaussian          & $0.79\pm0.06$   & $0.30\pm0.05$  & $0.79\pm0.10$   & - & 0.035    \\ 

% {2x2pt SOMPZ $\Lambda$CDM 4 bins hyperrank}          & $0.80\pm0.04$   & $0.31\pm0.03$  & $0.79\pm0.07$   & -    \\ 

SOMPZ+WZ & $\Lambda$CDM & 6  & Gauss.          & $0.28\pm 0.03$ & $0.79\pm 0.03$     & $0.82\pm 0.06$   & - & 0.065   & 2.41$\sigma$ \\ 
% SOMPZ & $\Lambda$CDM & 6 bins & Gaussian           & $0.78\pm 0.08$   & $0.31\pm 0.04$  & $0.77\pm 0.09$   & - & 0.065    \\ 

SOMPZ+WZ &  $w$CDM & 4  & Fixed          & $0.29 \pm 0.04$ &  $0.79 \pm 0.06$  &  $0.81 \pm  0.08$ &      $-1.2 \pm 0.3$ & 0.032 & 0.46$\sigma$ \\

SOMPZ+WZ &  $w$CDM & 4  & Gauss.          & $0.29\pm 0.04$  & $0.79\pm 0.06$   & $0.81\pm 0.07$ &  $-1.0 \pm 0.3$ & 0.035  & 0.46$\sigma$  \\ 

SOMPZ+WZ &  $w$CDM & 6  & Fixed           & $0.30 \pm 0.04$ & $0.78 \pm 0.04$   & $0.78 \pm  0.06$  &     $-0.9 \pm 0.3$ & 0.012 & 2.29$\sigma$\\

% SOMPZ &  $w$CDM & 4 bins & Gaussian            & $0.76\pm 0.07$ & $0.30\pm 0.05$  & $0.77\pm 0.10$ &  $-1.0 \pm 0.3$ & 0.035    \\ 
SOMPZ+WZ &  $w$CDM & 6  & Gauss.           &  $0.31\pm 0.03$  & $0.83\pm 0.04$ & $0.82\pm 0.05$ &  $-0.7 \pm 0.2$ & 0.059   & 2.21$\sigma$ \\ 
% SOMPZ &  $w$CDM & 6 bins & Gaussian           & $0.78\pm 0.08$ &  $0.31\pm 0.04$  & $0.77\pm 0.09$ &  $-0.7 \pm 0.2$ & 0.059    \\ 

\hline
DNF+WZ &  $\Lambda$CDM & 4  & Fixed                      & $0.32\pm0.04$  & $0.78\pm0.04$   & $0.76\pm0.07$   & -   & 0.019 & 1.0$\sigma$\\ 
DNF+WZ &  $w$CDM & 4  & Fixed                              & $0.32\pm0.05$   & $0.78\pm0.05$   & $0.76\pm0.07$   & $-1.0 \pm 0.3$   & 0.024  & -\\ 

%wCDM 4 bins   -1.2 \pm 0.3
%wCDM 4 bins free mag   -1.0 \pm 0.3
%wCDM 6 bins   -0.9 \pm 0.3
%wCDM 6 bins free mag   -0.7 \pm 0.2
%$\Lambda$CDM 4 bins hyperrank   -1.0000 0.0000
%DNF $\Lambda$CDM 4 bins   -1.0000 0.0000
%DNF $\Lambda$CDM 4 bins free mag   -1.0000 0.0000
%DNF wCDM 4 bins   -1.0306 0.3135
%

\end{tabular}
\label{tab:table_results}
\end{table*}

\subsubsection{Fiducial results: 4 bins, fixed magnification and comparison with DNF results}

The first cosmological constraints we analyse are the ones obtained assuming a $\Lambda$CDM cosmology, using 4 lens bins and fixed magnification parameters. The decision on which set of results will be quoted as ''fiducial'' for this work had to be made before conducting any cosmological analysis on data. We initially planned to only run the fiducial analyses with fixed magnification, as in \cite{y3-2x2maglimforecast}. The choice between 4 or 6 lens bins would depend on the $p$-value criteria: if the $\Lambda$CDM, 6 bins, fixed magnification scenario were to yield a $p$-value above the specified threshold, then we would favour that configuration. This analysis though did not fulfil our $p$-value criteria ($p$-value = 0.008, see Table \ref{tab:table_results}), similarly as for the analysis ran with the same settings but using the fiducial redshift distributions from DNF; hence, we do not show those results here. We then chose as fiducial the $\Lambda$CDM, 4 bins, fixed magnification analysis, which is equivalent to the ''fiducial'' setup assumed in \cite{y3-2x2maglimforecast}, which also allows us to compare our results directly to the ones obtained using the DNF+WZ \nz . The posterior on the cosmological parameters $\Omega_{\rm m}$, and $S_8$ is shown in the left panel of Fig. \ref{fig:contours}; the marginalised mean values of $S_8$, $\Omega_{\rm m}$, and $\sigma_8$, along with the 68\% confidence intervals, are: 
\begin{linenomath} 
 % See https://tex.stackexchange.com/questions/461186
\begin{align}
\Omega_{\rm m} &{} =   0.30\pm 0.04,  \\%0.27^{0.05}_{-0.04} ()\\
\sigma_8 &{} = 0.81\pm 0.07,  \\
S_8 &{} =  0.81\pm 0.04.  
\end{align}
\end{linenomath}
\noindent 
%
%\textbf{This value is larger than the $p-$value ($p-$value = 0.019) obtained for the same type of analysis using the DNF+WZ \nz } .
The PPD goodness-of-fit test for this analysis results into $p-$value=0.029, well above our threshold (see also Table \ref{tab:table_results}). In the left panel of Fig. \ref{fig:contours} we also compare our results with the constraints obtained using the fiducial DNF+WZ \nz . The size of the posteriors is similar for the two cases, but the two posteriors are slightly shifted; the distance between the posteriors' peaks in the 2D $\Omega_{\rm m}-S_8$ plane is $d \sim 0.4\sigma$. In DES Y3 we impose a $0.3\sigma$ threshold for differences in the $\Omega_{\rm m}-S_8$ plane induced by different analysis choices, as larger statistical distances would indicate the presence of systematic uncertainties unaccounted for; these results would apparently violate this criteria. We note, however, that the (arbitrary) $0.3\sigma$ threshold adopted by DES refers to differences in the $\Omega_{\rm m}-S_8$ plane when \textit{noiseless} theory data vectors are assumed. In the presence of noisy data vectors these differences can become larger, without invalidating our criteria. Having said this, a $d \sim 0.4\sigma$ difference nonetheless show the large impact a different redshift calibration of the lens sample can have on the cosmological constraints. This is somewhat different from the results obtained for the source sample $n(z)$ \citep{y3-cosmicshear1}, where uncertainties in the redshift calibration had a negligible impact on the cosmological constraints.

In Section \ref{sec:wzunc} we explained how for the combination of the two methods we marginalise over possible functional forms of the unknown galaxy-matter bias of the \maglim\ sample, by means of the systematic function \sys\ in our clustering model. The prior on the parameters \textbf{s} is inferred from the clustering auto-correlation. We tested the impact on the redshift distributions of using a broader prior (the same used in \citealt*{y3-sompz}) in Section \ref{sec:results}. We have tested the impact of using these \nz\ for the cosmological inference, and found that there is no change in constraining power and no shift for $\Omega_{\rm m}$, but there is a shift on $S_8$ such to overlap with the fiducial results from DNF+WZ. Therefore it is clear that the information carried by the auto-correlation is crucial in our cosmological analysis.

\subsubsection{4 and 6 bins, free magnification} \label{sec:magnif}

% \begin{figure*}
% \includegraphics[width=\textwidth]{Figures/mag_plot.pdf}
%      \caption{Posterior distributions of magnification coefficients for 4 and 6 bins. The grey bands represent 68 per cent and 95 per cent confidence interval of the the prior values from \texttt{Balrog}; the colored lines represent the posteriors from the $\Lambda$CDM analyses, comparing SOMPZ+WZ and DNF+WZ posteriors.}
%      \label{fig:mag}
% \end{figure*}

As supplementary analyses, we then proceed to relax the fixed priors on the magnification parameters for the lens sample. Instead of fixing them to the values estimated from \cite{y3-2x2ptmagnification} (as done in the previous section), we leave them as free parameters, using Gaussian priors. In short, \cite{y3-2x2ptmagnification} estimate the magnification parameters using \texttt{Balrog}, by injecting fake galaxies into the wide field with and without applying a small magnification; the difference between the number of galaxies passing the selection in the two cases is then used to estimate the magnification parameters of the sample. These parameters come with a small uncertainty, which is however ignored in the fiducial analysis, as the magnification parameters are assumed to be fixed to the mean \texttt{Balrog} value. The central values and the uncertainties are reported in Table \ref{tab:parameters} in Appendix \ref{app:params}. One of the main reasons the DES Y3 fiducial analysis did not vary the magnification parameters was merely computational, as 4 (or 6) additional parameters lengthen the parameter inference process. In principle there is no reason to doubt these estimates. Differences might be caused by the fact that the \texttt{Balrog} injections do not completely sample the full DES Y3 footprint, or in case our injections were not fully representative of the DES sample we are analysing. 

When varying these parameters in our analyses, we find that the $p-$value computed using PPD indicates a good fit of the model to the data not only for the 4 bins case, but also for 6 bins case (see Table \ref{tab:table_results}). Adding the last 2 lens bins significantly improves the constraining power on $\Omega_{\rm m}$ by 30\% compared to the 4 bins case, whereas the constraints on $S_8$ are 20\% tighter.  % This is a general fact - at fixed analyses choices, the $p-$values of the analyses using the SOMP+WZ $n(z)$ are always larger than the ones obtained using the DNF+WZ $n(z)$.

% \subsubsection{Galaxy-matter bias prior from WZ auto-correlation}

% We tested the impact on the $\LambdaCDM$ cosmological parameters of using the same broad prior on the $Sys(\textbf{s})$ function describing the galaxy-matter bias as was done for the WL sample \citep{y3-sourcewz}. In this work we used more informative values computed from the clustering auto-correlation of the \maglim\ sample. In Section \ref{sec:wzunc} was explained how the
% SHOW NZ WITH AND WITHOUT AUTOCORRELATION SOMETIME BEFORE THIS POINT

\subsection{wCDM Results}
\begin{figure}
\includegraphics[width=0.49\textwidth]{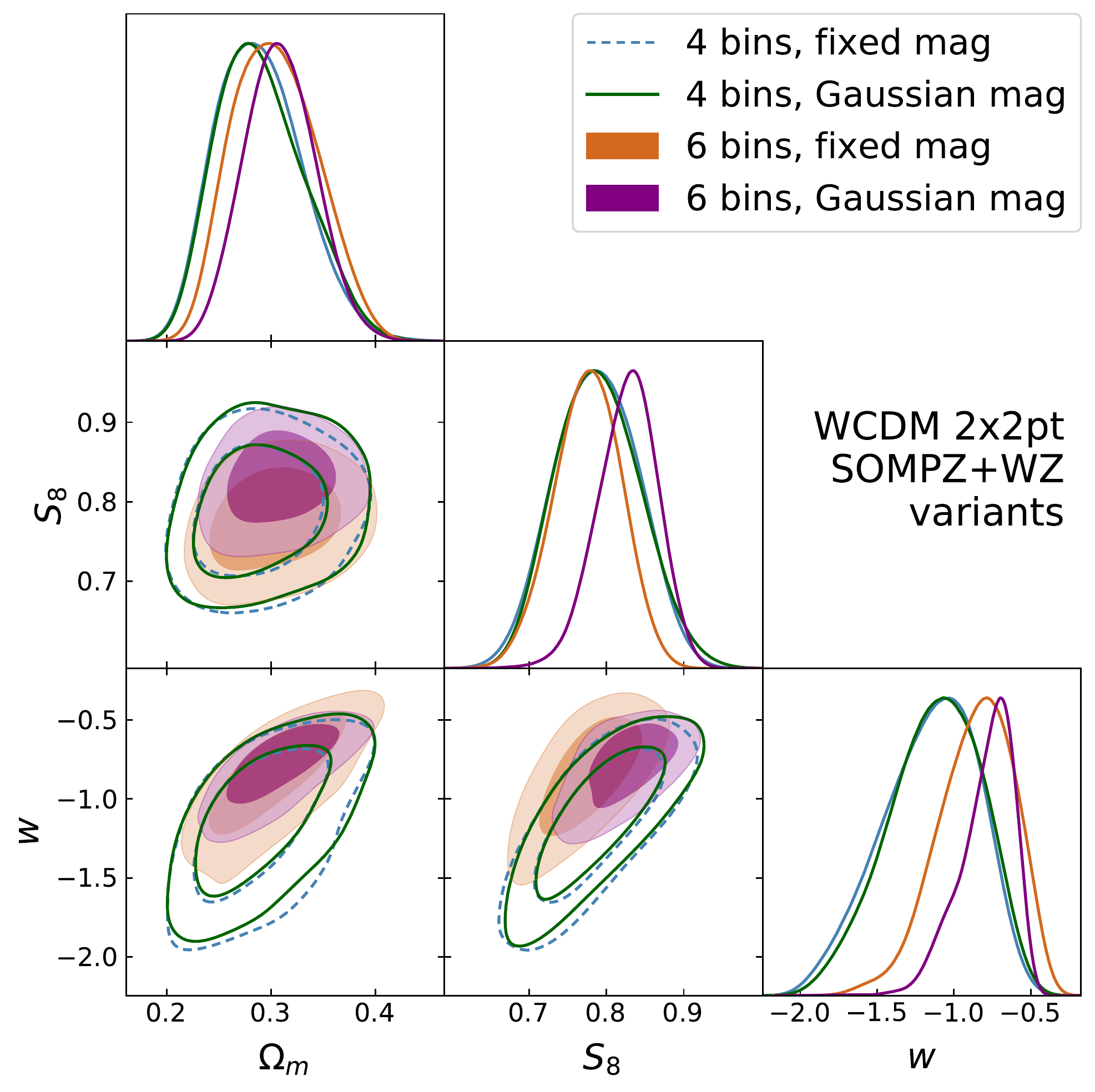}
     \caption{Posterior distributions of the cosmological parameters $\Omega_{\rm m}$, and $S_8$ and $w$ for four different cases: 1) wCDM, 4 bins and fixed magnification parameters; 2) wCDM, 6 bins and fixed magnification parameters, 3) wCDM, 4 bins and free magnification parameters; 4) wCDM, 6 bins and free magnification parameters.  The 2D marginalised contours in these figures show the 68 per cent and 95 per cent confidence levels. We note that the posteriors of $w$ for the 6 bins cases are partially affected by the prior edge ($w \in [-2,-0.33]$, Table \ref{tab:parameters}); see text for more details.
     }
     \label{fig:2x2pt_wcdm}
\end{figure}

We then proceed to analyse the results obtained with $w$CDM, for all four cases: 4 and 6 bins, fixed and free magnification, as described in the previous section. Parameter posteriors are shown in Fig. \ref{fig:2x2pt_wcdm}, whereas p-values and parameters constraints are reported in Table \ref{tab:table_results}. All the reported p-values are above our $p=0.01$ threshold. 

In general, the 2x2pt constraints on $w$ are loose and affected by the prior ($-2<w<-0.3$), but compatible with a $\Lambda$CDM scenario. With respect to $\Lambda$CDM 4 bins case, freeing $w$ loosens the constraint on $S_8$ (both with fixed and with free magnification) by $\sim30\%$, while leaves it unvaried for $\Omega_m$.  For the 6 bins, we are unable to directly compare to the fixed magnification case, but for the free magnification the constraint on $S_8$ is $\sim25\%$ looser, while, similarly to the 4 bins case, it is unvaried for $\Omega_{\rm m}$.

Passing from the 4 bins to the 6 bins configuration, besides increasing the constraints on $S_8$, also the constraints on $w$ improves (by $\sim$20\%), although part of the improvement is due to the posterior partially hitting the prior edge. 

Freeing the magnification parameters slightly shifts $w$ towards the upper edge of the prior ($w=-0.3$), and $S_8$ slightly towards higher values, due to a degeneracy between $w$, $S_8$, and the magnification parameters of the two highest lens bins, which are now fairly broad (see Table \ref{tab:parameters}). Such a shift is not present in the case of 4 bins, as the Gaussian priors used for the first 4 magnification parameters are much tighter.

%. For the more constraining 6 bins case this results in the posterior on $S_8$ shifting towards larger values. The constraints on $\Omega_{\rm m}$ are unaltered in all four scenarios. 
%Freeing the magnification parameters shifts $w$ towards the upper edge of the prior ($w=-0.3$), and contrary to the $\Lambda$CDM case, leaves unaltered the constraints on $S_8$. This indicates that in the $w$CDM case, the values of the magnification parameters are more degenerate with $w$ rather than with $S_8$.

%\begin{figure*}
%\includegraphics[width=\textwidth]{Figures/xlens.pdf}
%     \caption{X lens
%     }
%     \label{fig:xlens}
%\end{figure*}
%
%
%
%\subsubsection{X lens}
%% - 1 plot with X lens
%As mentioned earlier, the cosmological analysis run with 6 bins with the fiducial %redshift distributions did not meet the criteria set for unblinding. An in depth %investigation highlighted that the effect could be parametrised by adding a factor %called "$X_lens$" which tends to 1 at low redshift, but drives to lower values in the %last two bins. The cause of this could be various, and should be searched for in %observational systematics not fully accounted for, or..\giulia{finish this}
%The unblinding criteria is met, but the $X_{lens}$ posterior is still not centered %around 1 in bin 5 and 6, suggesting that this effect has little to do with the fit of %our data to the $\LambdaCDM$ model. 
%
%
%

\subsection{Statistical distance to Planck}

We compute here the statistical distances between our cosmological constraints and the early Universe ones from the \textit{Planck} satellite \citep{aghanim2020planck}. To this aim, we used the algorithm presented in \cite{Raveri2021}, which estimates the probability of tension between parameters via Monte Carlo approximation. In particular, the probability of tension between parameters can be expressed as follows:

\begin{equation} \label{Eq:ParameterDifferencePDF}
\mathcal{P}(\Delta \theta) = \int_{V_p} \mathcal{P}_A(\theta) \mathcal{P}_B(\theta-\Delta \theta) d\theta,
\end{equation}
where $V_p$ represents the prior volume, while $\mathcal{P}_A$ and $\mathcal{P}_B$ represent two posterior parameter distributions under study. The probability of having a shift in the parameter space is described by the parameter shifts density:
\begin{equation} \label{Eq:ParamShiftProbability}
\Delta = \int_{\mathcal{P}(\Delta\theta)>\mathcal{P}(0)} \mathcal{P}(\Delta\theta) \, d\Delta\theta,
\end{equation}
This refers to the posterior portion beyond the constant probability contour for no shift, $\Delta\theta=0$. The integration in Eq. (\ref{Eq:ParamShiftProbability}) is performed via Monte Carlo techniques.

The comparison between the results has been performed considering all the parameters shared by our analyses and \textit{Planck}. The values are reported in the last column of Table \ref{tab:table_results}; we find no sign of significant tension ($<3\sigma$) in any of the analysis setups considered. In particular, we find that for the 4 bins case for $\Lambda$CDM (both fixed and free magnification) there is good agreement ($1.15\sigma, 1.11\sigma$), similarly for wCDM with 4 bins we have $0.46\sigma$ for both fixed and free magnification. For the 6 bins cases the values are larger ($2.2-2.4\sigma$), but still below the $3\sigma$ threshold.

\section{Conclusions}
\label{sec:conclusions}

In this paper, we presented an alternative calibration of the \maglim\ lens sample 
redshift distributions from the Dark Energy Survey (DES) first three years of data (Y3). This new method, which has already been applied to the DES Y3 weak lensing sample \citep*{y3-sompz}, is based on a combination of a Self-Organising Maps (SOMPZ) based scheme and clustering redshifts (WZ) to estimate redshift distributions and inherent uncertainties. The original redshift calibration of the \maglim\ sample (and cosmological results obtained adopting that calibration) have been originally presented in \cite{y3-2x2ptaltlensresults}, and has been based on the photo-$z$ code DNF \citep{DNF2016} and WZ constraints \citep{y3-lenswz}. The methodology presented in this paper is meant to be more accurate than the original one. First, the SOMPZ method allows a better control over all the potential sources of uncertainties affecting the estimates compared to DNF; second, the clustering constraints (WZ) are incorporated through a rigorous joint likelihood framework which allows to draw \nz\ samples conditioned on both clustering and photometric measurements, improving the \nz\ estimates (e.g., the final ``SOMPZ+WZ'' \nz\ have a smaller scatter, or uncertainty, compared to the SOMPZ ones, see Figure \ref{fig:nz_hmc_dataa}). 

We described in detail the methodology followed to produce the alternative \maglim\ \nz\ based on the SOMPZ+WZ approach, together with a detailed report on the main systematics dominating our calibration error budget. Our redshift uncertainties, in particular, are dominated by the impact of sample variance on the SOMPZ estimate (due to the limited area spanned by the deep field sample used in the calibration) and by the effect of the redshift evolution of the galaxy-matter bias of the \maglim\ sample on the WZ constraints. We then compared our SOMPZ+WZ \nz\ with the fiducial DNF+WZ \nz\ estimates; the means and widths of the 6 \maglim\ tomographic bins show moderate statistical distances, with the largest deviation of $2.7\sigma$ in bin 2 (see Table \ref{tab:dnf_values}). We also found the uncertainties on mean of the redshift distributions of the SOMPZ+WZ method to be slightly larger than the ones of the DNF+WZ method, due to a more conservative calibration of the tails of the redshift distributions. On the other hand, we found the two methods to have a similar constraining power on the widths of the distributions.

We then proceeded investigating the impact on the cosmological constraints of our new redshift calibration. In particular, we used the DES Y3 galaxy-galaxy lensing and galaxy clustering measurements \citep{y3-gglensing,y3-galaxyclustering} (a.k.a. 2x2pt), and the \nz\ from this work, and compared to the results from  \cite{y3-2x2ptaltlensresults}. In the ``fiducial'' configuration, which involves using the first 4 lens bins and assuming a $\Lambda$CDM cosmology, we obtained as marginalised mean values $\Omega_{\rm m} =   0.30\pm 0.04$, $\sigma_8  = 0.81\pm 0.07 $ and $S_8 =  0.81\pm 0.04$. We noted a $\sim 0.4\sigma$ shift in the $\Omega_{\rm}-S_8$ plane compared to the \cite{y3-2x2ptaltlensresults} results, but no change in terms of constraining power. The shift indicates that the redshift calibration of the lens sample plays a key role on cosmological constraints from the 2x2pt analysis, contrary to the redshift calibration of the source sample \citep{y3-cosmicshear1}. Subsequently, we explored different analysis setups; we tested the case where all the 6 \maglim\ redshift bins were included, a scenario where the magnification coefficients of the lens sample were marginalised during the inference, and last, we assumed a $w$CDM cosmology. We found that the inclusion of the last two redshift bins of the \maglim\ sample help improving the constraints on $\Omega_{\rm m}$ by $\sim 25\%$, and on $S_8$ by $\sim 20\%$.

We also compared our results to the cosmological constraints from \textit{Planck} \citep{aghanim2020planck}, finding a no-tension of $1.15 \sigma$  between the results when 4 lens bins where considered. We did find a statistical distance of $2.41 \sigma$ in $\Lambda$CDM with free magnification coefficients when including in the analysis the two high redshift bins ($z>0.85$), which have not been included in the fiducial DES Y3 analysis \citep{y3-2x2ptaltlensresults}.

As a final comment, despite the SOMPZ+WZ method's ability to produce \nz\ samples capturing the redshift uncertainties of our estimates, we could not efficiently marginalise over these realisation during the cosmological inference, due to computational constraints. Our marginalisation strategy followed the one adopted in  \cite{y3-2x2ptaltlensresults}: we adopted the mean of the SOMPZ+WZ samples as our fiducial \nz, and marginalised over a shift in the mean and a stretch of the width of the distribution, using as priors the variances in the mean and widths of the SOMPZ+WZ \nz\ samples. While this strategy was deemed sufficient for this current work, we plan to implement the full marginalisation scheme for subsequent analyses of the lens samples with DES Y6 data.

\section*{Acknowledgements}

The project leading to these results have received funding from ''la Caixa'' Foundation (ID 100010434), under the fellowship LCF/BQ/DI17/11620053 and has received funding from the European Union's Horizon 2020 research and innovation programme under the Marie Sk\l{}odowska-Curie grant agreement No. 713673.

Funding for the DES Projects has been provided by the U.S. Department of Energy, the U.S. National Science Foundation, the Ministry of Science and Education of Spain, 
the Science and Technology Facilities Council of the United Kingdom, the Higher Education Funding Council for England, the National Center for Supercomputing 
Applications at the University of Illinois at Urbana-Champaign, the Kavli Institute of Cosmological Physics at the University of Chicago, 
the Center for Cosmology and Astro-Particle Physics at the Ohio State University,
the Mitchell Institute for Fundamental Physics and Astronomy at Texas A\&M University, Financiadora de Estudos e Projetos, 
Funda{\c c}{\~a}o Carlos Chagas Filho de Amparo {\`a} Pesquisa do Estado do Rio de Janeiro, Conselho Nacional de Desenvolvimento Cient{\'i}fico e Tecnol{\'o}gico and 
the Minist{\'e}rio da Ci{\^e}ncia, Tecnologia e Inova{\c c}{\~a}o, the Deutsche Forschungsgemeinschaft and the Collaborating Institutions in the Dark Energy Survey.

The Collaborating Institutions are Argonne National Laboratory, the University of California at Santa Cruz, the University of Cambridge, Centro de Investigaciones Energ{\'e}ticas, 
Medioambientales y Tecnol{\'o}gicas-Madrid, the University of Chicago, University College London, the DES-Brazil Consortium, the University of Edinburgh, 
the Eidgen{\"o}ssische Technische Hochschule (ETH) Z{\"u}rich, 
Fermi National Accelerator Laboratory, the University of Illinois at Urbana-Champaign, the Institut de Ci{\`e}ncies de l'Espai (IEEC/CSIC), 
the Institut de F{\'i}sica d'Altes Energies, Lawrence Berkeley National Laboratory, the Ludwig-Maximilians Universit{\"a}t M{\"u}nchen and the associated Excellence Cluster Universe, 
the University of Michigan, the National Optical Astronomy Observatory, the University of Nottingham, The Ohio State University, the University of Pennsylvania, the University of Portsmouth, 
SLAC National Accelerator Laboratory, Stanford University, the University of Sussex, Texas A\&M University, and the OzDES Membership Consortium.

Based in part on observations at Cerro Tololo Inter-American Observatory at NSF's NOIRLab (NOIRLab Prop. ID 2012B-0001; PI: J. Frieman), which is managed by the Association of Universities for Research in Astronomy (AURA) under a cooperative agreement with the National Science Foundation.

The DES data management system is supported by the National Science Foundation under Grant Numbers AST-1138766 and AST-1536171.
The DES participants from Spanish institutions are partially supported by MINECO under grants AYA2015-71825, ESP2015-66861, FPA2015-68048, SEV-2016-0588, SEV-2016-0597, and MDM-2015-0509, 
some of which include ERDF funds from the European Union. IFAE is partially funded by the CERCA program of the Generalitat de Catalunya.
Research leading to these results has received funding from the European Research
Council under the European Union's Seventh Framework Program (FP7/2007-2013) including ERC grant agreements 240672, 291329, and 306478.
We  acknowledge support from the Brazilian Instituto Nacional de Ci\^encia
e Tecnologia (INCT) e-Universe (CNPq grant 465376/2014-2).

This manuscript has been authored by Fermi Research Alliance, LLC under Contract No. DE-AC02-07CH11359 with the U.S. Department of Energy, Office of Science, Office of High Energy Physics.

Funding for the Sloan Digital Sky Survey IV has been provided by the Alfred P. Sloan Foundation, the U.S. Department of Energy Office of Science, and the Participating Institutions. SDSS acknowledges support and resources from the Center for High-Performance Computing at the University of Utah. The SDSS web site is www.sdss.org.

SDSS is managed by the Astrophysical Research Consortium for the Participating Institutions of the SDSS Collaboration including the Brazilian Participation Group, the Carnegie Institution for Science, Carnegie Mellon University, Center for Astrophysics | Harvard \& Smithsonian (CfA), the Chilean Participation Group, the French Participation Group, Instituto de Astrofisica de Canarias, The Johns Hopkins University, Kavli Institute for the Physics and Mathematics of the Universe (IPMU) / University of Tokyo, the Korean Participation Group, Lawrence Berkeley National Laboratory, Leibniz Institut f{\"u}r Astrophysik Potsdam (AIP), Max-Planck-Institut f{\"u}r Astronomie (MPIA Heidelberg), Max-Planck-Institut f{\"u}r Astrophysik (MPA Garching), Max-Planck-Institut f{\"u}r Extraterrestrische Physik (MPE), National Astronomical Observatories of China, New Mexico State University, New York University, University of Notre Dame, Observatorio Nacional / MCTI, The Ohio State University, Pennsylvania State University, Shanghai Astronomical Observatory, United Kingdom Participation Group, Universidad Nacional Aut\'{o}noma de M\'{e}xico, University of Arizona, University of Colorado Boulder, University of Oxford, University of Portsmouth, University of Utah, University of Virginia, University of Washington, University of Wisconsin, Vanderbilt University, and Yale University.

\section{Data Availability}
The DES Y3 data products used in this work, as well as the full ensemble of DES Y3 \maglim\ sample redshift distributions described by this work, are publicly available at https://des.ncsa.illinois.edu/releases. As cosmology likelihood sampling software we use \texttt{cosmosis}, available at https://github.com/joezuntz/cosmosis.

%%%%%%%%%%%%%%%%%%%%%%%%%%%%%%%%%%%%%%%%%%%%%%%%%%


\bibliographystyle{mn2e_2author_arxiv_amp.bst}

% \bibliographystyle{mnras_2author}

% \bibliography{export-bibtex,des_y3kp}
\bibliography{references}

\begin{thebibliography}{99}
\providecommand{\natexlab}[1]{#1}
\providecommand{\url}[1]{\texttt{#1}}
\providecommand{\urlprefix}{URL }
\providecommand{\eprint}[1][]{\url{#1}}

\bibitem[{{Abbott} et~al.(2018){Abbott} \& {Abdalla} et~al.}]{DES_DR1}
{Abbott}, T.~M.~C., {Abdalla}, F.~B., {Allam}, S., et~al., 2018, \apjs, 239, 2,
  18, \eprint arXiv:{1801.03181}

\bibitem[{{Abolfathi} et~al.(2018){Abolfathi} \& {Aguado}
  et~al.}]{Abolfathi2017}
{Abolfathi}, B., {Aguado}, D.~S., {Aguilar}, G., et~al., 2018, \apjs, 235, 2,
  42, \eprint arXiv:{1707.09322}

\bibitem[{Aghanim et~al.(2020)Aghanim \& Akrami et~al.}]{aghanim2020planck}
Aghanim, N., Akrami, Y., Ashdown, M., et~al., 2020, Astronomy \& Astrophysics,
  641, A6

\bibitem[{{Ahumada} et~al.(2020){Ahumada} \& {Allende Prieto} et~al.}]{dr16}
{Ahumada}, R., {Allende Prieto}, C., {Almeida}, A., et~al., 2020, \apjs, 249,
  1, 3, \eprint arXiv:{1912.02905}

\bibitem[{{Aihara} et~al.(2018){Aihara} \& {Arimoto} et~al.}]{Aihara2018}
{Aihara}, H., {Arimoto}, N., {Armstrong}, R., et~al., 2018, \pasj, 70, S4,
  \eprint arXiv:{1704.05858}

\bibitem[{{Alam} et~al.(2021){Alam} \& {de Mattia} et~al.}]{Alam2020}
{Alam}, S., {de Mattia}, A., {Tamone}, A., et~al., 2021, \mnras, 504, 4, 4667,
  \eprint arXiv:{2007.09004}

\bibitem[{{Alarcon} et~al.(2021){Alarcon} \& {Gaztanaga} et~al.}]{Alarcon2020}
{Alarcon}, A., {Gaztanaga}, E., {Eriksen}, M., et~al., 2021, \mnras, 501, 4,
  6103, \eprint arXiv:{2007.11132}

\bibitem[{{Amon} et~al.(2022){Amon} \& {Gruen} et~al.}]{y3-cosmicshear1}
{Amon}, A., {Gruen}, D., {Troxel}, M.~A., et~al., 2022, \prd, 105, 2, 023514,
  \eprint arXiv:{2105.13543}

\bibitem[{{Becker}(2013)}]{Becker2013}
{Becker}, M.~R., 2013, \mnras, 435, 115

\bibitem[{{Behroozi} et~al.(2013){Behroozi} \& {Wechsler} \&
  {Wu}}]{Behroozi2013}
{Behroozi}, P.~S., {Wechsler}, R.~H., {Wu}, H.-Y., 2013, \apj, 762, 2, 109,
  \eprint arXiv:{1110.4372}

\bibitem[{{Benjamin} et~al.(2013){Benjamin} \& {Van Waerbeke}
  et~al.}]{benjamin2013}
{Benjamin}, J., {Van Waerbeke}, L., {Heymans}, C., et~al., 2013, \mnras, 431,
  2, 1547, \eprint arXiv:{1212.3327}

\bibitem[{{Blanton} et~al.(2017){Blanton} \& {Bershady} et~al.}]{Blanton2017}
{Blanton}, M.~R., {Bershady}, M.~A., {Abolfathi}, B., et~al., 2017, \aj, 154,
  1, 28, \eprint arXiv:{1703.00052}

\bibitem[{{Bonnett} et~al.(2016){Bonnett} \& {Troxel} et~al.}]{dessv-photoz}
{Bonnett}, C., {Troxel}, M.~A., {Hartley}, W., et~al., 2016, \prd, 94, 4,
  042005, \eprint arXiv:{1507.05909}

\bibitem[{{Buchs} et~al.(2019){Buchs} \& {Davis} et~al.}]{y3-sompzbuzzard}
{Buchs}, R., {Davis}, C., {Gruen}, D., et~al., 2019, \mnras, 489, 1, 820,
  \eprint arXiv:{1901.05005}

\bibitem[{{Cawthon} et~al.(2022){Cawthon} \& {Elvin-Poole} et~al.}]{y3-lenswz}
{Cawthon}, R., {Elvin-Poole}, J., {Porredon}, A., et~al., 2022, \mnras, 513, 4,
  5517, \eprint arXiv:{2012.12826}

\bibitem[{{Conroy} et~al.(2006){Conroy} \& {Wechsler} \&
  {Kravtsov}}]{Conroy2006}
{Conroy}, C., {Wechsler}, R.~H., {Kravtsov}, A.~V., 2006, \apj, 647, 1, 201,
  \eprint arXiv:{astro-ph/0512234}

\bibitem[{{Cordero} et~al.(2022){Cordero} \& {Harrison} et~al.}]{y3-hyperrank}
{Cordero}, J.~P., {Harrison}, I., {Rollins}, R.~P., et~al., 2022, \mnras, 511,
  2, 2170, \eprint arXiv:{2109.09636}

\bibitem[{{Cunha} et~al.(2012){Cunha} \& {Huterer} \& {Busha} \&
  {Wechsler}}]{Cunha2012}
{Cunha}, C.~E., {Huterer}, D., {Busha}, M.~T., {Wechsler}, R.~H., 2012, \mnras,
  423, 1, 909, \eprint arXiv:{1109.5691}

\bibitem[{{Davis} et~al.(2017){Davis} \& {Gatti} et~al.}]{Davis2017}
{Davis}, C., {Gatti}, M., {Vielzeuf}, P., et~al., 2017, arXiv e-prints,
  arXiv:1710.02517, \eprint arXiv:{1710.02517}

\bibitem[{{Davis} \& {Peebles}(1983)}]{DavisPeebles1983}
{Davis}, M., {Peebles}, P.~J.~E., 1983, \apj, 267, 465

\bibitem[{{Dawson} et~al.(2016){Dawson} \& {Kneib} et~al.}]{eboss2016}
{Dawson}, K.~S., {Kneib}, J.-P., {Percival}, W.~J., et~al., 2016, \aj, 151, 2,
  44, \eprint arXiv:{1508.04473}

\bibitem[{{Dawson} et~al.(2013){Dawson} \& {Schlegel} et~al.}]{boss}
{Dawson}, K.~S., {Schlegel}, D.~J., {Ahn}, C.~P., et~al., 2013, \aj, 145, 1,
  10, \eprint arXiv:{1208.0022}

\bibitem[{{De Vicente} et~al.(2016){De Vicente} \& {S{\'a}nchez} \&
  {Sevilla-Noarbe}}]{DNF2016}
{De Vicente}, J., {S{\'a}nchez}, E., {Sevilla-Noarbe}, I., 2016, \mnras, 459,
  3, 3078, \eprint arXiv:{1511.07623}

\bibitem[{{DeRose} et~al.(2019){DeRose} \& {Wechsler} et~al.}]{DeRose2018}
{DeRose}, J., {Wechsler}, R.~H., {Becker}, M.~R., et~al., 2019, arXiv e-prints,
  arXiv:1901.02401, \eprint arXiv:{1901.02401}

\bibitem[{{DeRose} et~al.(2022){DeRose} \& {Wechsler}
  et~al.}]{y3-simvalidation}
{DeRose}, J., {Wechsler}, R.~H., {Becker}, M.~R., et~al., 2022, \prd, 105, 12,
  123520, \eprint arXiv:{2105.13547}

\bibitem[{{DES Collaboration}(2022)}]{y3-3x2ptkp}
{DES Collaboration}, 2022, \prd, 105, 2, 023520, \eprint arXiv:{2105.13549}

\bibitem[{{Doux} et~al.(2021){Doux} \& {Baxter} et~al.}]{Doux2021}
{Doux}, C., {Baxter}, E., {Lemos}, P., et~al., 2021, \mnras, 503, 2, 2688

\bibitem[{{Eisenstein} et~al.(2011){Eisenstein} \& {Weinberg}
  et~al.}]{Eisenstein2011}
{Eisenstein}, D.~J., {Weinberg}, D.~H., {Agol}, E., et~al., 2011, \aj, 142, 3,
  72, \eprint arXiv:{1101.1529}

\bibitem[{{Elvin-Poole} et~al.(2021)}]{y3-2x2ptmagnification}
{Elvin-Poole}, J., et~al., 2021, To be submitted to MNRAS

\bibitem[{{Eriksen} et~al.(2019){Eriksen} \& {Alarcon} et~al.}]{Eriksen2019}
{Eriksen}, M., {Alarcon}, A., {Gaztanaga}, E., et~al., 2019, \mnras, 484, 3,
  4200, \eprint arXiv:{1809.04375}

\bibitem[{{Everett} et~al.(2022){Everett} \& {Yanny} et~al.}]{y3-balrog}
{Everett}, S., {Yanny}, B., {Kuropatkin}, N., et~al., 2022, \apjs, 258, 1, 15,
  \eprint arXiv:{2012.12825}

\bibitem[{{Flaugher} et~al.(2015){Flaugher} \& {Diehl} et~al.}]{Flaugher2015}
{Flaugher}, B., {Diehl}, H.~T., {Honscheid}, K., et~al., 2015, \aj, 150, 150,
  \eprint arXiv:{1504.02900}

\bibitem[{{Garilli} et~al.(2014){Garilli} \& {Guzzo} et~al.}]{Garilli2014}
{Garilli}, B., {Guzzo}, L., {Scodeggio}, M., et~al., 2014, VizieR Online Data
  Catalog, J/A+A/562/A23

\bibitem[{{Gatti} et~al.(2022){Gatti} \& {Giannini} et~al.}]{y3-sourcewz}
{Gatti}, M., {Giannini}, G., {Bernstein}, G.~M., et~al., 2022, \mnras, 510, 1,
  1223, \eprint arXiv:{2012.08569}

\bibitem[{{Gatti} et~al.(2021){Gatti} \& {Sheldon} et~al.}]{y3-shapecatalog}
{Gatti}, M., {Sheldon}, E., {Amon}, A., et~al., 2021, \mnras, 504, 3, 4312,
  \eprint arXiv:{2011.03408}

\bibitem[{{Gatti} et~al.(2018){Gatti} \& {Vielzeuf} et~al.}]{Gatti2018}
{Gatti}, M., {Vielzeuf}, P., {Davis}, C., et~al., 2018, \mnras, 477, 1664,
  \eprint arXiv:{1709.00992}

\bibitem[{{G{\'o}rski} \& {Hivon}(2011)}]{healpix}
{G{\'o}rski}, K.~M., {Hivon}, E., 2011, {HEALPix: Hierarchical Equal Area
  isoLatitude Pixelization of a sphere}, Astrophysics Source Code Library,
  record ascl:1107.018, \eprint ascl:{1107.018}

\bibitem[{{Gschwend} et~al.(2018){Gschwend} \& {Rossel} et~al.}]{Gschwend2018}
{Gschwend}, J., {Rossel}, A.~C., {Ogando}, R.~L.~C., et~al., 2018, Astronomy
  and Computing, 25, 58, \eprint arXiv:{1708.05643}

\bibitem[{{Gunn} et~al.(2006){Gunn} \& {Siegmund} et~al.}]{Gunn2006}
{Gunn}, J.~E., {Siegmund}, W.~A., {Mannery}, E.~J., et~al., 2006, \aj, 131, 4,
  2332, \eprint arXiv:{astro-ph/0602326}

\bibitem[{{Handley} et~al.(2015{\natexlab{a}}){Handley} \& {Hobson} \&
  {Lasenby}}]{poly1}
{Handley}, W.~J., {Hobson}, M.~P., {Lasenby}, A.~N., 2015{\natexlab{a}},
  \mnras, 450, L61

\bibitem[{{Handley} et~al.(2015{\natexlab{b}}){Handley} \& {Hobson} \&
  {Lasenby}}]{poly2}
{Handley}, W.~J., {Hobson}, M.~P., {Lasenby}, A.~N., 2015{\natexlab{b}},
  \mnras, 453, 4, 4384

\bibitem[{{Hartley} et~al.(2022){Hartley} \& {Choi} et~al.}]{y3-deepfields}
{Hartley}, W.~G., {Choi}, A., {Amon}, A., et~al., 2022, \mnras, 509, 3, 3547,
  \eprint arXiv:{2012.12824}

\bibitem[{{Hildebrandt} et~al.(2021){Hildebrandt} \& {van den Busch}
  et~al.}]{Hildebrandt2020}
{Hildebrandt}, H., {van den Busch}, J.~L., {Wright}, A.~H., et~al., 2021, \aap,
  647, A124, \eprint arXiv:{2007.15635}

\bibitem[{{Hildebrandt} et~al.(2017){Hildebrandt} \& {Viola}
  et~al.}]{Hildebrandt2017}
{Hildebrandt}, H., {Viola}, M., {Heymans}, C., et~al., 2017, \mnras, 465, 1454,
  \eprint arXiv:{1606.05338}

\bibitem[{{Hoyle} et~al.(2018){Hoyle} \& {Gruen} et~al.}]{desy1-photoz}
{Hoyle}, B., {Gruen}, D., {Bernstein}, G.~M., et~al., 2018, \mnras, 478, 1,
  592, \eprint arXiv:{1708.01532}

\bibitem[{{Huff} \& {Mandelbaum}(2017)}]{HuffMcal2017}
{Huff}, E., {Mandelbaum}, R., 2017, arXiv e-prints, arXiv:1702.02600, \eprint
  arXiv:{1702.02600}

\bibitem[{{Huterer} et~al.(2013){Huterer} \& {Cunha} \& {Fang}}]{huterer2013}
{Huterer}, D., {Cunha}, C.~E., {Fang}, W., 2013, \mnras, 432, 4, 2945, \eprint
  arXiv:{1211.1015}

\bibitem[{{Huterer} et~al.(2006){Huterer} \& {Takada} \& {Bernstein} \&
  {Jain}}]{Huterer2006}
{Huterer}, D., {Takada}, M., {Bernstein}, G., {Jain}, B., 2006, \mnras, 366, 1,
  101, \eprint arXiv:{astro-ph/0506030}

\bibitem[{{Jarvis} et~al.(2021){Jarvis} \& {Bernstein} et~al.}]{y3-piff}
{Jarvis}, M., {Bernstein}, G.~M., {Amon}, A., et~al., 2021, \mnras, 501, 1,
  1282, \eprint arXiv:{2011.03409}

\bibitem[{{Jarvis} et~al.(2013){Jarvis} \& {Bonfield} et~al.}]{Jarvis2013}
{Jarvis}, M.~J., {Bonfield}, D.~G., {Bruce}, V.~A., et~al., 2013, \mnras, 428,
  1281, \eprint arXiv:{1206.4263}

\bibitem[{{Johnson} et~al.(2017){Johnson} \& {Blake} et~al.}]{Johnson2017}
{Johnson}, A., {Blake}, C., {Amon}, A., et~al., 2017, \mnras, 465, 4118,
  \eprint arXiv:{1611.07578}

\bibitem[{{Joudaki} et~al.(2020){Joudaki} \& {Hildebrandt}
  et~al.}]{Joudaki2019}
{Joudaki}, S., {Hildebrandt}, H., {Traykova}, D., et~al., 2020, \aap, 638, L1,
  \eprint arXiv:{1906.09262}

\bibitem[{Kohonen(1982)}]{kohonensom}
Kohonen, T., 1982, Biological Cybernetics, 43, 1, 59, ISSN 0340-1200

\bibitem[{{Krause} et~al.(2021){Krause} \& {Fang} et~al.}]{krause}
{Krause}, E., {Fang}, X., {Pandey}, S., et~al., 2021, arXiv e-prints,
  arXiv:2105.13548, \eprint arXiv:{2105.13548}

\bibitem[{{Kuijken} et~al.(2015){Kuijken} \& {Heymans} et~al.}]{Kuijken2015}
{Kuijken}, K., {Heymans}, C., {Hildebrandt}, H., et~al., 2015, \mnras, 454, 4,
  3500, \eprint arXiv:{1507.00738}

\bibitem[{{Laigle} et~al.(2016){Laigle} \& {McCracken} et~al.}]{Laigle2016}
{Laigle}, C., {McCracken}, H.~J., {Ilbert}, O., et~al., 2016, \apjs, 224, 24

\bibitem[{{Laureijs} et~al.(2011){Laureijs} \& {Amiaux} et~al.}]{Laureijs2011}
{Laureijs}, R., {Amiaux}, J., {Arduini}, S., et~al., 2011, arXiv e-prints,
  arXiv:1110.3193, \eprint arXiv:{1110.3193}

\bibitem[{{Le F{\`e}vre} et~al.(2013){Le F{\`e}vre} \& {Cassata} et~al.}]{vvds}
{Le F{\`e}vre}, O., {Cassata}, P., {Cucciati}, O., et~al., 2013, \aap, 559,
  A14, \eprint arXiv:{1307.0545}

\bibitem[{{Lehmann} et~al.(2017){Lehmann} \& {Mao} \& {Becker} \& {Skillman} \&
  {Wechsler}}]{Lehmann2017}
{Lehmann}, B.~V., {Mao}, Y.-Y., {Becker}, M.~R., {Skillman}, S.~W., {Wechsler},
  R.~H., 2017, \apj, 834, 37, \eprint arXiv:{1510.05651}

\bibitem[{{Lidman} et~al.(2020){Lidman} \& {Tucker} et~al.}]{Lidman2020}
{Lidman}, C., {Tucker}, B.~E., {Davis}, T.~M., et~al., 2020, \mnras, 496, 1,
  19, \eprint arXiv:{2006.00449}

\bibitem[{{Lilly} et~al.(2009){Lilly} \& {Le Brun} et~al.}]{Lilly09zcosmos}
{Lilly}, S.~J., {Le Brun}, V., {Maier}, C., et~al., 2009, \apjs, 184, 2, 218

\bibitem[{{Limber}(1953)}]{Limber}
{Limber}, D.~N., 1953, \apj, 117, 134

\bibitem[{{LSST Science Collaboration} et~al.(2009){LSST Science Collaboration}
  \& {Abell} et~al.}]{Abell2009}
{LSST Science Collaboration}, {Abell}, P.~A., {Allison}, J., et~al., 2009,
  arXiv e-prints, arXiv:0912.0201, \eprint arXiv:{0912.0201}

\bibitem[{{Lupton} et~al.(1999){Lupton} \& {Gunn} \& {Szalay}}]{lupton1999}
{Lupton}, R.~H., {Gunn}, J.~E., {Szalay}, A.~S., 1999, \aj, 118, 1406, \eprint
  arXiv:{astro-ph/9903081}

\bibitem[{{MacCrann} et~al.(2022){MacCrann} \& {Becker} et~al.}]{y3-imagesims}
{MacCrann}, N., {Becker}, M.~R., {McCullough}, J., et~al., 2022, \mnras, 509,
  3, 3371, \eprint arXiv:{2012.08567}

\bibitem[{{Masters} et~al.(2017){Masters} \& {Stern} et~al.}]{Masters2017}
{Masters}, D.~C., {Stern}, D.~K., {Cohen}, J.~G., et~al., 2017, \apj, 841, 111

\bibitem[{{Masters} et~al.(2019){Masters} \& {Stern} et~al.}]{C3R2_DR2}
{Masters}, D.~C., {Stern}, D.~K., {Cohen}, J.~G., et~al., 2019, \apj, 877, 2,
  81, \eprint arXiv:{1904.06394}

\bibitem[{{McCracken} et~al.(2012){McCracken} \& {Milvang-Jensen}
  et~al.}]{McCracken2012}
{McCracken}, H.~J., {Milvang-Jensen}, B., {Dunlop}, J., et~al., 2012, \aap,
  544, A156, \eprint arXiv:{1204.6586}

\bibitem[{{McQuinn} \& {White}(2013)}]{McQuinn2013}
{McQuinn}, M., {White}, M., 2013, \mnras, 433, 2857, \eprint arXiv:{1302.0857}

\bibitem[{{M{\'e}nard} et~al.(2013){M{\'e}nard} \& {Scranton}
  et~al.}]{Menard2013}
{M{\'e}nard}, B., {Scranton}, R., {Schmidt}, S., et~al., 2013, arXiv e-prints,
  arXiv:1303.4722, \eprint arXiv:{1303.4722}

\bibitem[{{Morganson} et~al.(2018){Morganson} \& {Gruendl}
  et~al.}]{Morganson2018}
{Morganson}, E., {Gruendl}, R.~A., {Menanteau}, F., et~al., 2018, \pasp, 130,
  989, 074501, \eprint arXiv:{1801.03177}

\bibitem[{{Morrison} et~al.(2017){Morrison} \& {Hildebrandt}
  et~al.}]{Morrison2017}
{Morrison}, C.~B., {Hildebrandt}, H., {Schmidt}, S.~J., et~al., 2017, \mnras,
  467, 3576, \eprint arXiv:{1609.09085}

\bibitem[{{Myles} et~al.(2020){Myles} \& {Alarcon} et~al.}]{y3-sompz}
{Myles}, J., {Alarcon}, A., {Amon}, A., et~al., 2020, arXiv e-prints,
  arXiv:2012.08566, \eprint arXiv:{2012.08566}

\bibitem[{{Myles} et~al.(2023){Myles} \& {Gruen} et~al.}]{pitpz}
{Myles}, J., {Gruen}, D., {Amon}, A., et~al., 2023, \mnras, 519, 2, 1792,
  \eprint arXiv:{2210.03130}

\bibitem[{{Newman}(2008)}]{Newman2008}
{Newman}, J.~A., 2008, \apj, 684, 88, \eprint arXiv:{0805.1409}

\bibitem[{{Porredon} et~al.(2021{\natexlab{a}}){Porredon} \& {Crocce}
  et~al.}]{y3-2x2ptaltlensresults}
{Porredon}, A., {Crocce}, M., {Elvin-Poole}, J., et~al., 2021{\natexlab{a}},
  arXiv e-prints, arXiv:2105.13546, \eprint arXiv:{2105.13546}

\bibitem[{{Porredon} et~al.(2021{\natexlab{b}}){Porredon} \& {Crocce}
  et~al.}]{y3-2x2maglimforecast}
{Porredon}, A., {Crocce}, M., {Fosalba}, P., et~al., 2021{\natexlab{b}}, \prd,
  103, 4, 043503, \eprint arXiv:{2011.03411}

\bibitem[{{Prakash} et~al.(2016){Prakash} \& {Licquia} et~al.}]{prakash16}
{Prakash}, A., {Licquia}, T.~C., {Newman}, J.~A., et~al., 2016, \apjs, 224, 2,
  34, \eprint arXiv:{1508.04478}

\bibitem[{{Prat} et~al.(2022){Prat} \& {Blazek} et~al.}]{y3-gglensing}
{Prat}, J., {Blazek}, J., {S{\'a}nchez}, C., et~al., 2022, \prd, 105, 8,
  083528, \eprint arXiv:{2105.13541}

\bibitem[{{Raichoor} et~al.(2017){Raichoor} \& {Comparat} et~al.}]{raichoor17}
{Raichoor}, A., {Comparat}, J., {Delubac}, T., et~al., 2017, \mnras, 471, 4,
  3955, \eprint arXiv:{1704.00338}

\bibitem[{{Raveri} \& {Doux}(2021)}]{Raveri2021}
{Raveri}, M., {Doux}, C., 2021, \prd, 104, 4, 043504, \eprint
  arXiv:{2105.03324}

\bibitem[{{Reid} et~al.(2016){Reid} \& {Ho} et~al.}]{reid16}
{Reid}, B., {Ho}, S., {Padmanabhan}, N., et~al., 2016, \mnras, 455, 2, 1553,
  \eprint arXiv:{1509.06529}

\bibitem[{{Rodr{\'\i}guez-Monroy} et~al.(2022){Rodr{\'\i}guez-Monroy} \&
  {Weaverdyck} et~al.}]{y3-galaxyclustering}
{Rodr{\'\i}guez-Monroy}, M., {Weaverdyck}, N., {Elvin-Poole}, J., et~al., 2022,
  \mnras, 511, 2, 2665, \eprint arXiv:{2105.13540}

\bibitem[{{Rozo} et~al.(2016){Rozo} \& {Rykoff} et~al.}]{Rozo2016}
{Rozo}, E., {Rykoff}, E.~S., {Abate}, A., et~al., 2016, \mnras, 461, 1431,
  \eprint arXiv:{1507.05460}

\bibitem[{{S{\'a}nchez} et~al.(2022){S{\'a}nchez} \& {Prat}
  et~al.}]{y3-shearratio}
{S{\'a}nchez}, C., {Prat}, J., {Zacharegkas}, G., et~al., 2022, \prd, 105, 8,
  083529, \eprint arXiv:{2105.13542}

\bibitem[{{S{\'a}nchez} et~al.(2020){S{\'a}nchez} \& {Raveri} \& {Alarcon} \&
  {Bernstein}}]{Sanchez2020}
{S{\'a}nchez}, C., {Raveri}, M., {Alarcon}, A., {Bernstein}, G.~M., 2020,
  \mnras, 498, 2, 2984, \eprint arXiv:{2004.09542}

\bibitem[{{Scodeggio} et~al.(2018){Scodeggio} \& {Guzzo}
  et~al.}]{scodeggio2018}
{Scodeggio}, M., {Guzzo}, L., {Garilli}, B., et~al., 2018, \aap, 609, A84,
  \eprint arXiv:{1611.07048}

\bibitem[{{Scottez} et~al.(2018){Scottez} \& {Benoit-L{\'e}vy} \& {Coupon} \&
  {Ilbert} \& {Mellier}}]{Scottez2018}
{Scottez}, V., {Benoit-L{\'e}vy}, A., {Coupon}, J., {Ilbert}, O., {Mellier},
  Y., 2018, \mnras, 474, 3921, \eprint arXiv:{1705.02629}

\bibitem[{{Secco} et~al.(2022){Secco} \& {Samuroff} et~al.}]{y3-cosmicshear2}
{Secco}, L.~F., {Samuroff}, S., {Krause}, E., et~al., 2022, \prd, 105, 2,
  023515, \eprint arXiv:{2105.13544}

\bibitem[{{Sevilla} et~al.(2011){Sevilla} \& {Armstrong} et~al.}]{Sevilla2011}
{Sevilla}, I., {Armstrong}, R., {Bertin}, E., et~al., 2011, arXiv e-prints,
  arXiv:1109.6741, \eprint arXiv:{1109.6741}

\bibitem[{{Sevilla-Noarbe} et~al.(2021){Sevilla-Noarbe} \& {Bechtol}
  et~al.}]{y3-gold}
{Sevilla-Noarbe}, I., {Bechtol}, K., {Carrasco Kind}, M., et~al., 2021, \apjs,
  254, 2, 24, \eprint arXiv:{2011.03407}

\bibitem[{{Sheldon} \& {Huff}(2017)}]{SheldonMcal2017}
{Sheldon}, E.~S., {Huff}, E.~M., 2017, \apj, 841, 24, \eprint
  arXiv:{1702.02601}

\bibitem[{{Smee} et~al.(2013){Smee} \& {Gunn} et~al.}]{Smee2013}
{Smee}, S.~A., {Gunn}, J.~E., {Uomoto}, A., et~al., 2013, \aj, 146, 2, 32,
  \eprint arXiv:{1208.2233}

\bibitem[{{Springel}(2005)}]{springel2005}
{Springel}, V., 2005, \mnras, 364, 1105, \eprint arXiv:{astro-ph/0505010}

\bibitem[{{Suchyta} et~al.(2016){Suchyta} \& {Huff} et~al.}]{Suchyta2016}
{Suchyta}, E., {Huff}, E.~M., {Aleksi{\'c}}, J., et~al., 2016, \mnras, 457, 786

\bibitem[{{Tessore} \& {Harrison}(2020)}]{tessore}
{Tessore}, N., {Harrison}, I., 2020, The Open Journal of Astrophysics, 3, 1, 6,
  \eprint arXiv:{2003.11558}

\bibitem[{{van den Busch} et~al.(2020){van den Busch} \& {Hildebrandt}
  et~al.}]{vandenBusch2020}
{van den Busch}, J.~L., {Hildebrandt}, H., {Wright}, A.~H., et~al., 2020, arXiv
  e-prints, arXiv:2007.01846, \eprint arXiv:{2007.01846}

\bibitem[{{Wright} et~al.(2020){Wright} \& {Hildebrandt} et~al.}]{wright2020}
{Wright}, A.~H., {Hildebrandt}, H., {van den Busch}, J.~L., et~al., 2020, \aap,
  640, L14, \eprint arXiv:{2005.04207}

\bibitem[{{Zuntz} et~al.(2015){Zuntz} \& {Paterno} et~al.}]{cosmosis}
{Zuntz}, J., {Paterno}, M., {Jennings}, E., et~al., 2015, Astronomy and
  Computing, 12, 45, \eprint arXiv:{1409.3409}

\end{thebibliography}

%%%%%%%%%%%%%%%%%%%%%%%%%%%%%%%%%%%%%%%%%%%%%%%%%%

%%%%%%%%%%%%%%%%% APPENDICES %%%%%%%%%%%%%%%%%%%%%
\appendix
\section{\maglim sample in simulations}\label{app:buzz_maglim}
\begin{figure*}
    \centering
    \includegraphics[width=18cm]{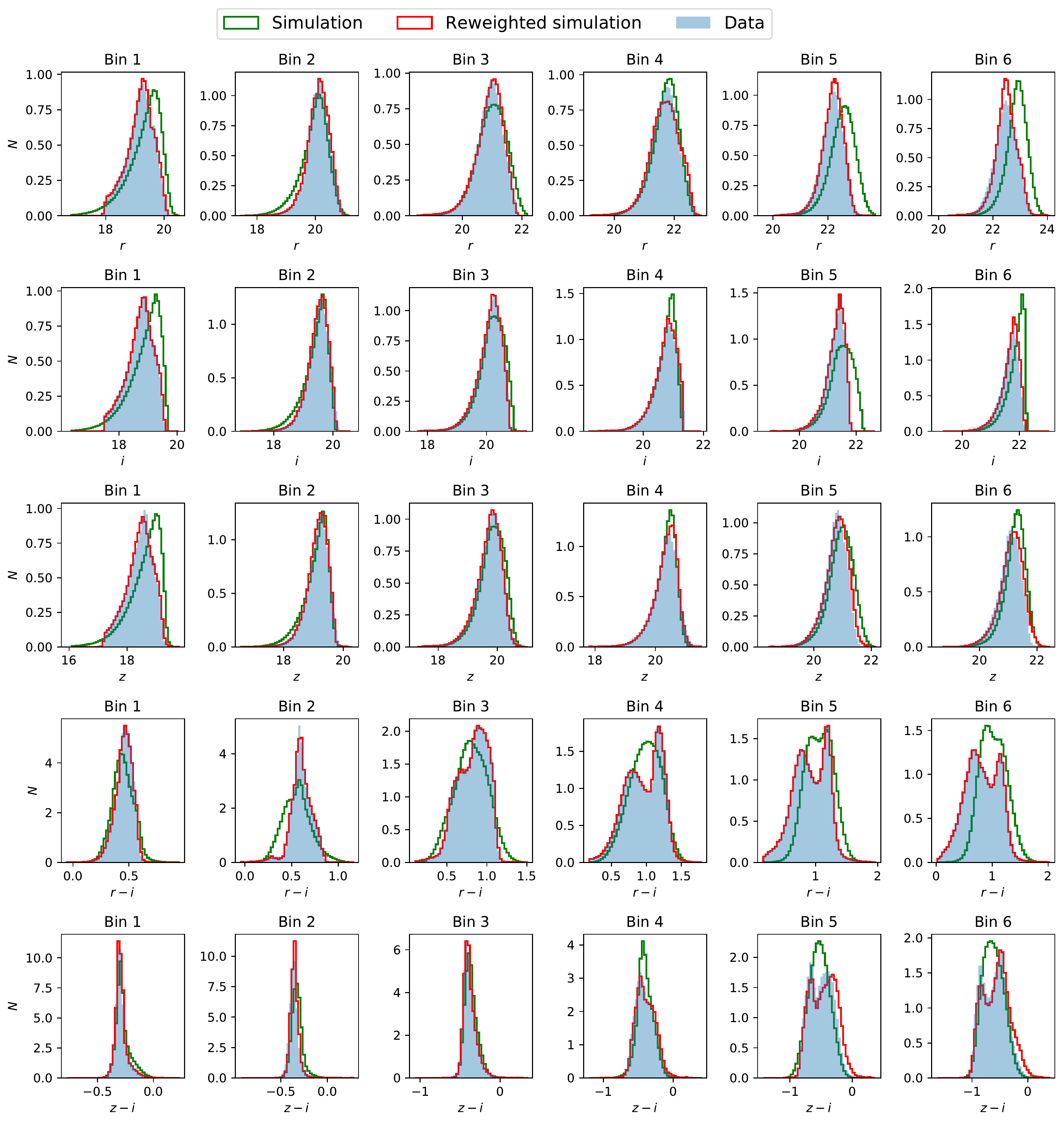}
    \caption{Comparison of $riz$-band magnitudes and $r-i$, $z-i$ colors of the 6 bins of the \maglim sample, between data (blue) and simulations, before (green) and after re-weighting (red). The re-weighting process has proven successful in yielding magnitude distributions that closely resemble those observed in the actual data.}
    \label{fig:mockmaglim}
\end{figure*}

Due to the small but existing differences in magnitude/color space between the Buzzard simulation and the DES data \citep{DeRose2018}, we expect the simulated sample to not be a perfect copy of the data sample, although we do not expect this to have a sensible impact on any of the conclusions drawn in this work.
The direct application of the fiducial \maglim selection (Eq.~\ref{eq:maglim_sel}) to the Buzzard catalog leads to slightly different number densities and color distributions with respect to data. We therefore re-define a more adequate \maglim selection for Buzzard, with the goal of achieving the same number density as the data sample. The new Buzzard \maglim selection is a piece-wise linear selection in redshift and magnitude, similar to Eq.~\ref{eq:maglim_sel} but with coefficients re-defined by minimising the quadratic sum of the difference in number density with the values in data, for each tomographic bin, in order to avoid discontinuities in the selection. Such a re-defined selection guarantees similar number densities as the data sample. We then ensure similar color distributions by an additional re-weighting procedure of the mock catalog, so as to resemble the color distributions of the data sample. In particular, we iteratively re-weight based on $i$, $r$ magnitudes and $i$-$r$ colors, with the final distributions matching closely the data ones, as shown in Figure \ref{fig:mockmaglim}.

The new \maglim selection in Buzzard for each tomographic bin is then the following: 
\begin{itemize}
    \item Bin 1:\quad  $i < $ 2.017 $ * z_{\rm mean}$ + 18.882
    \item Bin 2:\quad  $i < $ 2.687 $ * z_{\rm mean}$ + 18.614
    \item Bin 3:\quad  $i < $ 5.705 $ * z_{\rm mean}$ + 16.954
    \item Bin 4:\quad  $i < $ 2.399 $ * z_{\rm mean}$ + 19.268
    \item Bin 5:\quad  $i < $ 9.455 $ * z_{\rm mean}$ + 13.271
    \item Bin 6:\quad  $i < $-0.960 $* z_{\rm mean}$ + 23.165
\end{itemize}

We list in Table \ref{tab:buzz_numbdens} the number densities of \maglim in Buzzard, obtained with the fiducial selection and with the adapted in simulations. 

\begin{table*}
    \centering
    \begin{tabular}{c c c c c c c}
  \textbf{Number density}  & Bin 1 & Bin 2 & Bin 3 & Bin 4 & Bin 5 & Bin 6 \\

\hline
Before (Fiducial \maglim selection) & 1.10 & 0.90 & 1.12 & 0.97 & 0.69 & 0.76\\
After (Buzzard \maglim selection) & 0.98 & 0.99 & 0.99 & 0.99 & 0.99 & 0.98 \\

        \hline

    \end{tabular}
    \caption{Number densities of the \maglim sample in Buzzard as obtained with the fiducial \maglim selection, and with the one adapted for Buzzard. }
    \label{tab:buzz_numbdens}
 \end{table*}

\section{Validation in simulations}\label{sec:resultssims}

% cosa da mostrare:
% \begin{itemize}
%     \item N(z) di sompz+wz 
%     \item discutere dell'impatto di wz 
%     \item discutere dell'impatto del prior sulla bias evolution del maglim sample (forse)
%     \item far vedere che la <z> and std true e quelle di sompz+wz sono compatible within uncertainties
%     \item introdurre in un paragrafo la parametrizzazione degli errori (consideriamo shift, shift+stretch e hyprerrank) e far vedere che in simulations ottieni cose sensate, validando la metodologia
% \end{itemize}

% nei dati piu' o meno farei vedere le stesse cose; aggiungendo un paragrafo sulle differenze con la simulazione (confrontando le distribuzioni delle mean/std, facendo notare che i dati hanno piu' varianza)
% inoltre devi confrontarti con dnf, sia a livello di n(z) (con un plot), sia a livello di <z> e std (plot/tabella), sia a livello di constraints.
% \label{sec:results}

The validity of our methodology and pipeline has been tested in the Buzzard N-body simulation, introduced in Section \ref{sec:buzz_}. The measurements of redshift distributions using both phenotypes and clustering were validated in simulations to ensure unbiased estimates with respect to the true redshift distributions. The \maglim\ sample has been recreated in the Buzzard simulations, as described in Section \ref{sec:data}. The sample selection has been altered to reproduce as faithfully as possible the number density and color distributions of the data.

\begin{figure}
\centering
\includegraphics[width=\linewidth]{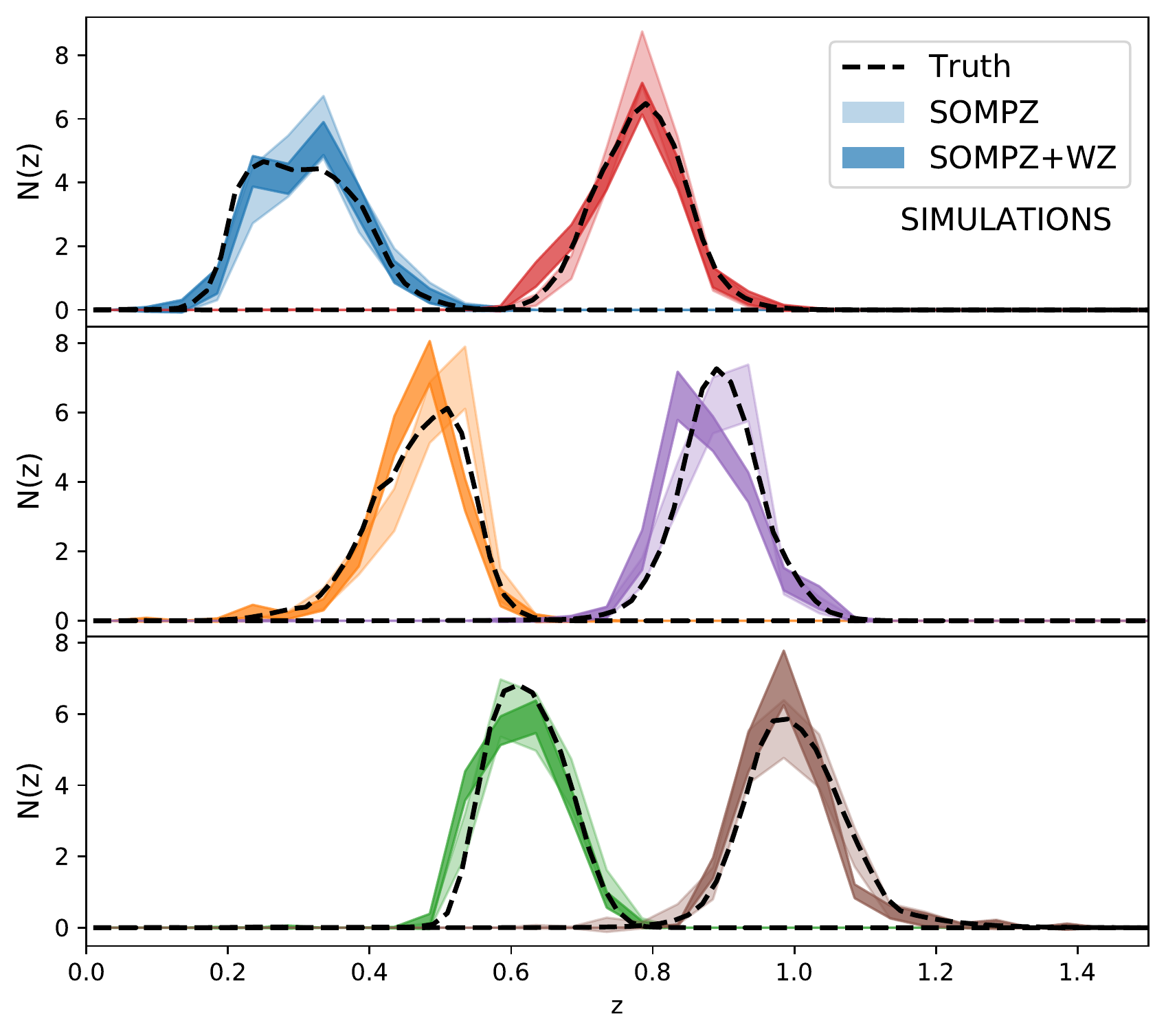}
\caption{Estimated $n(z)$ in four tomographic bins using a 12x12 cell deep SOM and 32x32 cell wide SOM trained on Buzzard simulations. In the top row we have bin 1 and 4, in the middle row bin 2 and 5, and in the bottom row bin 3 and 6. The Redshift sample used here has 100000 galaxies drawn from 1.38 deg$^2$, such that after the \maglim\ selection it yields $\sim15000$ unique galaxies, which is the same order of magnitude as the redshift samples in data, see Table \ref{tab:specz}. The deep sample is drawn from three fields of size  $3.32$, $3.29$, and $1.94$ deg$^2$, respectively from the Buzzard simulated sky catalog. The black dashed line marks the true value, the transparent bands are the \textit{3sDir} set of \nz\ and the solid bands are the realisations once combined with clustering redshifts. We can appreciate the effect of the combined likelihood, resulting in distributions more constrained in terms of shape, and still consistent with the truth. \label{fig:nz_ensemble_buzz}}
\end{figure}

As described in section \ref{sec:3sdir}, we generated 300 simulated deep field realizations that we used to estimate the SOMPZ method uncertainty, which we report in Table \ref{tab:unc}, and add into our overall error budget. Here we illustrate that the uncertainty predicted by the \textit{3sDir} and the \textit{3sDir}+WZ models is consistent with the true \nz in one of these simulated realizations.
We start by selecting one of these simulated realizations, which includes the four deep fields and their corresponding Balrog and redshift samples. We then proceeded to perform the \textit{3sDir} analytical sample variance estimation for that one specific realisation. The geometry and resolution of the SOM used in simulations are the same as the ones used in data. There are two differences between our simulated scenario and real data: 1) we use the true redshifts from the Buzzard simulations; 2) we assume all redshift information comes from one of our four deep fields. This latter point matches the modeling assumption of \textit{3sDir}, which also assumes that the redshift information only comes from one out of four fields. This is a conservative choice that inflates the modeled error due to sample variance in real data for the term p(z|c), and it avoids modeling the highly non-trivial selection function of spectroscopic samples coming from fields other than the COSMOS field. We note that the sample variance contribution to the color distribution p(c) is modeled correctly as coming from all 4 fields. The SOMPZ redshift distributions, and their uncertainties estimated through the \textit{3sDir} method, are in agreement with the true distribution, as shown in Figure \ref{fig:nz_ensemble_buzz}. In Table \ref{tab:buzz} we summarise the mean and width of the simulated \nz of the SOMPZ and SOMPZ+WZ methods in each tomographic bin, and of the true \nz, together with the respective statistical distances from the truth.

\begin{table*}
    \centering
    \begin{tabular}{c l c c c c c c} 
        
          & & Bin 1 & Bin 2 & Bin 3 & Bin 4 & Bin 5 & Bin 6\\
         & & z $\in$ [0.2, 0.4] &  z $\in$ [0.4, 0.55] &  z $\in$ [0.55, 0.7] &  z $\in$ [0.7, 0.85] &  z $\in$ [0.85, 0.95] &  z $\in$ [0.95, 1.05] \\
      
         % & & & Mean \\
        \hline
        \hline
        <z> & SOMPZ & 0.319 $\pm\ $ 0.009 & 0.484 $\pm\ $ 0.007 & 0.623 $\pm\ $ 0.006 & 0.784 $\pm\ $ 0.006 & 0.891 $\pm\ $ 0.007 & 0.993 $\pm\ $ 0.010\\

        & SOMPZ+WZ &  0.313 $\pm\ $ 0.008 &  0.466 $\pm\ $ 0.006 &  0.613 $\pm\ $ 0.005 &  0.774 $\pm\ $ 0.007 &  0.876 $\pm\ $ 0.007 &  0.988 $\pm\ $ 0.007\\
        \hline
        $\Delta_{<z>}$ & SOMPZ &  1.46 & 2.55 & 0.20 & 0.28 & 0.55 & 1.08\\
        & SOMPZ+WZ & 0.83& 0.42& 1.81& 1.28& 2.49& 2.10\\
        %0.0145 & 0.0100 & 0.0086 & 0.0084& 0.0093 & 0.0115\\
        \\
        % & & & Width \\

        \hline
        \hline
        $\sigma_z$ & SOMPZ & 0.075 $\pm\ $ 0.010 & 0.064 $\pm\ $ 0.007 & 0.062 $\pm\ $ 0.006 & 0.056 $\pm\ $ 0.005 & 0.060 $\pm\ $ 0.005 & 0.068 $\pm\ $ 0.007 \\
        & SOMPZ + WZ  &  0.077 $ \pm\ $ 0.005 &  0.057 $ \pm\ $ 0.005 &  0.064 $ \pm\ $ 0.004 &  0.068 $ \pm\ $ 0.005 &  0.064 $ \pm\ $ 0.005 &  0.060 $ \pm\ $ 0.003\\
        \hline
        
        $\Delta_{\sigma_z}$ & SOMPZ &  0.53 & 0.59 & 1.17 & 1.42 & 0.79 & 0.10\\
         & SOMPZ+WZ &  0.46 & 2.08 & 2.13 & 0.93 & 1.50 & 2.09\\
    \end{tabular}
    \caption{SIMULATIONS: Summary of values for center values for mean (top panel) and width (bottom panel) for the \nz\ distributions as measured in the Buzzard simulations. The values related to SOMPZ and SOMPZ+WZ refer to Figure \ref{fig:nz_ensemble_buzz}. Note that the uncertainties quoted here only include sample variance and shot noise.  }
    \label{tab:buzz}
 \end{table*}

We also repeated in simulations the same procedure as for data also for the WZ estimates. We created a mock BOSS/eBOSS catalog to use as a reference sample. As in data, also in simulations the BOSS/eBOSS sample is divided into 50 bins spanning the $0.1<z<1.1$ range of the catalog (width $\Delta z = 0.02$). Before proceeding with combining the SOMPZ and WZ information through the combined likelihood, the compatibility between SOMPZ and WZ was checked. This was tested by inferring the windowed means and widths of the WZ and SOMPZ redshift estimates, following \cite*{y3-sourcewz}. The window has been determined such that magnification effects related to the WZ measurements can be neglected. As for WZ, we used a ``simple'' estimator for the redshift distribution, inverting Eq. \ref{wursys} and ignoring magnification effects (this is possible as we are considering only windowed quantities). The means and widths computed in this way for the two methods were compatible within statistical (and systematic) errors, hence the SOMPZ and WZ could be combined together.

The posterior obtained in simulations from multiplying the two likelihoods is shown in Figure \ref{fig:nz_ensemble_buzz}, in which the effect of the combination immediately stands out: the additional information from clustering redshifts places a tight constraint on the shape of the \nz, while still being in agreement with the true distribution. This larger constraining power derives from the fact that in clustering the number density for each redshift bin correlates across neighbouring bins, which restrains the joint likelihood to prefer smoother realisations and reject the ones with more uncorrelated values of clustering.\\

As the second phase of the validation process, a full 2x2pt cosmological analysis was performed. We utilised the datavector consisting of the two point measurements from the Buzzard simulations and the redshift distributions obtained from the SOMPZ+WZ method, obtained as described in the previous paragraph. We considered both $\Lambda$CDM and $w$CDM models, fixing magnification parameters and including all 6 \maglim\ tomographic bins. Additionally, we fixed the source galaxies redshift distributions, to ensure any deviation from the true parameter values of the simulation would be caused by the lens \nz alone. 
The mean values of $S_8$, $\Omega_{\rm m}$ (and $w$), with their respective $68\%$ confidence intervals, are:
\begin{itemize}
    \item $\Lambda$CDM: $S_8$ = 0.73 $\pm$ 0.18, $\Omega_{\rm m}$ = 0.31 $\pm$ 0.07;
    \item $w$CDM: $S_8$ =  0.71 $\pm$ 0.18, $\Omega_{\rm m}$ = 0.30 $\pm$ 0.08,  $w$ = -1.3 $\pm$ 0.4.
\end{itemize}

\begin{figure}
    \centering
    \includegraphics[width=\linewidth]{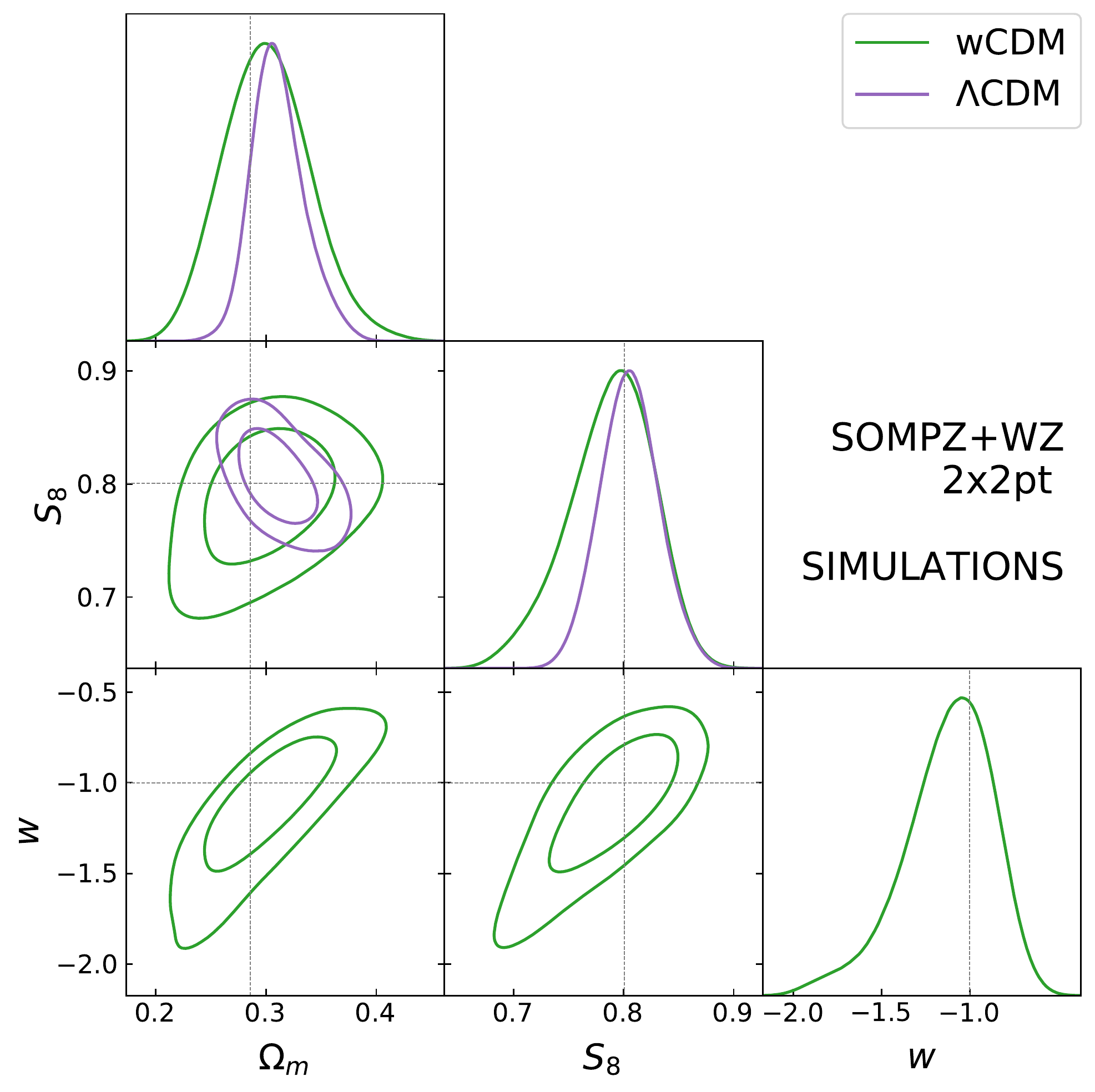}
    \caption{Posterior distributions of the cosmological parameters $\Omega_{\rm m}$, $S_8$, and $w$ for the $\Lambda$CDM and $w$CDM analyses. These have been run with 6 bins and fixed magnification parameters.}
    \label{fig:buzz_post}
\end{figure}

For both analyses, the posterior distributions successfully recovered the input parameters (see Section \ref{sec:data}), as displayed in Figure \ref{fig:buzz_post}.

% \begin{figure*}
% \centering
% \begin{subfigure}{.5\textwidth}
%     \centering
%     \includegraphics[width=\linewidth]{Figures/sompz_wz_means.png}
%   \label{fig:sub1}
% \end{subfigure}%
% \begin{subfigure}{.5\textwidth}
%     \centering
%     \includegraphics[width=\linewidth]{Figures/sompz_wz_nz.png}
%   \label{fig:sub2}
% \end{subfigure}
% \caption{A figure with two subfigures}
% \label{fig:compatibility}
% \end{figure*}

%We could exploit the information from its auto-correlation to limit the systematic uncertainties related to the redshift evolution of the galaxy-matter bias, by including it as a prior of the Sys function, as described in Section \ref{sec:wzunc}. The only difference between the data of course is that we used as center of the prior values not the value computed in data, but in Buzzard. We show the estimated galaxy-matter bias in simulations for the tomographic bin 4 of the \maglim\ sample in sub-figure a) of Figure Fig.~\ref{fig:autocorr_maglim}.

%With respect to the source galaxies, we can notice a evident difference: WZ is also able to reduce the uncertainty on the mean for the case of \maglim. 
%\giulia{could this be due to the fact that we correct for the autocorr? check in buzzard the unc on mean and witdh, with and without correcting for the auto-correlation.}
%The 3sDIR+WZ realisations, which account for shot noise and sample variance, are in agreement with the true n(z) distributions, thus confirming the solidity of our whole pipeline. 

\section{Cosmological parameters}\label{app:params}

In Table \ref{tab:parameters} are listed all the cosmological parameters included in our fiducial analysis.

\begin{table}
	%		\small
	\centering	
	\caption{The parameters and their priors used in the
		fiducial  \maglim $\Lambda$CDM and $w$CDM analyses. The parameter $w$ is fixed to $-1$ in $\Lambda$CDM.  Square brackets denote a flat prior, while parentheses denote a Gaussian prior of the form $\mathcal{N}(\mu,\sigma)$.} 
	\vspace{-0.2cm}
	\begin{tabular}{ccc}
		\hline
		Parameter & Fiducial & Prior\\\hline
		\multicolumn{3}{c}{\textbf{Cosmology}} \\
		$\Omega_{\rm m}$ &  0.3 &[0.1, 0.9] \\ 
		$A_{\rm s} 10^{9}$ & 2.19 & [$0.5$, $5.0$]  \\ 
		$n_{\rm s}$ & 0.97 & [0.87, 1.07]  \\
		$w$ &  -1.0  &[-2, -0.33]   \\
		$\Omega_{\rm b}$ & 0.048 &[0.03, 0.07]  \\
		$h_0$  & 0.69  &[0.55, 0.91]   \\
		$\Omega_\nu h^2 10^3$ & 0.83 & [0.6, 6.44]
		\\\hline

		\multicolumn{3}{c}{\textbf{Linear galaxy bias  } }	 \\
		$b_{i}$  & $1.5, 1.8, 1.8, 1.9, 2.3, 2.3$ & [0.8,3.0]\\\hline
		
% 		\multicolumn{3}{c}{\textbf{ Non-linear galaxy bias  } }\Tstrut	 \\
% 		$b_1^{i}\sigma_8 $  & $1.43, 1.43, 1.43, 1.69, 1.69, 1.69 $ & [0.67,3.0]\\
% 		$b_2^{i}\sigma_8^2 $  & $0.16, 0.16, 0.16, 0.36, 0.36, 0.36 $ & [-4.2, 4.2]\\\hline
		
		\multicolumn{3}{c}{\textbf{Lens
				magnification } }  \\
		$C_{1} $ & 0.43 & (0.43, 0.51)\\ 
		$C_{2} $ & 0.30 & (0.30, 0.48)\\
		$C_{3} $ & 1.75 & (1.75, 0.39) \\
		$C_{4} $ &  1.94 & (1.94,  0.35)\\
	    $C_{5} $ &  1.56  & (1.56, 0.71) \\
		$C_{6} $ &  2.96 & (2.96,  0.95) \\
		\hline

		\multicolumn{3}{c}{\textbf{Lens photo-z}}	 \\
		$\Delta z^1_{\rm l}$ & 0.0 & ($
		0.0, 0.0164$)\\ $\Delta z^2_{\rm l}$ & 0.0 & ($
		0.0, 0.0100 $)\\ $\Delta z^3_{\rm l}$ & 0.0 & ($
		0.0, 0.0085 $)\\ $\Delta z^4_{\rm l}$ & 0.0 & ($
		0.0, 0.0084 $)\\ $\Delta z^5_{\rm l}$ & 0.0 & ($
		0.0, 0.0094 $)\\ $\Delta z^6_{\rm l}$ & 0.0 & ($
		0.0, 0.0116 $)\\ $\sigma z^1_{\rm l}$ & 1.0 & ($
		1.0, 0.0639 $)\\ $\sigma z^2_{\rm l}$ & 1.0 & ($
		1.0, 0.0624 $)\\ $\sigma z^3_{\rm l}$ & 1.0 & ($
		1.0, 0.0315 $)\\ $\sigma z^4_{\rm l}$ & 1.0 &
		($1.0, 0.0409 $)\\ $\sigma z^5_{\rm l}$ & 1.0 &
		($1.0, 0.0515 $)\\ $\sigma z^6_{\rm l}$ & 1.0 &
		($1.0, 0.0650 $) \\\hline  
		
		\multicolumn{3}{c}{{\bf
				Intrinsic alignment}} \\
		$a_{i}$ ($i\in [1,2]$)   & 0.7, -1.36 &  [$-5,5$ ]\\
		$\eta_{i}$  ($i\in [1,2]$) & -1.7, -2.5  & [$-5,5$ ]\\
		$b_{\mathrm{TA}}$   & 1.0  & [$0,2$] \\
		$z_0$ & 0.62   &  Fixed\\
		\hline
		\multicolumn{3}{c}{{\bf Source photo-z}} \\
		$\Delta z^1_{\rm s}$  & 0.0  & ($0.0, 0.018$) \\
		$\Delta z^2_{\rm s}$  & 0.0  & ($0.0, 0.013$) \\
		$\Delta z^3_{\rm s}$  & 0.0  & ($0.0, 0.006$) \\
		$\Delta z^4_{\rm s}$  & 0.0  & ($0.0, 0.013$) \\
		\hline
		\multicolumn{3}{c}{{\bf Shear calibration}} \\
		$m^1$ & -0.006  & ($-0.006, 0.008$)\\
		$m^2$ & -0.010  & ($-0.010, 0.013$)\\
		$m^3$ & -0.026  & ($-0.026, 0.009$)\\
		$m^4$ & -0.032  & ($-0.032, 0.012$)\\
		\hline
	\end{tabular}\label{tab:parameters}
\end{table}

%  alpha_1 = gaussian 1.22 0.256
% alpha_2 = gaussian 1.15 0.240
% alpha_3 = gaussian 1.875 0.196
% alpha_4 = gaussian 1.97 0.173
% alpha_5 = gaussian 1.78 0.357
% alpha_6 = gaussian 2.48 0.475

% \section{Impact of using the informative prior on the galaxy-matter bias from the \maglim\ auto-correlation}\label{app:autocorr}

% We tested the impact on the $\LambdaCDM$ cosmological parameters of using the same broad prior on the $Sys(\textbf{s})$ function describing the galaxy-matter bias as was done for the WL sample \citep{y3-sourcewz}. In this work we used more informative values computed from the clustering auto-correlation of the \maglim\ sample. In Section \ref{sec:wzunc} was explained how the 

\section{Redshift uncertainties sampling strategy}\label{sampling}

How redshift uncertainties are propagated in the cosmological analysis can have an impact on the final result. In this section we discuss different strategies to marginalise over the redshift uncertainties of our sample during the cosmological inference. Because we have can rely on a full ensemble of \nz\ shapes capturing our redshift uncertainties, we can compare three different sampling methods: 

\begin{itemize}
    \item \textbf{Shift}: we compress the realisations by computing their average, and marginalise over a shift on the mean;
    \item \textbf{Shift and stretch}: we compress the realisations by computing their average, and marginalise over both a shift on the mean and on a stretch on the width;
    \item \textbf{Full shape}: we provide as input all the produced realisations and we rank them by one of their properties using the \textit{Hyperrank} method \citep{y3-hyperrank}, marginalising over the full shape of the distributions.
\end{itemize}

\begin{figure}
    \centering
    \includegraphics[width=\linewidth]{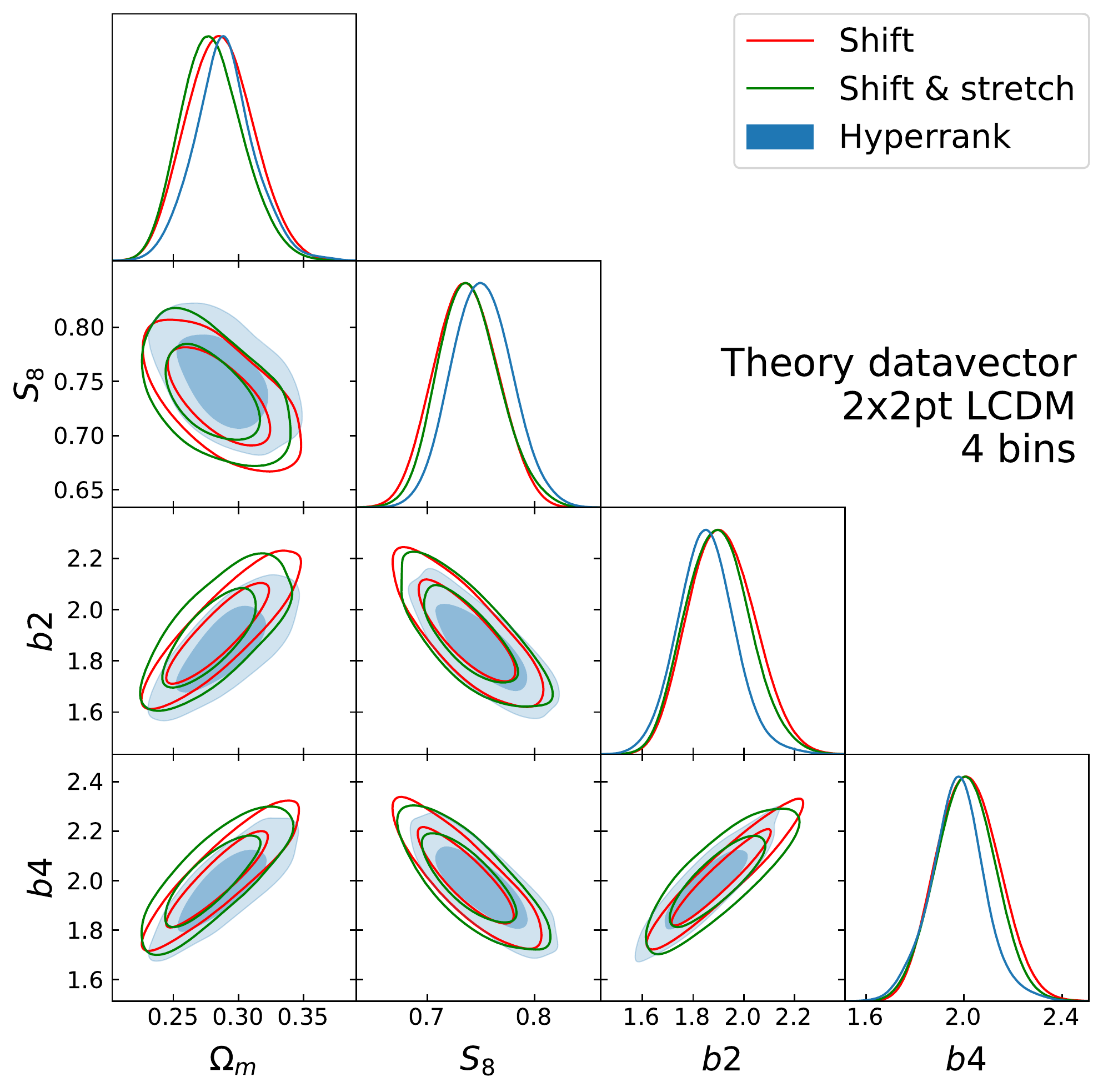}
    \caption{Posterior distributions of the cosmological parameters $\Omega_{\rm m}$, $S_8$, and two out of four of the galaxy-matter biases ($b_2$, $b_4$) for the $\Lambda$CDM analysis involving 4 bins and fixed magnification parameters. These analyses have been obtained assuming a theoretical datavector and adopting different marginalisation schemes on the redshift distribution of the lens sample.}
    \label{fig:biases_4b}
\end{figure}

Using only \textit{shifts} is the methodology usually adopted to model redshift uncertainties in weak lensing sample, as the weak lensing kernel is mostly sensitive to the mean of the redshift distributions. On the other hand, clustering and galaxy-galaxy lensing measurements are also very sensitive to the width of the lens redshift distributions; therefore, the \textit{shift and stretch} approach is preferred. The full shape marginalisation, in theory, is more accurate, because it accounts for the uncertainties in the higher order moments of the distribution; however, depending on the science case, it might not make a huge impact on the final constraints. The full shape marginalisation is implemented via hyperrank \citep{y3-hyperrank}, which is an algorithm that orders realisations of the ensemble according to a parameter, which facilitates the sampling and marginalization over the \nz\ ensemble within the cosmological likelihood Markov chains. Hyperrank was also implemented for the WL sources, although it had a negligible impact on the results. The quantity chosen for the ranking in that case was the mean. We decided for this case it would be more appropriate to perform the optimised ranking of the realisation by the $68\%$ sigma rather than the mean, and we tested it indeed improved the performance of the sampling.
%Since the mean and width of the \nz\ estimated in this work are uncorrelated
To test the different sampling strategies, we built a synthetic noiseless data vector based on theory predictions at fixed cosmology and we used as \nz\ the realisations average of the SOMPZ+WZ estimates in data. We then marginalised over redshift uncertainties using the three approaches aforementioned. We performed this test both using 4 or 6 lens bins, although here we are just going to show the posteriors obtained with 4 bins as they are not qualitatively different from the ones with 6 bins. The results of this test are shown in Figure \ref{fig:biases_4b}, where we show the posterior of $\sigma_8$, $\Omega_{\rm m}$ and for sake of simplicity, two out of the four galaxy-matter linear biases.

 Focusing on the shift and shift+stretch contours, one can notice that the width of the contour in the direction perpendicular to the degeneration axis is larger for the shift+stretch. This is related to impact of the additional marginalisation over the width of the distributions. One caveat is that in our marginalisation scheme (as adopted in the main DES Y3 2x2pt analysis), we are implicitly neglecting correlations between the uncertainties in the mean and widths of the distributions, which usually show a certain degree of correlation (from $\sim10\%$ to $\sim30\%$, depending from the tomographic bin). These are neglected, which might translate in a slight overestimation of our constraints. When marginalising over the uncertainties using the hyperrank framework, on the other hand, such correlations are implicitly accounted for.  Indeed, one can notice that the hyperrank posteriors are slightly tighter than the shift or shift-stretch posteriors.

%
%\begin{figure}
%    \centering
%    \includegraphics[width=\linewidth]{Figures/omegamsigma8_6bins.png}
%    \caption{Omega M vs sigma 8, 6 bins, LCDM}
%    \label{fig:om_sigma8_4bins}
%\end{figure}

Unfortunately, we did not manage to successfully apply hyperrank to the data. When performing the cosmological analysis on data using hyperrank, we found significantly less smooth posteriors compared to our tests on simulations. A similar behaviour has also been found when applying hyperrank to the DES Y3 source sample \cite{y3-cosmicshear1}, and it has been interpreted as a consequence of a possible larger degree of complexity of the redshift distributions of our data compared to simulations. We attempted both to artificially smooth our \nz\ and to increase the number of samples from the SOMPZ+WZ method, without reaching a satisfactory level. Due to the very high computational cost of running a cosmological chain using hyperrank, we could only test a few different levels of smoothing before deciding to abandon hyperrank for the present work, and choose the shift+stretch as photo-$z$ uncertainty marginalisation methodology. For DES Y6, we plan to apply several tools that will speed up our cosmological inference, enabling more tests on hyperrank, which has great potential and whose implementation is a goal for the DES Y6 analysis.

\subsection{Cosmological constraints with clipped \nz\ tails}\label{sect:notails}
\begin{figure}
    \centering
    \includegraphics[width=\linewidth]{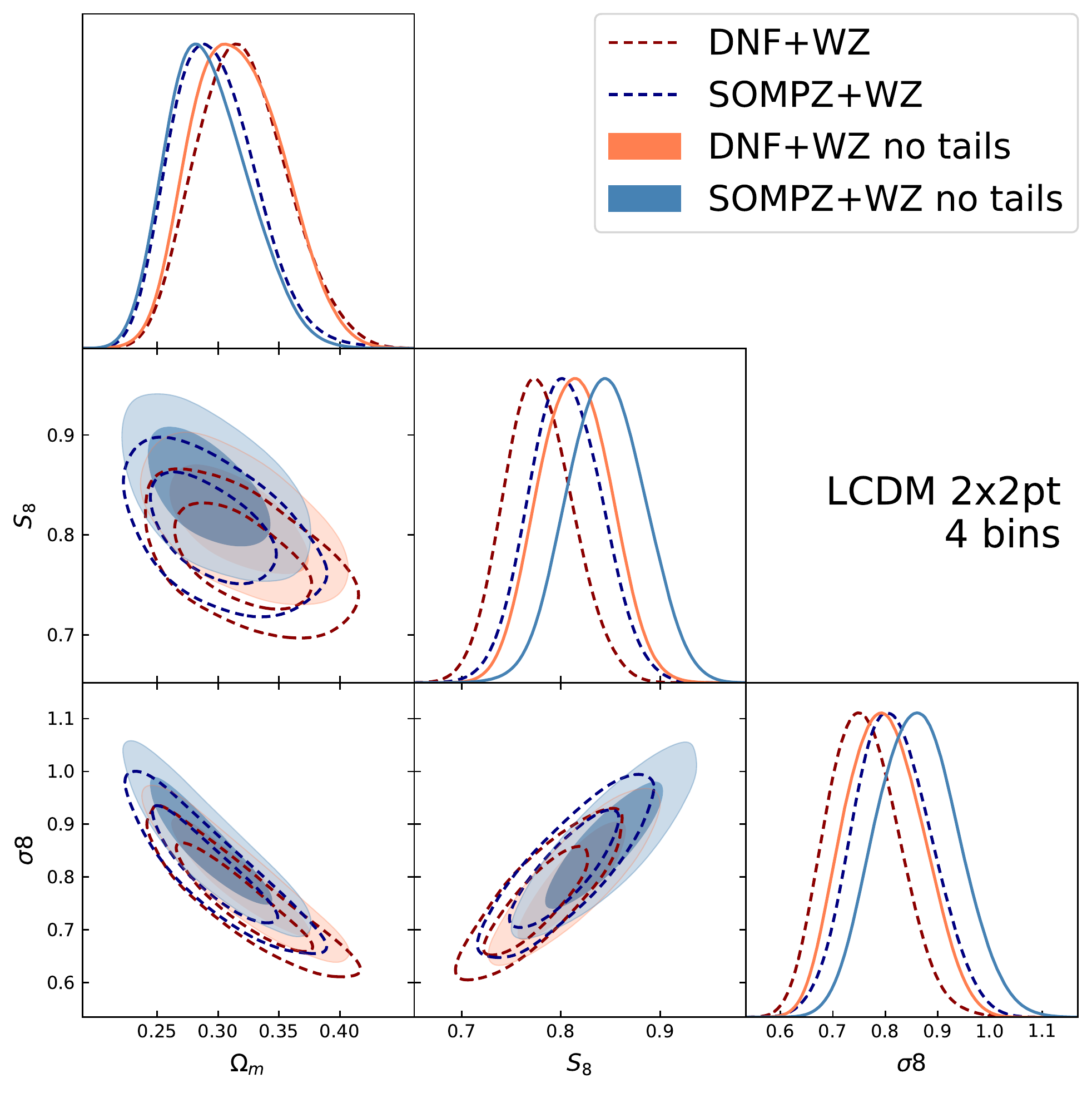}
    \caption{Same as the left panel of Fig. \ref{fig:contours}, but with two additional posteriors overplotted representing the constraints obtained using the redshift distributions with ``clipped'' tails.}
    \label{fig:notails}
\end{figure}

Here we test whether the difference between DNF+WZ and SOMPZ+WZ constraints (Fig. \ref{fig:contours}) were only due to the different treatment of redshift outliers and of the tails of the redshift distributions. We artificially removed the tails from the DNF+WZ and SOMPZ+WZ \nz\, (i.e., we set the distributions to zero), and repeated our cosmological analysis. We used as definition of the tails the same interval used to calibrate the DNF distribution with the WZ constraints adopted in \cite{y3-2x2ptaltlensresults}. Results for the $\Lambda$CDM case, 4 bins and fixed magnification are shown in Fig. \ref{fig:notails}. By removing the tails, both posteriors are shifted, which means that the calibration of the tails of the redshift distribution is important for our cosmological analysis. Since the two posteriors are shifted but they still do not overlap, we can assume that the differences in the bulk of the redshift distributions inferred by two methods is also crucially driving the differences at the constraints level seen in Fig. \ref{fig:contours}.

%%%%%%%%%%%%%%%%%%%%%%%%%%%%%%%%%%%%%%%%%%%%%%%%%%

% Don't change these lines
\bsp	% typesetting comment
\section*{Affiliations}
$^{1}$ Department of Astronomy and Astrophysics, University of Chicago, Chicago, IL 60637, USA\\
$^{2}$ Kavli Institute for Cosmological Physics, University of Chicago, Chicago, IL 60637, USA\\
$^{3}$ Institut de F\'{\i}sica d'Altes Energies (IFAE), The Barcelona Institute of Science and Technology, Campus UAB, 08193 Bellaterra (Barcelona) Spain\\
$^{4}$ Argonne National Laboratory, 9700 South Cass Avenue, Lemont, IL 60439, USA\\
$^{5}$ Department of Physics and Astronomy, University of Pennsylvania, Philadelphia, PA 19104, USA\\
$^{6}$ Center for Cosmology and Astro-Particle Physics, The Ohio State University, Columbus, OH 43210, USA\\
$^{7}$ Department of Physics, The Ohio State University, Columbus, OH 43210, USA\\
$^{8}$ Institut d'Estudis Espacials de Catalunya (IEEC), 08034 Barcelona, Spain\\
$^{9}$ Institute of Space Sciences (ICE, CSIC),  Campus UAB, Carrer de Can Magrans, s/n,  08193 Barcelona, Spain\\
$^{10}$ Physics Department, William Jewell College, Liberty, MO, 64068\\
$^{11}$ Department of Physics, Stanford University, 382 Via Pueblo Mall, Stanford, CA 94305, USA\\
$^{12}$ Kavli Institute for Particle Astrophysics \& Cosmology, P. O. Box 2450, Stanford University, Stanford, CA 94305, USA\\
$^{13}$ SLAC National Accelerator Laboratory, Menlo Park, CA 94025, USA\\
$^{14}$ Fermi National Accelerator Laboratory, P. O. Box 500, Batavia, IL 60510, USA\\
$^{15}$ Institute of Astronomy, University of Cambridge, Madingley Road, Cambridge CB3 0HA, UK\\
$^{16}$ Kavli Institute for Cosmology, University of Cambridge, Madingley Road, Cambridge CB3 0HA, UK\\
$^{17}$ Department of Physics, University of Michigan, Ann Arbor, MI 48109, USA\\
$^{18}$ Institute for Astronomy, University of Hawai'i, 2680 Woodlawn Drive, Honolulu, HI 96822, USA\\
$^{19}$ Physics Department, 2320 Chamberlin Hall, University of Wisconsin-Madison, 1150 University Avenue Madison, WI  53706-1390\\
$^{20}$ Department of Physics, Northeastern University, Boston, MA 02115, USA\\
$^{21}$ Instituto de F\'{i}sica Te\'orica, Universidade Estadual Paulista, S\~ao Paulo, Brazil\\
$^{22}$ Laborat\'orio Interinstitucional de e-Astronomia - LIneA, Rua Gal. Jos\'e Cristino 77, Rio de Janeiro, RJ - 20921-400, Brazil\\
$^{23}$ Department of Physics, Carnegie Mellon University, Pittsburgh, Pennsylvania 15312, USA\\
$^{24}$ Instituto de Astrofisica de Canarias, E-38205 La Laguna, Tenerife, Spain\\
$^{25}$ Universidad de La Laguna, Dpto. Astrofísica, E-38206 La Laguna, Tenerife, Spain\\
$^{26}$ Center for Astrophysical Surveys, National Center for Supercomputing Applications, 1205 West Clark St., Urbana, IL 61801, USA\\
$^{27}$ Department of Astronomy, University of Illinois at Urbana-Champaign, 1002 W. Green Street, Urbana, IL 61801, USA\\
$^{28}$ California Institute of Technology, 1200 East California Blvd, MC 249-17, Pasadena, CA 91125, USA\\
$^{29}$ Jodrell Bank Center for Astrophysics, School of Physics and Astronomy, University of Manchester, Oxford Road, Manchester, M13 9PL, UK\\
$^{30}$ Centro de Investigaciones Energ\'eticas, Medioambientales y Tecnol\'ogicas (CIEMAT), Madrid, Spain\\
$^{31}$ Lawrence Berkeley National Laboratory, 1 Cyclotron Road, Berkeley, CA 94720, USA\\
$^{32}$ NSF AI Planning Institute for Physics of the Future, Carnegie Mellon University, Pittsburgh, PA 15213, USA\\
$^{33}$ Jet Propulsion Laboratory, California Institute of Technology, 4800 Oak Grove Dr., Pasadena, CA 91109, USA\\
$^{34}$ Department of Astronomy, University of California, Berkeley,  501 Campbell Hall, Berkeley, CA 94720, USA\\
$^{35}$ Department of Astronomy/Steward Observatory, University of Arizona, 933 North Cherry Avenue, Tucson, AZ 85721-0065, USA\\
$^{36}$ Departments of Statistics and Data Science, University of Texas at Austin, Austin, TX 78757, USA\\
$^{37}$ University Observatory, Faculty of Physics, Ludwig-Maximilians-Universit\"at, Scheinerstr. 1, 81679 Munich, Germany\\
$^{38}$ Observat\'orio Nacional, Rua Gal. Jos\'e Cristino 77, Rio de Janeiro, RJ - 20921-400, Brazil\\
$^{39}$ School of Physics and Astronomy, Cardiff University, CF24 3AA, UK\\
$^{40}$ Department of Astronomy, University of Geneva, ch. d'\'Ecogia 16, CH-1290 Versoix, Switzerland\\
$^{41}$ Department of Physics \& Astronomy, University College London, Gower Street, London, WC1E 6BT, UK\\
$^{42}$ Department of Physics and Astronomy, Pevensey Building, University of Sussex, Brighton, BN1 9QH, UK\\
$^{43}$ Department of Applied Mathematics and Theoretical Physics, University of Cambridge, Cambridge CB3 0WA, UK\\
$^{44}$ Perimeter Institute for Theoretical Physics, 31 Caroline St. North, Waterloo, ON N2L 2Y5, Canada\\
$^{45}$ Brookhaven National Laboratory, Bldg 510, Upton, NY 11973, USA\\
$^{46}$ Department of Physics, Duke University Durham, NC 27708, USA\\
$^{47}$ George P. and Cynthia Woods Mitchell Institute for Fundamental Physics and Astronomy, and Department of Physics and Astronomy, Texas A\&M University, College Station, TX 77843,  USA\\
$^{48}$ Cerro Tololo Inter-American Observatory, NSF's National\\ Optical-Infrared Astronomy Research Laboratory, Casilla 603, La Serena, Chile\\
$^{49}$ Institute of Cosmology and Gravitation, University of Portsmouth, Portsmouth, PO1 3FX, UK\\
$^{50}$ CNRS, UMR 7095, Institut d'Astrophysique de Paris, F-75014, Paris, France\\
$^{51}$ Sorbonne Universit\'es, UPMC Univ Paris 06, UMR 7095, Institut d'Astrophysique de Paris, F-75014, Paris, France\\
$^{52}$ Astronomy Unit, Department of Physics, University of Trieste, via Tiepolo 11, I-34131 Trieste, Italy\\
$^{53}$ INAF-Osservatorio Astronomico di Trieste, via G. B. Tiepolo 11, I-34143 Trieste, Italy\\
$^{54}$ Institute for Fundamental Physics of the Universe, Via Beirut 2, 34014 Trieste, Italy\\
$^{55}$ Hamburger Sternwarte, Universit\"{a}t Hamburg, Gojenbergsweg 112, 21029 Hamburg, Germany\\
$^{56}$ Department of Physics, IIT Hyderabad, Kandi, Telangana 502285, India\\
$^{57}$ Institute of Theoretical Astrophysics, University of Oslo. P.O. Box 1029 Blindern, NO-0315 Oslo, Norway\\
$^{58}$ Instituto de Fisica Teorica UAM/CSIC, Universidad Autonoma de Madrid, 28049 Madrid, Spain\\
$^{59}$ Department of Astronomy, University of Michigan, Ann Arbor, MI 48109, USA\\
$^{60}$ School of Mathematics and Physics, University of Queensland,  Brisbane, QLD 4072, Australia\\
$^{61}$ Santa Cruz Institute for Particle Physics, Santa Cruz, CA 95064, USA\\
$^{62}$ Center for Astrophysics $\vert$ Harvard \& Smithsonian, 60 Garden Street, Cambridge, MA 02138, USA\\
$^{63}$ Australian Astronomical Optics, Macquarie University, North Ryde, NSW 2113, Australia\\
$^{64}$ Lowell Observatory, 1400 Mars Hill Rd, Flagstaff, AZ 86001, USA\\
$^{65}$ Centre for Gravitational Astrophysics, College of Science, The Australian National University, ACT 2601, Australia\\
$^{66}$ The Research School of Astronomy and Astrophysics, Australian National University, ACT 2601, Australia\\
$^{67}$ Departamento de F\'isica Matem\'atica, Instituto de F\'isica, Universidade de S\~ao Paulo, CP 66318, S\~ao Paulo, SP, 05314-970, Brazil\\
$^{68}$ Department of Astrophysical Sciences, Princeton University, Peyton Hall, Princeton, NJ 08544, USA\\
$^{69}$ Instituci\'o Catalana de Recerca i Estudis Avan\c{c}ats, E-08010 Barcelona, Spain\\
$^{70}$ School of Physics and Astronomy, University of Southampton,  Southampton, SO17 1BJ, UK\\
$^{71}$ Computer Science and Mathematics Division, Oak Ridge National Laboratory, Oak Ridge, TN 37831\\
$^{72}$ Institute of Astronomy, University of Cambridge, Madingley Road, Cambridge CB3 0HA, UK\\

\label{lastpage}

\end{document}